\documentclass[a4paper,10pt]{article}

\usepackage[pdftex]{color,graphicx,hyperref}
\usepackage{amsmath,amsfonts,amssymb}
\usepackage{fullpage}
\usepackage{setspace}
\usepackage{booktabs}
\usepackage[labelfont=bf,labelsep=period]{caption}
\usepackage{subcaption}
\usepackage{cite}
\usepackage[utf8]{inputenc}
\usepackage[affil-it]{authblk}

\newcommand{\eref}[1]{Eq.~(\ref{#1})}
\newcommand{\esref}[2]{Eqs.~(\ref{#1})-(\ref{#2})}

\newcommand{\fref}[1]{Fig.~\ref{#1}}

\newcommand{\srefsi}[1]{SI Section~S{#1}}
\newcommand{\frefsi}[1]{SI Fig.~S{#1}}
\newcommand{\fsrefsi}[2]{SI Figs.~S{#1}-S{#2}}
\newcommand{\trefsi}[1]{SI Table~S{#1}}
\newcommand{\tsrefsi}[2]{SI Tables~S{#1}-S{#2}}

\graphicspath{ {./figures/} }

\def\ptau{\tau} 
\def\pnu{\nu} 
\def\p0{p} 
\def\prtau{\tau_r} 
\def\prnu{\nu_r} 
\def\mod{\dot{o}} 
\def\Prxt{P_{x, t}} 
\def\Lt{L_t} 
\def\Drxt{D_{x, t}} 
\def\Cr{C} 
\def\Wlevy{W_{\mathrm{levy}} } 
\def\Wdiff{W_{\mathrm{diff}} } 
\def\Wrepl{W_{\mathrm{repl}} } 

\title{Dynamics of ranking}

\author[1,2,3,*]{Gerardo I\~{n}iguez}
\author[4]{Carlos Pineda}
\author[3,5,6]{Carlos Gershenson}
\author[7,8,1,*]{Albert-László Barabási}

\affil[1]{\small{Department of Network and Data Science, Central European University, 1100 Vienna, Austria}}
\affil[2]{\small{Department of Computer Science, Aalto University School of Science, 00076 Aalto, Finland}}
\affil[3]{\small{Centro de Ciencias de la Complejidad, Universidad Nacional Auton\'{o}ma de M\'{e}xico, 04510 Ciudad de México, Mexico}}
\affil[4]{\small{Instituto de Física, Universidad Nacional Auton\'{o}ma de M\'{e}xico, 04510 Ciudad de México, Mexico}}
\affil[5]{\small{Instituto de Investigaciones en Matemáticas Aplicadas y en Sistemas, Universidad Nacional Auton\'{o}ma de M\'{e}xico, 04510 Ciudad de México, Mexico}}
\affil[6]{\small{Lakeside Labs GmbH, Lakeside Park B04, 9020 Klagenfurt am Wörthersee, Austria}}
\affil[7]{\small{Network Science Institute, Center for Complex Network Research \& Department of Physics, Northeastern University, 02115 Boston, MA, USA}}
\affil[8]{\small{Channing Division of Network Medicine \& Department of Medicine, Brigham and Women’s Hospital, Harvard Medical School, 02215 Boston, MA, USA}}
\affil[*]{\small{Corresponding author email: iniguezg@ceu.edu, alb@neu.edu}}

\date{}

\begin{document}
\maketitle
\begin{abstract} 
Virtually anything can be and is ranked; people,  institutions,  countries,  words,  genes. Rankings reduce complex systems to ordered lists,  reflecting the ability of their elements to perform relevant functions, and are being used from socioeconomic policy to knowledge extraction. A century of research has found regularities when temporal rank data is aggregated.  Far less is known, however, about how rankings change in time.  Here we explore the dynamics of 30 rankings in natural, social, economic, and infrastructural systems, comprising millions of elements and timescales from minutes to centuries. We find that the flux of new elements determines the stability of a ranking: for high flux only the top of the list is stable,  otherwise top and bottom are equally stable. We show that two basic mechanisms --- displacement and replacement of elements --- capture empirical ranking dynamics. The model uncovers two regimes of behavior; fast and large rank changes,  or slow diffusion. Our results indicate that the balance between robustness and adaptability in ranked systems might be governed by simple random processes irrespective of system details.
\end{abstract}

\section*{Introduction}

Rankings are everywhere. From country development indices, academic indicators
and candidate poll numbers to music charts and sports scoreboards, rankings are
key to how humans measure and make sense of the
world~\cite{erdi2019ranking,langville2012s}. The ubiquity of rankings stems
from the generality of their definition: Reducing the (often high-dimensional)
complexity of a system to a few or even a single measurable quantity of
interest~\cite{diamond1997guns,turchin2018quantitative}, dubbed score, leads to
an ordered list where elements are ranked, typically from highest to lowest
score. Rankings are, in this sense, a proxy of relevance or fitness to perform
a function in the system. Rankings are being used to identify the most
accomplished individuals or institutions, and to find the essential
pieces of knowledge or infrastructure in society~\cite{erdi2019ranking}. Since
rankings often determine who gets access to resources (education,
jobs, funds), they play a role in the formation of social
hierarchies~\cite{posfai2018talent,kawakatsu2021emergence} and the potential
rise of systematic inequality~\cite{clauset2015systematic}.

The statistical properties of ranking lists have caught the attention of
natural and social scientists for more than a century. A heavy-tailed decay of
score with rank, commonly known as Zipf's
law~\cite{zipf1949human,newman2005power}, has been systematically observed in
the ranking of cities by population~\cite{auerbach1913gesetz,rosen1980size},
words and phrases by frequency of
use~\cite{booth1967law,ha2002extension,cancho2003least,corominas2011emergence,dodds2011temporal,cocho2015rank},
companies by size~\cite{lucas1978size,stanley1996scaling,axtell2001zipf}, and
many features of the Internet~\cite{adamic2002zipf}. Zipf's law appears even in
the score-rank distributions of natural
systems, such as earthquakes~\cite{ogata1993analysis,sornette1996rank}, DNA
sequences~\cite{mantegna1995systematic}, and metabolic
networks~\cite{wagner2001small}. Rankings have also proven useful when
analyzing productivity and impact in science and the
arts~\cite{radicchi2009diffusion, clauset2015systematic, sinatra2016quantifying, fraiberger2018quantifying, janosov2020elites},
in human urban
mobility~\cite{gonzalez2008understanding,noulas2012tale,alessandretti2018evidence},
epidemic spreading by influentials~\cite{gu2017ranking}, and the development
pathways of entire countries~\cite{hidalgo2009building}. Recently, studies of
language use~\cite{cocho2015rank,morales2018rank}, sports
performance~\cite{morales2016generic} and many biological and socioeconomic
rankings~\cite{martinez2009universality} have strengthened the notion of
universality suggested by Zipf's law:
Despite microscopic differences in elements, scores, and types of
interaction, the aggregate, macroscopic properties of ranking lists are
remarkably similar throughout nature and society.

The similarity of score-rank distributions across systems raises the question
of the existence of simple generative mechanisms behind
them. While mechanisms of proportional growth~\cite{simon1955class},
cumulative advantage~\cite{price1976general}, and preferential
attachment~\cite{barabasi1999emergence} are often used to explain the
heavy-tailed distributions of ranking lists at single points of time~\cite{maillart2008empirical,dodds2020allotaxonometry}, they fail to
reproduce the way elements actually move in rank~\cite{blumm2012dynamics}, such
as the sudden changes in city size throughout
history~\cite{batty2006rank,verbavatz2020growth}. Here we report on the existence of generic
features of rank dynamics over a wide array of systems, from individuals to countries, and spatio-temporal
scales, from minutes to centuries. By measuring
the flux of elements across ranking
lists~\cite{gerlach2016similarity,pechenick2017language,garcia2018ranking}, we identify a continuum ranging from systems where highly ranked elements are more stable than the rest, to systems
where the least relevant elements are also stable. We show
that simple mechanisms relying on fluxes generated by displacement and replacement of
elements can account for
all observed patterns of rank stability. A
model based on these ingredients uncovers two regimes in rank
dynamics, a fast regime driven by long jumps in rank space, and
a slow one driven by diffusion.

\section*{Results} 

\begin{figure}[t] 
\centering
\includegraphics[width=\textwidth]{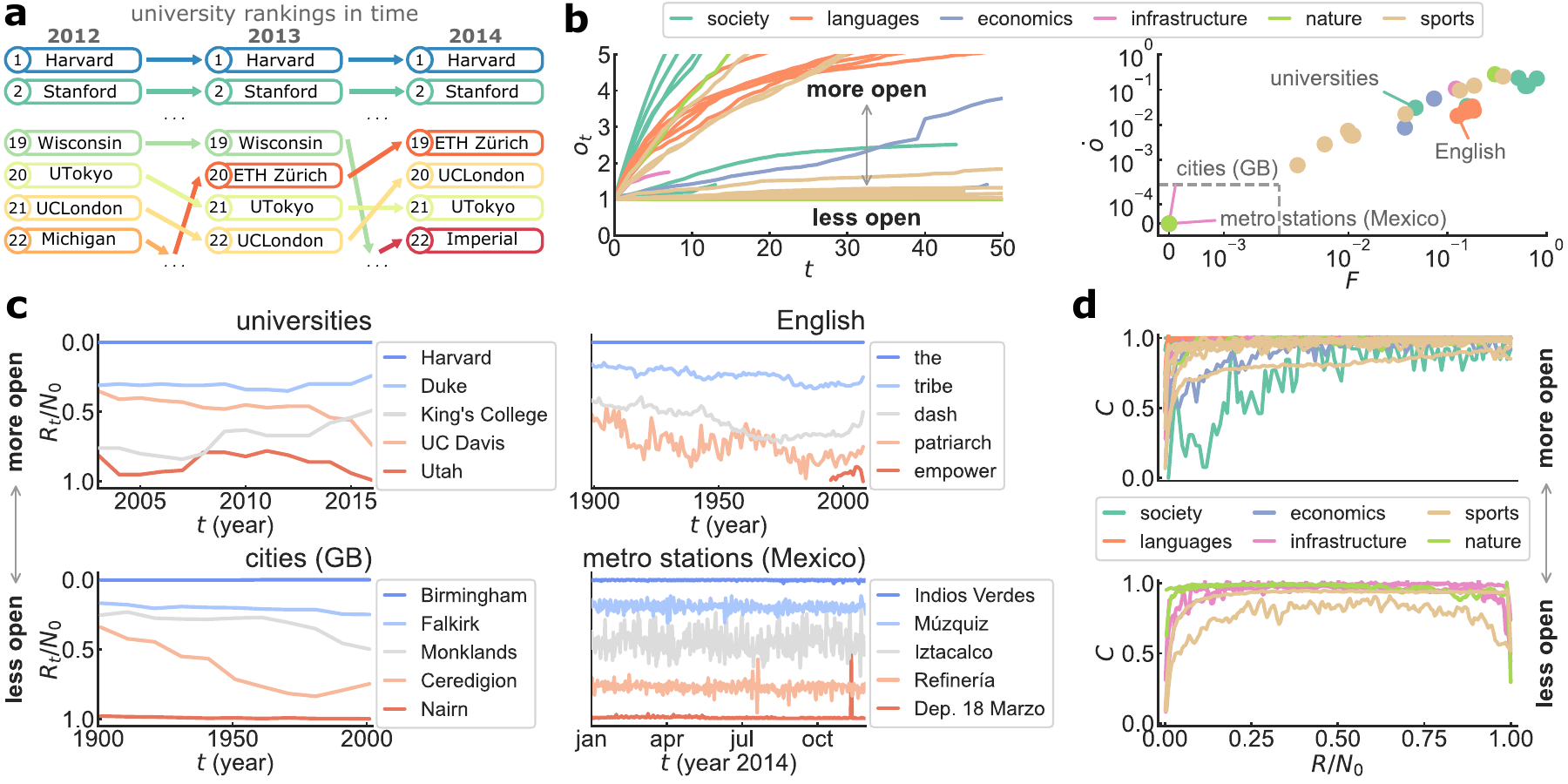}
\caption{\small {\bf Ranking lists in nature and society show generic patterns in their dynamics.}
{\bf (a)} Yearly top ranking of universities worldwide according to ARWU
score~\cite{liu2005academic}. Elements in the system change rank as their
scores evolve in time. {\bf (b)} (Left) Rank turnover $o_t$ at time $t$ for studied systems, defined as the number
$N_t$ of elements ever seen in the ranking list up to $t$ relative to list size $N_0$ (see \frefsi{5}).
(Right) Correlation between mean
turnover rate $\mod$ and mean flux $F$ (average probability
that an element enters or leaves the list). Ranking lists form a continuum
from the most open systems ($F, \mod \sim 1$) to the less open ($F,
\mod \sim 0$; for values see \trefsi{2}). The area between dashed lines has linear scales to show closed systems with $F = \mod = 0$.
{\bf (c)} Time series of rank $R_t / N_0$ occupied by
elements across the ranking list in selected systems (all datasets in \frefsi{2}). In the least open
systems available, the top and bottom of the ranking list are stable. In
open systems, only the top is stable. {\bf (d)} Rank change $\Cr$ (average
probability that element at rank $R$ changes between $t-1$ and $t$) across
ranking lists (see \frefsi{6}), for $F \geq 0.01$ (top) and $F < 0.01$ (bottom).
The stable top and bottom ranks of less open systems mean $\Cr$ is roughly symmetric. In open systems, $\Cr$ increases with rank $R$.
}
\label{fig:figure1}
\end{figure}

We gather 30 ranking lists in natural, social, economic, and infrastructural
systems. Data includes human and animal groups, languages, countries and cities,
universities, companies, transportation systems, online platforms, and sports, with no selection criteria other than having enough information for analysis (for data details see Supplementary Information [SI] Section S2 and Table S1).
Elements in each list are ranked by a measurable score that changes in time:
scientists by citations, businesses by revenue, regions by number of
earthquakes, players by points, etc. Size and temporal scales in the data vary
widely, from the number of people in 636 station entrances of the London
Underground every 15 minutes during a week in
2012~\cite{murcio2015identifying}, to the written frequency of 124k English
words every year since the 17th century~\cite{michel2011quantitative}.
Following an element's rank through time reveals systematic
patterns (\fref{fig:figure1}). For example, in the Academic Ranking of World
Universities (ARWU)~\cite{liu2005academic}, published yearly since 2003,
institutions like Harvard and Stanford maintain a high score, while institutions
down the list change rank frequently (\fref{fig:figure1}a).

Ranking lists typically have a fixed size $N_0$ (e.g., the Top 100
universities~\cite{liu2005academic}, the Fortune 500 companies~\cite{zhu2000multi}), so elements may enter or leave the
list at any of the $T$ observations $t = 0, \ldots, T-1$, allowing us to
measure the flux of elements across rank
boundaries~\cite{gerlach2016similarity,pechenick2017language,dodds2020allotaxonometry}
(for the observed values of $N_0$, $T$ see \trefsi{1}).
We introduce two time-dependent
measures of flux: the {\it rank turnover} $o_t = N_t / N_0$, representing the number $N_t$ of
elements ever seen in the ranking list until time $t$ relative to the list size
$N_0$, and the {\it rank flux} $F_t$, representing the probability that an element enters
or leaves the ranking list at time $t$. Rank turnover is a monotonic increasing
function indicating how fast new elements reach the list
(\fref{fig:figure1}b left; all datasets in \frefsi{5}). In turn, flux shows a
striking stationarity in time despite differences in temporal scales and
potential shocks to the system (\frefsi{3}). By averaging over time, the mean
turnover rate $\mod = (o_{T-1} - o_0) / (T-1)$ and the mean flux $F = \langle F_t \rangle$ turn out to be
highly correlated quantities that uncover a continuum of ranking lists
(\fref{fig:figure1}b right; values in \trefsi{2}). In one extreme, the most open
systems ($F, \mod \sim 1$) have elements that constantly enter and leave the
list. Less open systems ($F, \mod \sim 0$) have progressively lower turnout of
constituents. Five out of 30 ranking lists are completely
closed ($F = \mod = 0$), meaning no single new element is recorded during the
observation window.

The measures of rank turnover and flux reveal regularities in the stability of ranking
processes~\cite{blumm2012dynamics,ghoshal2011ranking}. We follow the time series of the rank $R_t$
occupied by a given element at time $t$~\cite{batty2006rank} (\fref{fig:figure1}c; all datasets in
\frefsi{2}). In most systems, highly ranked elements like Harvard University and the
English word `the' never change position, showcasing the correspondence between
rank stability and notions of relevance like academic
prestige~\cite{clauset2015systematic,sinatra2016quantifying}, grammatical
function~\cite{michel2011quantitative,cocho2015rank}, and underlying network structure~\cite{ghoshal2011ranking}.
As we go down the ranking list of open systems,
rank trajectories increasingly fluctuate in time. In the least open systems where
turnover and flux are low, however, low ranked elements are also
stable. In the ranking of British cities by population, for example, both Birmingham and
Nairn remain the most and least populated local authority areas throughout the
20th century~\cite{edwards2016city}. These findings uncover a more fine-grained
sense of rank stability: most systems have a stable top
ranking, but only the least open systems feature stable bottom ranks as well. The {\it
rank change} $\Cr$,
measured as the average probability that element at rank
$R$ changes between times $t-1$ and $t$, varies between an
approximately monotonic increasing
function of $R$ for open systems to a symmetric shape as systems become less open
(\fref{fig:figure1}d; all datasets in \frefsi{6}).

\begin{figure}[t] 
\centering
\includegraphics[width=\textwidth]{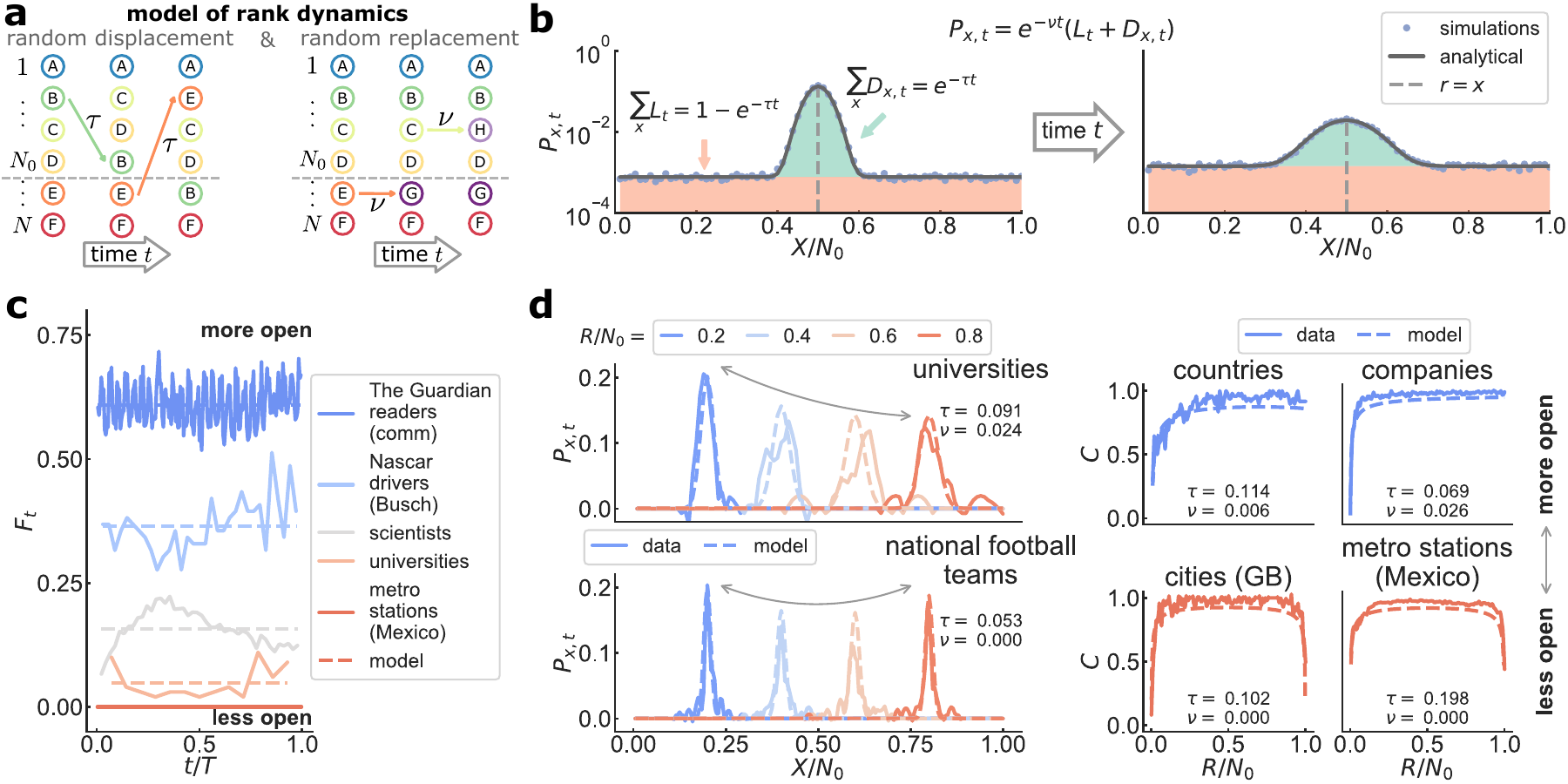}
\caption{\small {\bf Model of rank dynamics reproduces features of real-world ranking lists.}
{\bf (a)} Model of rank dynamics in system of $N$ elements and ranking list
size $N_0$. At time $t$, a random element is moved to a random rank with
probability $\ptau$. A random element is also replaced by a new element with
probability $\pnu$. {\bf (b)} Probability $\Prxt$ that element in rank
$r = R/N$ moves to $x = X/N$ after time $t$ (uppercase/lowercase symbols are integer/normalized ranks). Elements not replaced
diffuse around $x=r$ (with probability $\Drxt$) or perform Lévy
walks~\cite{shlesinger1993strange} (with probability $\Lt$). \eref{eq:diffEq}
recovers simulation results, shown here for $\ptau = 0.1$, $\pnu = 0.2$, $N =
100$, and $N_0 = 80$ at times $t = 1, 5$ (left/right plots), averaged over $10^5$ realizations.
{\bf (c)} Time series of rank flux $F_t$ over observation period $T$ for
data (lines), and mean flux $F$ from fitted model (dashes) (all datasets in \frefsi{3}; for fitting see \srefsi{5}). {\bf (d)} Probability $\Prxt$ for $t = 1$ and varying $r$ (left) and rank change $\Cr$ (right), shown for selected datasets (lines) and fitted model (dashes; $\ptau$ and $\pnu$ in plot) (empirical $\Prxt$ is passed through a Savitzky–Golay smoothing filter; see \fsrefsi{6}{9} and
\trefsi{2}). As systems become more open, we lose symmetry in the rank dependence of both $\Cr$ and the height of the diffusion peaks of $\Prxt$ (signaled by curved arrows). Data and model have similar qualitative behavior in all rank measures (for a systematic comparison see \frefsi{19}).
}
\label{fig:figure2}
\end{figure}

Since the stability patterns of an empirical ranking list (as measured by rank change
$\Cr$) can be systematically connected to the amount of elements flowing into
and out of the list,
we build a model of rank dynamics based solely on simple generative mechanisms of flux
(\fref{fig:figure2}). Without assuming system-specific features of elements or
their interactions, there are at least two ways to implement flux in rank
space. Smooth (but arbitrarily large) changes in the score of an element might
make it larger or smaller than other scores, causing elements to move across
ranks (the way some scientists gather more citations than
others~\cite{sinatra2016quantifying}, or how population size fluctuates due to
historical events~\cite{batty2006rank}). Regardless of score, elements might
also disappear from the list and be replaced by new elements: young athletes
enter competitions while old ones retire~\cite{morales2016generic}; new words
replace anachronisms due to cultural shifts~\cite{michel2011quantitative}. We
implement random mechanisms of displacement and replacement in a simple model
by considering a synthetic ranking list of length $N_0$ embedded within a larger system
of size $N \geq N_0$. At each time step of length $\Delta t = 1 / N$,
a randomly chosen element moves to a randomly selected rank with
probability $\ptau$, displacing others. At the same time, a randomly
chosen element gets replaced by a new one with probability $\pnu$, leaving other
ranks untouched. The dynamics involves all $N$ elements, but to mimic real ranking lists, we only consider
the top $N_0$ ranks when comparing with empirical data (\fref{fig:figure2}a; model details in
\srefsi{4}).

We solve the model analytically by introducing the displacement probability
$\Prxt$ that an element with rank $r = R/N$ gets displaced to rank $x = X/N$
after a time $t$ (\fref{fig:figure2}b; uppercase/lowercase symbols
denote integer/normalized ranks).
Since for small $\Delta x = 1 / N$ the probability that at time $t$ an
element has not yet been replaced is $e^{-\pnu t }$, we have 
\begin{equation}
\label{eq:dispProb}
\Prxt = e^{-\pnu t } ( \Lt + \Drxt ).
\end{equation}
Here, $\Lt = ( 1 - e^{-\ptau t } ) / N$ is the
(rank-independent) probability that up until time $t$ an element gets selected
and jumps to any other rank.  The length of jumps is uniformly distributed, so
they can be thought of as a Lévy random walk with step length exponent
0~\cite{shlesinger1993strange} (full derivation in
\srefsi{4}). The probability $\Drxt = D(x, t) \Delta x$ that the element in rank $r$ gets displaced to rank
$x$ after a time $t$ (due to Lévy walks of other elements) follows
approximately the diffusion-like equation
\begin{equation}
\label{eq:diffEq}
\frac{\partial D}{\partial t} = \alpha x(1 - x) \frac{\partial^2 D}{\partial x^2},
\end{equation}
where $\alpha = \ptau / N$. Since $\sum_x \Drxt = e^{-\ptau t }$, both $\Drxt$ and $D(x, t)$ are not conserved in time.
Instead of a standard diffusion equation, \eref{eq:diffEq} is equivalent to the
Wright-Fisher equation of random genetic drift in allele
populations~\cite{chen2010fundamental,epstein2010wright}.
The solution $D(x, t)$ of \eref{eq:diffEq} is well approximated
by a decaying Gaussian distribution with mean $r$ and standard deviation $\sqrt{ 2
\alpha r (1 - r) t }$, i.e. a diffusion kernel (\fref{fig:figure2}b).
Overall, local displacement makes elements slowly diffuse around their
initial rank, while Lévy walks and the replacement dynamics reduces exponentially the probability that old elements
remain in the ranking list.

An explicit expression for the displacement probability $\Prxt$ allows us to derive the mean flux
\begin{equation}
\label{eq:flux}
F = 1 - e^{-\pnu } [ \p0 + (1 - \p0) e^{-\ptau } ],
\end{equation}
and the mean turnover rate
\begin{equation}
\label{eq:openness}
\mod = \pnu \frac{ \pnu + \ptau }{ \pnu + \p0 \ptau },
\end{equation}
where $\p0 = N_0 / N$ is the length of the ranking list relative to system size
(see \srefsi{4}). In order to fit the model to each empirical ranking list, we
obtain $N_0$ from the data and approximate $N = N_{T-1}$ as the number of
distinct elements ever seen in the list during the observation period $T$, thus
fixing $\p0$ (values for all datasets in \tsrefsi{1}{2}). The remaining free
parameters $\ptau$ and $\pnu$ (regulating the mechanisms of displacement
and replacement) come from numerically solving \esref{eq:flux}{eq:openness}
with $F$ and $\mod$ fixed by the data (\fref{fig:figure2}c
and \fref{fig:figure1}b; for model fitting see \srefsi{5}).
The approximations in \esref{eq:flux}{eq:openness} introduce a small bias in the estimation of $\ptau$ (\frefsi{18}). Despite this bias,
the simple generative mechanisms of flux in the model are enough to recover the behavior of ranking lists as quantified by $\Prxt$ and $\Cr$
(\fref{fig:figure2}d and \frefsi{19}): When rank flux is low, both the top and bottom of the list are similarly stable and rank dynamics is mostly driven by an interplay between Lévy walks and diffusion.
As systems become more open, however, this
symmetry gets broken due to a growing flux of elements at the bottom of the
ranking list (see \frefsi{4}). Regardless of whether we rank people or animals,
words or countries, the pattern of stability across a ranking list is accurately emulated by random mechanisms of flux that disregard the microscopic details of the individual system.

\begin{figure}[t] 
\centering
\includegraphics[width=\textwidth]{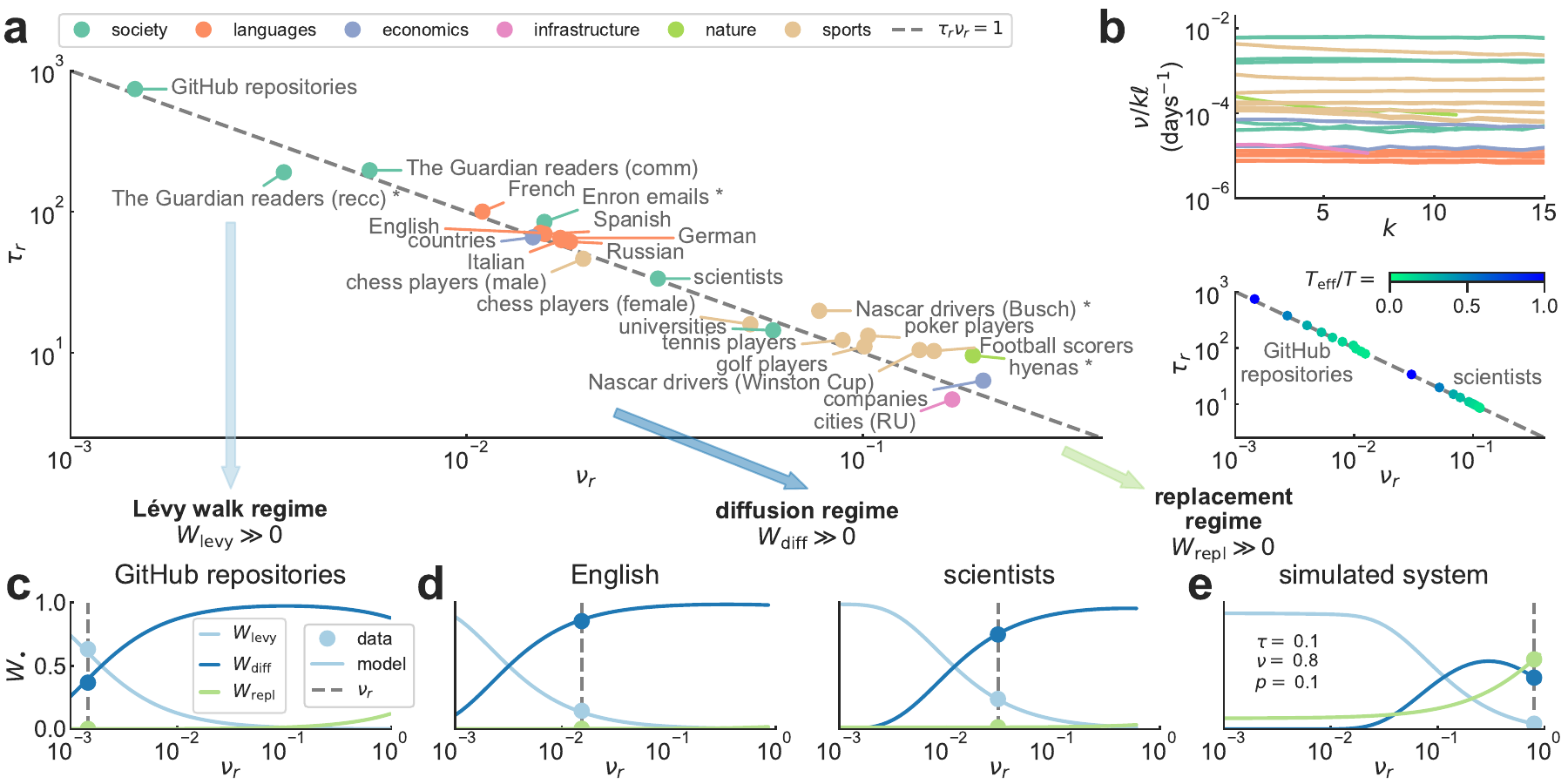}
\caption{\small {\bf Model uncovers regimes of dynamical behavior in open
ranking lists.}
{\bf (a)} Rescaled model parameters $\prtau$ and $\prnu$ in open ranking lists,
obtained from fitted parameters $\ptau$ and $\pnu$, relative ranking list size
$\p0$, and mean turnover rate $\mod$ [see \eref{eq:rescParams} and
\srefsi{5}; only systems with $\mod > 10^{-3}$ are shown]. Values collapse onto
the universal curve $\prtau \prnu = 1$, so an inverse relationship between
displacement and replacement is enough to emulate empirical rank dynamics
(asterisks denote datasets that are farther away from the universal curve than bootstrapped model simulations; see \frefsi{20}). {\bf (b)} (Top) Rate of element replacement $\pnu / k
\ell$ when subsampling data every $k$ observations of length $\ell$ (see
\trefsi{1} and \srefsi{5}). Online social systems have the largest rates, followed by sports and languages (\trefsi{2}). (Bottom) Parameters $\prtau$ and $\prnu$ for
$T_{\mathrm{eff}} = \lceil T / k \rceil $ subsampled observations (all datasets
in \frefsi{22}). By subsampling ranking dynamics, systems move downwards along
the universal curve while keeping a constant replacement rate. {\bf (c)--(e)}
Average probability that an element changes rank by Lévy walk
($\Wlevy$), diffusion ($\Wdiff$), or is replaced ($\Wrepl$) between consecutive
observations in the data. Probabilities shown both for selected datasets
(dots), and for the model moving along the curve $\prtau \prnu = 1$ with the
same $\p0$ and $\mod$ as the data (lines) (for rest of systems see
\frefsi{17}). The simulated system in (e) is the model itself for
given values of $\ptau$, $\pnu$, and $\p0$ (shown in plot). The model reveals a
crossover in real-world ranking lists between a regime dominated by
Lévy walks (b) to one driven by diffusion (c). Although not seen in
data, the model also predicts a third regime driven by replacement (d).
}
\label{fig:figure3}
\end{figure} 

The characterization of flux in ranking lists with mechanisms of displacement and replacement of elements reveals regimes of dynamical behavior that are not apparent from the data alone (\fref{fig:figure3}). By rescaling the fitted parameter values of the model as
\begin{equation}
\label{eq:rescParams}
\prtau = \frac{\ptau}{ \p0 (1 - \p0) \mod }
\quad \text{and} \quad
\prnu = \frac{ \pnu - \p0 \mod }{ \mod },
\end{equation}
most open ranking lists ($F, \mod > 0$) are predicted to follow the universal curve
\begin{equation}
\label{eq:univCurve}
\prtau \prnu = 1,
\end{equation}
which suggests that ranking dynamics are regulated by a single effective
parameter [\fref{fig:figure3}a; derivation in \srefsi{5}; for a discussion of the role of fluctuations on the validity of \eref{eq:univCurve} see \fsrefsi{18}{20}]. Even if,
potentially, displacement and replacement could appear in any relative
quantity, adjusting the model to observations of rank flux and turnover
(\fref{fig:figure1}b) leads to an inverse relationship between parameters regulating their generative
mechanisms. Real-world ranking lists lie in a spectrum where their dynamics is either mainly driven by
score changes that displace elements in rank (high $\prtau$ and low $\prnu$, like for GitHub
software repositories~\cite{vedres2019gendered} ranked by daily popularity), or by birth-death processes
triggering element replacement (low $\prtau$ and high $\prnu$, like for Fortune 500 companies~\cite{zhu2000multi} ranked by yearly revenue).
While the symmetry (or lack thereof) in rank change $C$ may seemingly imply two distinct classes of systems (see \fref{fig:figure1}d and \fref{fig:figure2}d), \eref{eq:univCurve} reveals the existence of a continuum of open ranking lists, which can be captured by a single model with a single effective parameter.

Data on empirical ranking lists is constrained by the average time
length $\ell$ between recorded observations, which varies from minutes to years
depending on the source and intended use of the rankings ($\ell$ for all
datasets is listed in \trefsi{1}). We explore such scoping effect by subsampling data
every $k$ observations (for details see \srefsi{5} and \fsrefsi{21}{22}). Longer
times between snapshots of the ranking list lead to an increase in rank flux,
turnover, and fitted parameters, such that the rate of element replacement
$\pnu / k \ell$ stays roughly constant (\fref{fig:figure3}b top). A conserved
replacement probability per unit time, robust to changes in sampling rate, is
yet another measure of rank stability: online social systems exchange elements frequently (e.g., the ranking
lists by daily popularity of both GitHub software repositories and of online
readers of the British newspaper The Guardian~\cite{thurman2008forums}), followed by sports, while languages are the most
stable (values for $k=1$ in \trefsi{2}).

The universal curve in \eref{eq:univCurve} displays three regimes in the
dynamics of open ranking lists, as measured by the average probabilities that,
between consecutive observations in the data, an element performs either a Lévy
walk [$\Wlevy = e^{-\pnu} (1 - e^{-\ptau} )$], changes rank by
diffusion [$\Wdiff = e^{-\pnu} e^{-\ptau}$], or is replaced [$\Wrepl
= 1 - e^{-\pnu}$], with $\Wlevy + \Wdiff + \Wrepl = 1$. In systems with the largest rank flux and turnover (GitHub repositories and The Guardian readers), elements tend to change
rank via long jumps, following a Lévy walk, where $\Wlevy > \Wdiff,
\Wrepl$ (\fref{fig:figure3}c). Here, long-range rank changes take elements in
and out of a short ranking list within a big system (low $\p0$), thus
generating large mean flux $F$ (see \trefsi{2}). Most datasets, like the yearly rankings of scientists by citations in American Physical Society journals~\cite{sinatra2016quantifying,sinatra2015century} and of countries by economic complexity~\cite{hidalgo2009building,hidalgo2007product},
belong instead to a diffusion regime with $\Wdiff > \Wlevy, \Wrepl$
(\fref{fig:figure3}d). In this regime, a local, diffusive rank dynamics is the result of
elements smoothly changing their scores and overcoming their neighbors in rank
space. Under subsampling, ranking lists move downwards along the universal curve, going from a state with a certain number of Lévy walks to one more driven by diffusion (\fref{fig:figure3}b bottom; all datasets in \frefsi{22}).

The model also predicts a third regime dominated by replacement ($\Wrepl > \Wlevy, \Wdiff$; \fref{fig:figure3}e), where elements are more likely to disappear than change rank.
Such ranking lists replace most constituents from one observation to the next, forming a highly fluctuating regime that we do not observe in empirical data.
To showcase the crossover between regimes, we simulate
the model along the universal curve of \eref{eq:univCurve} while keeping $\p0$
and $\mod$ fixed in \eref{eq:rescParams} (lines in
\fref{fig:figure3}c--e). These curves show how close systems are to a change of
regime, i.e. from one dominated by Lévy walks to one driven by diffusion.
When a ranking list is close to a regime boundary, external shocks (amounting to variations in parameters $\ptau$ and $\pnu$) may change the main mechanism behind rank dynamics, thus affecting the overall stability of the system.

\section*{Discussion} 

Ranking lists reduce the elements of high-dimensional complex systems into ordered values of a
summary statistic, allowing us to compare seemingly disparate phenomena in nature and
society~\cite{langville2012s,dodds2020allotaxonometry}. The diversity of their
components (people, animals, words, institutions, countries) stands in contrast
with the statistical regularity of score-rank distributions when aggregated
over time~\cite{newman2005power,martinez2009universality}. By exploring the
flux of elements of 30 ranking lists in natural, social, economic, and
infrastructural systems, we present evidence of generic temporal patterns of
rank dynamics. While open systems (large flux) keep the same elements only in top
ranks, less open systems (lower flux) also have stable bottom ranks, forming a
continuum of ranking lists explained by a single class of models. 
The model reveals two
regimes of dynamical behavior for systems with nonzero flux. Real-world ranking
lists are driven either by Lévy random walks~\cite{shlesinger1993strange} that
change the rank of elements abruptly, or by a more local, diffusive movement
similar to genetic drift~\cite{chen2010fundamental,epstein2010wright}, both
alongside a relatively low rate of element replacement independent from the frequency at which the ranking list is measured.

Our results suggest that, even though score distributions differ across
systems depending on what type of elements and interactions they have
(\frefsi{1}), ranking lists have similar stability features.
What are the underlying properties of the system that enhance
this similarity? An extension of our model explicitly considering
the links between score and rank may help further understand the experimental
evidence in this area, like the recent observation that the stability of
crowdsourced rankings depends on the magnitude difference between quality
scores~\cite{burghardt2020origins}. It is also interesting to consider the
observed deviations from the predictions of our model, even at the level of
ranks. Rank flux for languages is not constant but decreases over time
(\frefsi{3}), arguably due to the long observation period (over three
centuries; see \trefsi{1}), variations in word sampling across decades, or even
cognitive distortions at the societal scale~\cite{bollen2021historical}. The
rank-dependence of flux for very open systems (\frefsi{4}) and the slow decay
of inertia with long times we observe in most datasets (\frefsi{7}) might be better
reproduced by a non-uniform sampling of elements in the mechanisms of
displacement and replacement of the model. Finally, deviations in the data
(indicating a departure from the assumptions of randomness and stationarity
built into our model) could be used to detect shocks to the system larger than
expected statistical fluctuations, such as the sudden increase in rank flux of
Fortune 500 companies during financial crises (\frefsi{3}).

A more nuanced understanding of the generic features of ranking dynamics
might help us limit resource exhaustion in competitive environments, such as
information overload in online social platforms and prestige biases in
scientific publishing~\cite{sekara2018chaperone}, via better algorithmic rating
tools~\cite{ciampaglia2018algorithmic}. The observation of a systematic interplay between ``slower'' and ``faster'' ranking dynamics~\cite{oka2013exploring} (see \srefsi{6}) can be refined by exploring the relationship between ranking lists
associated to the same system, or by incorporating networked interactions that
lead to macroscopic ordering~\cite{batty2006rank,krapivsky2002statistics}, which may provide a deeper understanding of network centrality measures based on ranking~\cite{posfai2019consensus}.
Given that rankings often mediate access to resources via policy, similar
mechanisms to those explored here may play a role in finding better ways to
avoid social and economic disparity. In general, a better understanding of rank
dynamics is promising for regulating systems by adjusting their temporal
heterogeneity.

{\small
\subsubsection*{Data availability}
For data availability see \srefsi{2}. Non-public data is available from the authors upon reasonable request.

\subsubsection*{Code availability}
Code to reproduce the results of the paper is publicly available at \url{https://github.com/iniguezg/Farranks} \cite{iniguez2022dynamics}.
}

{\small
\subsubsection*{Acknowledgments}
In memory of Jorge Flores and Germinal Cocho. We acknowledge Jos{\'e} A. Morales and Sergio S{\'a}nchez for data handling at the start of the project. We are grateful for data provision to Gustavo Carreón, Syed
Haque, Kay Holekamp, Amiyaal Ilany, Márton Karsai, Raj Kumar Pan, Roberto
Murcio, and Roberta Sinatra. G.I. thanks Tiina Näsi for valuable suggestions.
G.I. acknowledges support from AFOSR (\#FA8655-20-1-7020), project EU H2020 Humane AI-net (\#952026), and CHIST-ERA project SAI (\#FWF I 5205-N).
C.P. and C.G. acknowledge
support by CONACyT (\#285754) and UNAM-PAPIIT (\#IG100518, IG101421, IN107919, and IV100120). C.G. was also supported by the PASPA program from UNAM-DGAPA.
A.-L.B. was supported by a EU H2020 SYNERGY grant (\#810115-DYNASNET), the John Templeton Foundation (\#61066), and AFOSR (\#FA9550-19-1-0354).

\subsubsection*{Author contributions statement}
G.I., C.P., C.G., and A.-L.B. designed the study. G.I. performed data analysis and model fitting. G.I. and C.P. derived analytical results and performed numerical simulations. G.I., C.P., C.G., and A.-L.B. wrote the paper.

\subsubsection*{Competing interest statement}
A.-L.B. is founder of Foodome, ScipherMedicine, and Datapolis, companies that explore the role of networks in health and urban environments. G.I. is founder of Predify, a data science consulting startup in Mexico.  C.P.  and C.G. declare no competing interest.
}

\end{document}


\begin{center}
{\LARGE Supplementary Information for}\\[0.7cm]
{\Large \textbf{Dynamics of ranking}}\\[0.5cm]
{\large G. Iñiguez, C. Pineda, C. Gershenson, and A.-L. Barabási}\\[0.7cm]
\end{center}

\addtocontents{toc}{\protect\setstretch{0.1}}
\tableofcontents
\clearpage
\section{Summary of notation} 
\label{sec:intro}
We consider a {\it ranking list} at times $t = 0, \ldots, T - 1$, that is, an
ordered set of $N_0$ elements (with $N_0$ constant in time) where each element
$i$ in the list at time $t$ has a {\it score} $s_i(t)$, and elements are
ordered across the list with decreasing score. Thus, elements and their order
may change throughout time. The most important element (with the largest score)
at time $t$ has {\it rank} $R_t = 1$; the least important element (with the
lowest score) has rank $R_t = N_0$, and the ranking is the particular order of
elements and scores across ranks $R = 1, \ldots, N_0$, which changes in time as
elements vary their scores. We consider the case where we do not have access to
the scores at all times, but only in a
discrete set of $T$ observations, separated in average by a real time interval
$\ell$ (measured in days)~\footnote{The time unit is arbitrary and
may be anything other than days.}. We also consider the case where we may not
have empirical data on the scores of all elements at all times, either because
elements enter/leave the ranking at some point in time, or because score data
is unavailable. Thus, there are $\Nt\geq N_0$ distinct elements that have ever been in
the ranking up to (and including) time $t$. If $\Nt = N_0$ for all $t$ the ranking list is {\it
closed}, since
elements do not leave or enter the ranking list, and if $\Nt > N_0$ the ranking list is
{\it open}. In what follows we characterize the temporal variability of ranking
lists across many systems by analysing the flow of elements in and out of the
ranking list as time goes by.

\section{Data description} 
\label{sec:data}
We analyse 30 datasets in a wide range of structural and temporal scales,
comprising several definitions of elements and scores. \tref{tab:datasets}
lists the observed system size $N_{T - 1}$ (number of distinct elements ever
seen in the ranking), ranking list size $N_0$, number of observations $T$, and
real time interval $\ell$ for all datasets
considered. 
\tref{tab:datasets} also includes a system classification based
on the nature of the elements in the system, and the corresponding definitions of
elements and scores for each system. 
Social systems
reflect human interactions
at the individual and organizational levels. Language datasets show how word
usage has changed across centuries. Economic rankings illustrate value at
different scales. Systems in the infrastructure category are specific to public
transport and city populations. Nature datasets gather information from
biological and geological phenomena. Finally, rankings in sports capture the
relative performance of players and teams according to sets of rules. We now
describe the datasets in more detail.

In their original state some datasets are not homogeneous in time and size,
since they do not have the same real time interval between any times $t$ and $t
+ 1$ or the same number of elements per observation (i.e. they have a variable
ranking list size). To consistently analyse all datasets, we crop the data to obtain
roughly homogeneous time intervals and have a constant ranking list size
across observations, while trying to retain as large $T$ and $N_0$ as possible.

\begin{table}[!ht]
\small
\noindent\makebox[\textwidth]{ \begin{tabular}{l l l | r r r r}
\toprule
& & & \multicolumn{4}{c}{Measure} \\
\cmidrule(l){4-7}
Dataset & Element & Score & $N_{T - 1}$ & $N_0$ & $T$ & $\ell$ (days) \\
\midrule
{\bf Society} & & & & & & \\
\cmidrule(l){1-1}
GitHub repositories~\cite{github2018} & repository & \# watchers & 450655 & 4773 & 727 & 1.00 \\
The Guardian readers (recc)~\cite{guardian2018} & person & avg \# recommends & 29165 & 753 & 182 & 1.00 \\
The Guardian readers (comm)~\cite{guardian2018} & person & \# comments & 18244 & 753 & 182 & 1.00 \\
Enron emails~\cite{enron2015} & person & \# emails & 4720 & 209 & 101 & 7.00 \\
Scientists~\cite{sinatra2015,sinatra2016} & person & \# citations & 2614 & 1041 & 45 & 365.25 \\
Universities~\cite{shanghai2016} & university & ARWU score~\cite{shanghai2018} & 140 & 100 & 14 & 365.23 \\
\midrule
{\bf Languages} & & & & & & \\
\cmidrule(l){1-1}
Russian~\cite{google2018,michel2011,cocho2015,morales2018} & word & frequency & 281346 & 35494 & 210 & 365.24 \\
Spanish~\cite{google2018,michel2011,cocho2015,morales2018} & word & frequency & 233323 & 31750 & 260 & 365.24 \\
German~\cite{google2018,michel2011,cocho2015,morales2018} & word & frequency & 195455 & 22661 & 262 & 365.25 \\
French~\cite{google2018,michel2011,cocho2015,morales2018} & word & frequency & 182507 & 17645 & 367 & 365.24 \\
Italian~\cite{google2018,michel2011,cocho2015,morales2018} & word & frequency & 139645 & 20879 & 244 & 365.24 \\
English~\cite{google2018,michel2011,cocho2015,morales2018} & word & frequency & 124464 & 17750 & 334 & 365.24 \\
\midrule
{\bf Economics} & & & & & & \\
\cmidrule(l){1-1}
Companies~\cite{fortune2005} & company & revenue & 1895 & 500 & 51 & 365.26 \\
Countries~\cite{atlas2018,hidalgo2009,hausmann2014} & country & complexity~\cite{atlas2018,hidalgo2009,hausmann2014} & 139 & 99 & 49 & 365.25 \\
\midrule
{\bf Infrastructure} & & & & & & \\
\cmidrule(l){1-1}
Cities (RU)~\cite{cottineau2016} & city & population & 1639 & 936 & 8 & 4383.00 \\
Metro stations (London)~\cite{murcio2015} & station & \# passengers & 636 & 636 & 69 & 0.0104 \\
Cities (GB)~\cite{edwards2016} & city & population & 458 & 458 & 11 & 3652.50 \\
Metro stations (Mexico) & station & \# passengers & 175 & 175 & 365 & 1.00 \\
\midrule
{\bf Nature} & & & & & & \\
\cmidrule(l){1-1}
Hyenas~\cite{ilany2015} & animal & association index & 303 & 43 & 23 & 365.23 \\
Regions JP (quake mag)~\cite{junec2018,karsai2012} & region & avg quake magnitude & 264 & 264 & 176 & 28.00 \\
Regions JP (quakes)~\cite{junec2018,karsai2012} & region & \# quakes & 264 & 264 & 176 & 28.00 \\
\midrule
{\bf Sports} & & & & & & \\
\cmidrule(l){1-1}
Chess players (male)~\cite{fide2018} & person & Elo rating~\cite{elo2018} & 16568 & 13500 & 46 & 30.44 \\
Chess players (female)~\cite{fide2018} & person & Elo rating~\cite{elo2018} & 16539 & 12681 & 46 & 30.44 \\
Poker players~\cite{poker2018} & person & GPI score~\cite{gpi2018} & 9799 & 1795 & 221 & 7.04 \\
Tennis players~\cite{tennis2018} & person & ATP points~\cite{atp2018} & 4793 & 1600 & 400 & 7.00 \\
Golf players~\cite{golf2018} & person & OWGR points~\cite{owgr2018} & 3632 & 1150 & 768 & 7.00 \\
Football scorers~\cite{football2018} & person & \# goals~\cite{fwr2018} & 2397 & 400 & 53 & 7.04 \\
NASCAR drivers (Busch)~\cite{nascar2018} & person & NASCAR points & 676 & 76 & 34 & 365.24 \\
NASCAR drivers (Winston Cup)~\cite{nascar2018} & person & NASCAR points & 272 & 50 & 35 & 365.26 \\
National football teams~\cite{teams2018} & team & FIFA points~\cite{fifa2018} & 210 & 200 & 71 & 30.44 \\
\bottomrule
\end{tabular}}
\caption{
\small {\bf Datasets used in this study}.
Characteristics of the available datasets, including the observed system size
$N_{T - 1}$, ranking list size $N_0$, number of observations $T$, and real time
interval between observations $\ell$. The table
includes a system type based on the nature of the elements in the
ranking list, as well as the corresponding definitions of elements and scores
for each system, and references regarding the dataset and score system. 
}
\label{tab:datasets}
\end{table}

\subsection{Society} 
\label{ssec:dataSoc}

\paragraph{GitHub repositories.}

GitHub~\cite{github2018} is perhaps the most popular web-based version control repository, mostly used for source code. This dataset contains daily rankings of repositories, based on the number of users watching each project, from April 1, 2012 to December 30, 2014.

\paragraph{The Guardian readers.} The Guardian~\cite{guardian2018} is a British
national daily newspaper, publishing online articles (news and opinion pieces)
in diverse subjects. Users registered to the website can post comments to some
of the articles, which readers may `recommend' (by clicking a `like' button
akin to those of social network sites). Focusing on the period from November 1,
2011 till May 1, 2012, we crawl articles appearing in three sections of The
Guardian (`politics', `sport', and `comment is free'), and rank users daily by
the number of comments they write (denoted `comm'), as well as by the average
number of recommends their comments receive (denoted `recc').

\paragraph{Enron emails.}

The Federal Energy Regulatory Commission, during its investigation of the
company Enron, made public about half a million emails from roughly 150 users,
mostly senior managers. A cleaned, current version of the dataset is available
online~\cite{enron2015}.
From this dataset we rank users (email accounts) by the number of emails
sent, on a weekly basis.

\paragraph{Scientists.}

We use yearly citation data from journals of the American Physical Society to
rank the most cited scientists by number of citations~\cite{sinatra2015,sinatra2016}.

\paragraph{Universities.}

ShangaiRanking Consultancy is an independent organization dedicated to higher education research. During the period 2003--2016~\cite{shanghai2016}, the organization ranked top universities according to the Academic Ranking of World Universities (ARWU) score, which considers several criteria such as number of students, publications, Nobel Prizes, Fields Medals, etc.~\cite{shanghai2018}.

\subsection{Languages} 
\label{ssec:dataLang}

We use a subset of the publicly available Google Books Ngram dataset~\cite{google2018,michel2011}, the result of the digitalization and conversion of millions of books in several languages (about $4\%$ of all published books until 2009). From this data we extract the frequency of words by year for Russian, Spanish, German, French, Italian, and English. The original dataset is case sensitive, so we merge words with different cases. We have already analyzed the temporal variability of ranks at several scales for these languages in previous studies~\cite{cocho2015,morales2018}.

\subsection{Economics} 
\label{ssec:dataEcon}

\paragraph{Companies.}

Since 1955, \emph{Fortune} magazine has compiled a yearly dataset of the top 500 corporations in the world based on yearly revenue. Here we use the freely accessible archives of 1955--2005~\cite{fortune2005}.

\paragraph{Countries.}

MIT's Observatory of Economic Complexity~\cite{atlas2018} has compiled a yearly ranking dataset of the economic complexity of countries between 1962 and 2015~\cite{hidalgo2009,hausmann2014}, where a complexity score reflects the diversity of exports of countries around the globe.

\subsection{Infrastructure} 
\label{ssec:dataStruct}

\paragraph{Cities.}

To obtain ranking datasets of cities by population, we use data from previous studies of Russia (RU)~\cite{cottineau2016} and Great Britain (GB)~\cite{edwards2016}. In Russia, urban population has decreased, while in most other countries it has increased. For Great Britain, population data corresponds to 63 primary urban areas from England, Scotland, and Wales between 1901 and 2011.

\paragraph{Metro stations.}

The London Underground has 270 stations, serving about 5 million passengers per
day. For London, the ranking dataset corresponds to aggregated Oyster
smart-card data of 636 station entrances in a week of 2012, considering
15-minute intervals~\cite{murcio2015}. For Mexico City (Mexico), we requested
and obtained data directly from \textit{Sistema de Transporte Colectivo Metro},
the part of Mexican government directing metro services, and used daily
entrance data for 175 stations during 2014.

\subsection{Nature} 
\label{ssec:dataNat}

\paragraph{Hyenas.}

A 23-year field study has monitored the social relationships of a spotted hyena population in Kenya~\cite{ilany2015}. Using this dataset, we rank individual hyenas according to the sum of an association index describing the strength of the relationships with the rest of the hyenas over a particular year.

\paragraph{Regions JP.}

The Japan University Network Earthquake Catalog openly records earthquake
events happening in Japan (JP)~\cite{junec2018}. Following a previous
study~\cite{karsai2012}, we only consider earthquakes with Richter magnitude larger
than $2.0$, happening between July 1, 1985 and December 31, 1998,
taking
main-shocks and after-shocks as separate events. We rank administrative regions
in Japan by both the monthly number of earthquakes happening in a region
(denoted `quakes'), and the average earthquake magnitude in a month (denoted
`quake mag').

\subsection{Sports} 
\label{ssec:dataSport}

\paragraph{Chess players.}

Every month, the Fédération Internationale des Échecs (FIDE, International
Chess Federation) ranks top male and female players~\cite{fide2018} based on
the Elo rating system~\cite{elo2018}.

\paragraph{Poker players.}

Weekly rankings of Poker players are listed online~\cite{poker2018}. Rankings
are based on the Global Poker Index (GPI) score~\cite{gpi2018}, which takes
into account players who participated in tournaments in the last 18 months,
including an aging factor and the scores obtained in different tournaments.

\paragraph{Tennis players.}

The Association of Tennis Professionals (ATP)~\cite{tennis2018} has an
intricate ranking system based on ATP points~\cite{atp2018}, which is updated
every week to rank male players in Singles.

\paragraph{Golf players.}

The Official World Golf Ranking (OWGR) publishes weekly lists of player
rankings, accumulating points achieved in tournaments in the previous two
years~\cite{golf2018}. To calculate OWGR points, the points achieved by each
player on all tournaments during these two years are averaged over the number
of tournaments involved~\cite{owgr2018}.

\paragraph{Football scorers.}

The Football World Rankings publishes a weekly list of top
scorers~\cite{football2018}, considering the number of goals scored by each
player in the previous year~\cite{fwr2018}.

\paragraph{NASCAR drivers.}

We use an open dataset containing the ranking of drivers of the National Association for Stock Car Auto Racing (NASCAR), both for the Busch Series and the Winston Cup~\cite{nascar2018}.

\paragraph{National football teams.}

The Fédération Internationale de Football Association (FIFA, International Federation of Association Football)~\cite{teams2018} publishes a monthly list of national teams ranked according to the points obtained in the previous four years, using several criteria to determine FIFA points~\cite{fifa2018}. Data for female teams has a lower temporal scale, so we restrict our analysis to male teams.

\subsection{Acknowledgments and data availability} 
\label{ssec:dataAck}

We acknowledge Jos{\'e} A. Morales and Sergio S{\'a}nchez for data handling in the initial stages of the project. We are grateful for data provision to Syed Haque (GitHub repositories and Enron emails), Raj Kumar Pan and author Gerardo I{\~n}iguez (The Guardian), Roberta Sinatra (Scientists), Roberto Murcio (Cities [RU and GB], Metro stations [London]), Gustavo Carreón (Metro stations [Mexico]), Amiyaal Ilany and Kay Holekamp (Hyenas), and Márton Karsai (Regions JP). All data is either openly accessible, can be crawled directly from the web, or can be requested directly from the previous researchers under reasonable conditions.


\section{Rank measures} 
\label{sec:measures}

\subsection{Fitness} 
\label{ssec:fitness}

\begin{figure}[t]
\centering
\includegraphics[width=\textwidth]{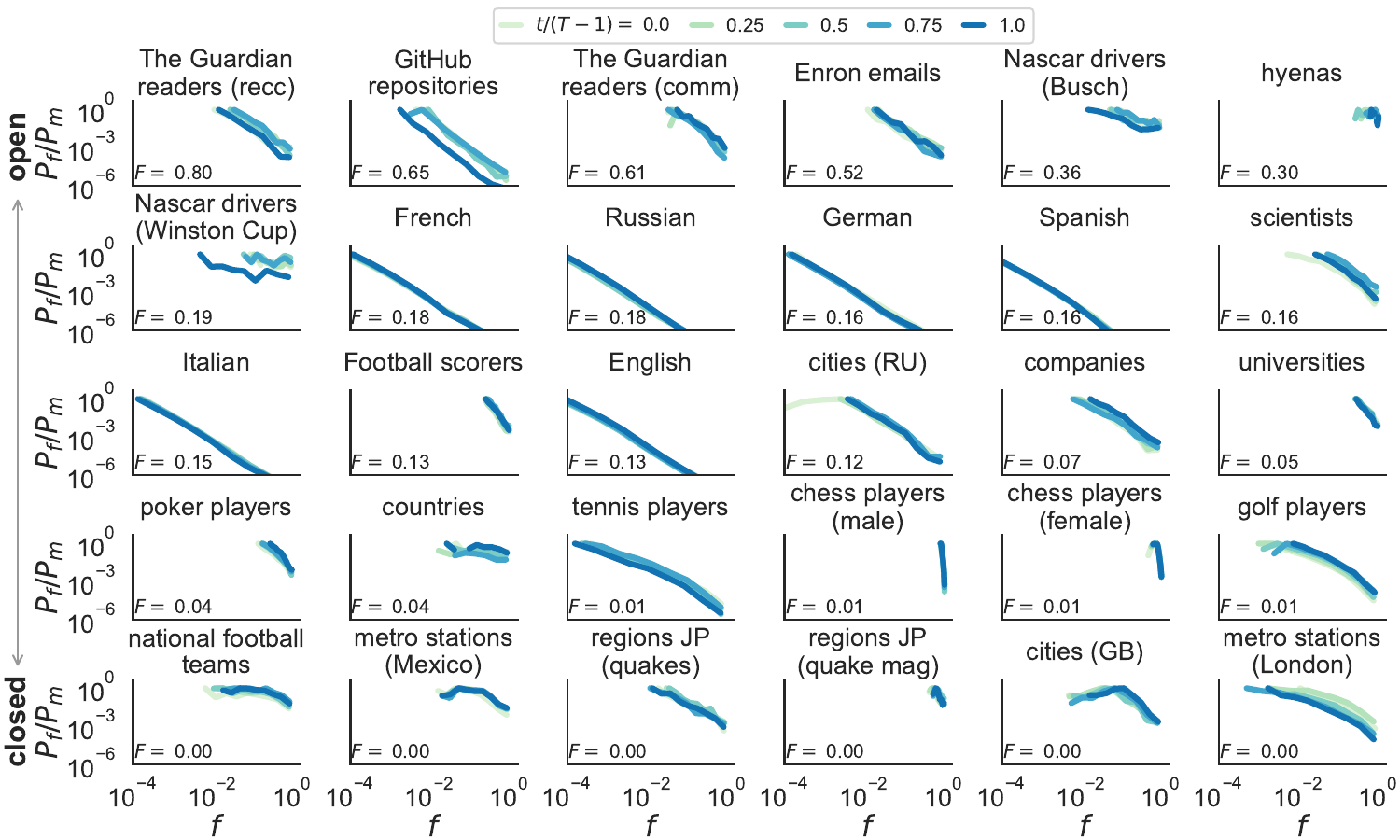}
\caption{
\small {\bf Fitness in open to closed ranking lists.}
Probability $P_f$ that a randomly selected element $i$ has fitness $f_i$ at time $t$,
normalized by the maximum value of $P_f$ at the time (denoted $P_m$). We plot $P_f$ for various times $t$ (colored
lines). To compare across observations $t = 0, \ldots, T-1$, we calculate
fitness as $f_i(t) = s_i(t) / s_m (t)$, with $s_i(t)$ the score of element $i$
and $s_m (t)$ the maximum score in the ranking at time $t$ [\eref{eq:fitness}].
Datasets are ordered from most open (upper row) to closed (lower row) according
to mean rank flux $F$ [\eref{eq:fluxMean}]. Fitness distributions are either
broad (spanning orders of magnitude) or constrained to a small interval, and
show wild variations across datasets.}
\label{fig:fitness}
\end{figure}

Scores $s_i(t)$ vary across elements and observations of the ranking list. To compare their distribution between times $t$, we normalize the score of element $i$ as the {\it fitness}
\begin{equation}
\label{eq:fitness}
f_i(t) = \frac{ s_i(t) }{ s_m (t) },
\end{equation}
with $s_m (t)$ the maximum score in the ranking at time $t$. We define $P_f$ as the probability that a randomly selected element $i$ has fitness $f_i$ at time $t$, and $P_m$ as the maximum value of $P_f$ at time $t$. The normalized fitness distribution $P_f / P_m$ for each time $t = 0, \ldots, T-1$ varies greatly between systems, having either a broad or narrow functional form (\fref{fig:fitness}). In most ranking lists, $P_f$ is relatively constant in time.

\subsection{Rank dynamics} 
\label{ssec:dynamics}

\begin{figure}[t]
\centering
\includegraphics[width=\textwidth]{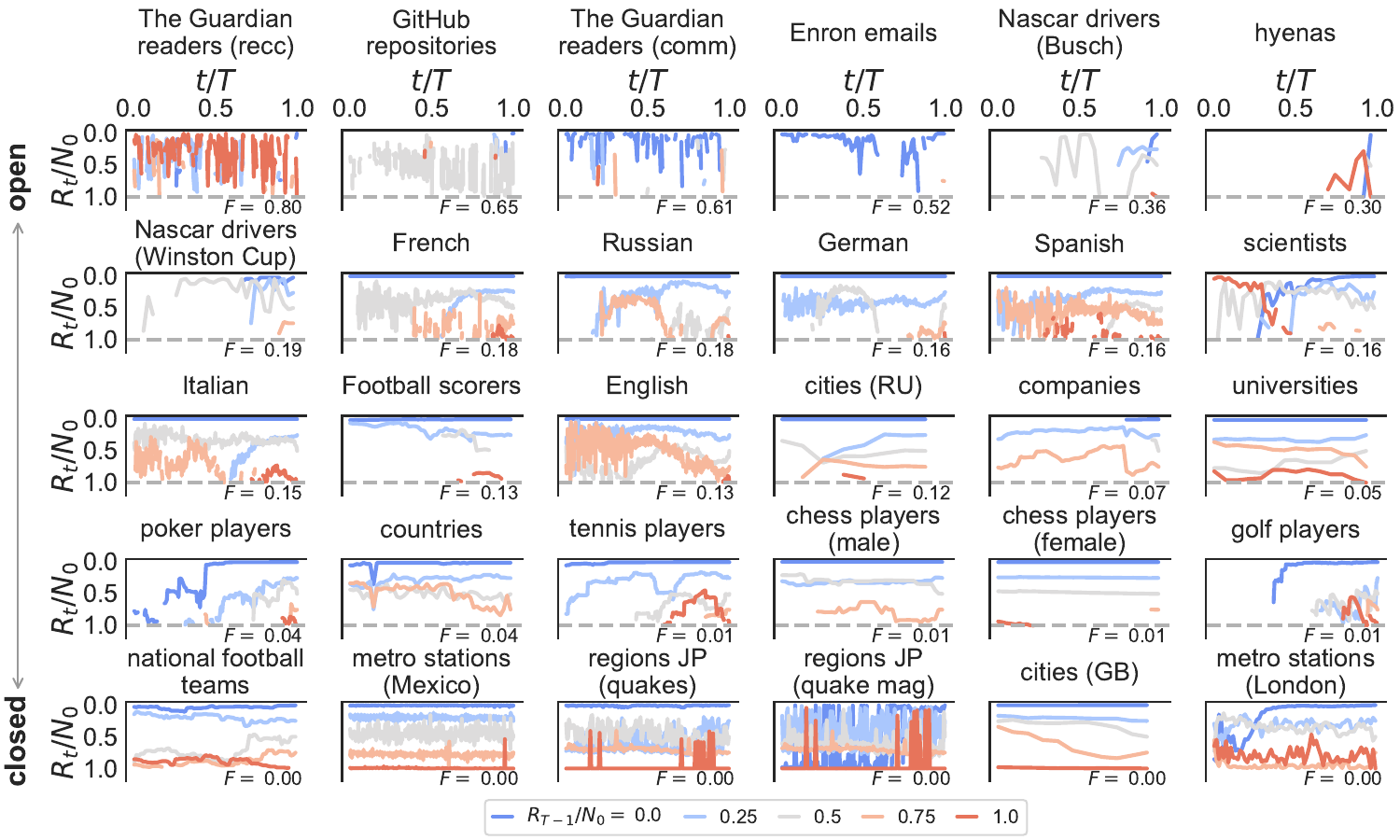}
\caption{\small {\bf Rank dynamics in open to closed ranking lists.} Normalized rank $R_t / N_0$ for selected elements in ranking as a function of normalized time $t / T$. Rank time series are colored according to the last observed rank of the element, $R_{T - 1} / N_0$. The lower part of plots ($ R > N_0$, below dashed lines) represents the unobserved part of an open system, containing elements without known score/rank at the time. Datasets are ordered from most open (upper row) to closed (lower row) according to mean rank flux $F$ [\eref{eq:fluxMean}]. Relatively open ranking lists are more stable at the top (low $R_t$) than at the bottom (high $R_t$). Closed ranking lists are more stable both at the top (low $R_t$) and bottom (high $R_t$), and less stable in the middle.}
\label{fig:dynamics}
\end{figure}

We consider the dynamics of rank $R_t = 1,
\ldots, N_0$ of a given element as a function of time $t$.
In most open ranking
lists ($\Nt > N_0$), the time series $R_t$ at the top of the ranking ($R_t \ll
N_0$) tend to be stable, while $R_t$ fluctuates more as we go down in ranking
(middle rows in \fref{fig:dynamics}). If at some time $t$ an element is
not in the ranking, its rank $R_t$ is not observable from empirical data (because
we do not know the corresponding score). For some open ranking lists even the
top of the ranking is not stable, since the time scale $T$ is larger than the
typical lifetime of an element in the system (see, e.g., upper rows in
\fref{fig:dynamics}, or datasets where people or companies have not been active
during the whole observation period). In closed ranking lists ($\Nt = N_0$),
both the top and the bottom of the ranking are stable, while the middle part of
the ranking shows more variation in $R_t$ (lower rows in \fref{fig:dynamics}).

\subsection{Rank flux} 
\label{ssec:flux}

\begin{figure}[t]
\centering
\includegraphics[width=\textwidth]{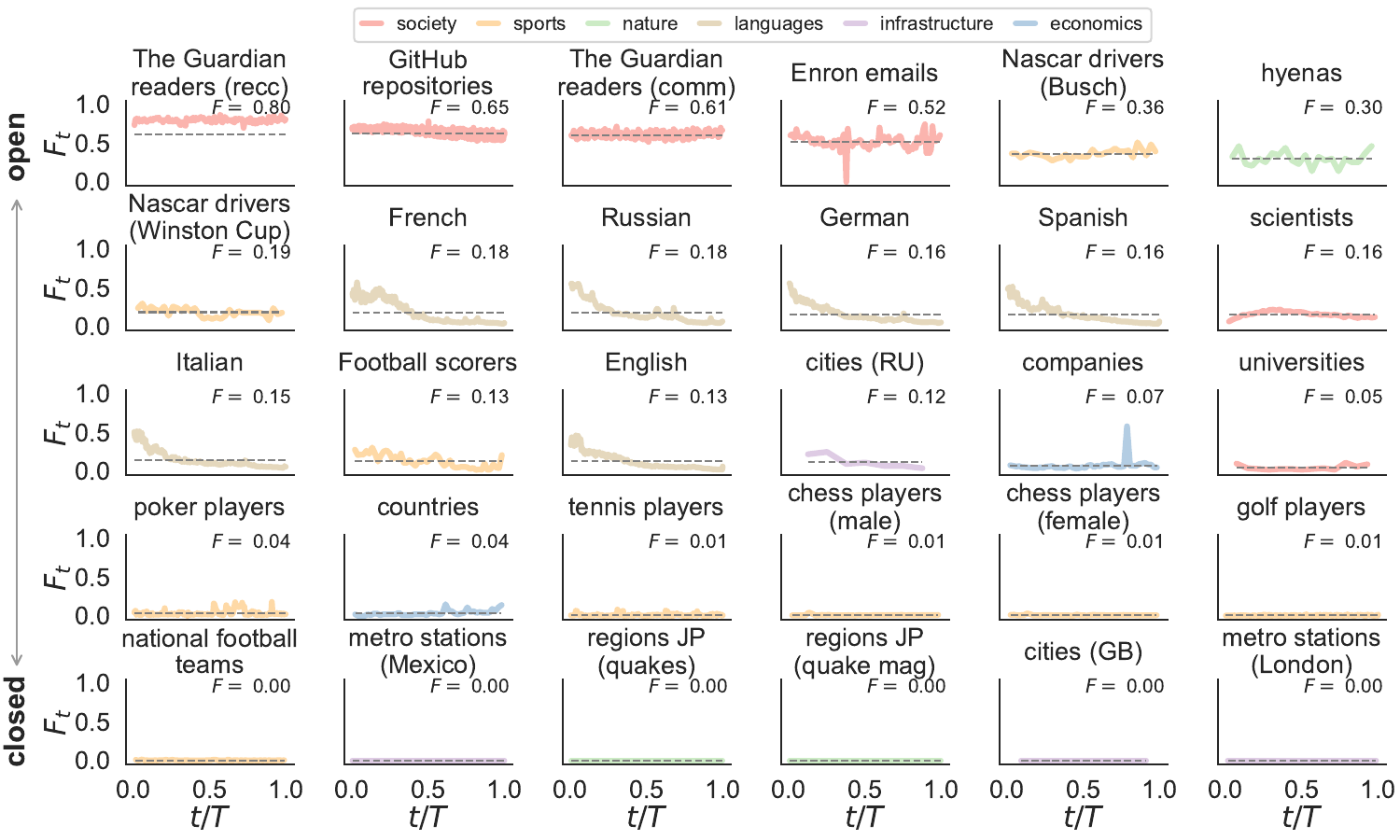}
\caption{\small {\bf Rank flux in open to closed ranking lists.} Flux $\Ft$ as
a function of normalized time $t / T$, calculated as the probability that an
element in (out of) the ranking list at time $t - 1$ leaves (enters) the
ranking at time $t$, averaged over all elements in the ranking at the time.
Flux is shown for data (continuous lines) and the model of \sref{sec:model}
with \eref{eq:fluxModel} for $s = N$ (dashed lines) (for parameter fitting see
\tref{tab:parameters} and \sref{sec:fitting}). Datasets are ordered from most
open (upper row) to closed (lower row) according to mean rank flux $F$
[\eref{eq:fluxMean}]. Flux is roughly constant over time, with occasional large
deviations from the mean. In languages, flux has a decreasing trend over time.
Closed ranking lists have zero flux, consistent with $\Nt = N_0$.
}
\label{fig:flux}
\end{figure}

In open ranking lists, elements flow into and out of the ranking (due to scores
not being measured all the time), meaning that an element may not have a
well-defined rank $R_t = 1, \ldots, N_0$ for all times $t = 0, \ldots, T-1$. We
define {\it rank flux} $\Ft$ as the probability that an element in (out of) the
ranking list at time $t - 1$ leaves (enters) the ranking at time $t$, averaged
over all elements in the ranking list~\footnote{Since the ranking size $N_0$ is
constant in time, the number of elements that enter and leave the ranking
between two consecutive observations are equal.}. For most open ranking lists,
flux is roughly constant throughout time (upper rows in \fref{fig:flux}), i.e.
$\Ft$ does not considerably deviate from its mean over time,
\begin{equation}
\label{eq:fluxMean}
F = \frac{1}{T - 1} \sum_{t=1}^{T - 1} \Ft.
\end{equation}
The constant trend in rank flux is sometimes disrupted by large deviations
when a larger or smaller number of elements enter/leave the ranking list. In
language datasets, $\Ft$ has a decreasing trend over time (\fref{fig:flux}),
arguably due to the exceptionally long observation period (centuries instead of
days/years, see \tref{tab:datasets}), or to a very heterogeneous sampling of words
throughout observations.
Closed ranking lists have zero flux, since $\Nt = N_0$ for all times $t$. Values of the
constant flux $F$ in all datasets are reproduced by the model of
\sref{sec:model} [\eref{eq:fluxModel}], except for the most open ranking list
explored [The Guardian readers (recc)], where the fitting process is less precise (see \sref{sec:fitting}).

\begin{figure}[t] 
\centering
\includegraphics[width=\textwidth]{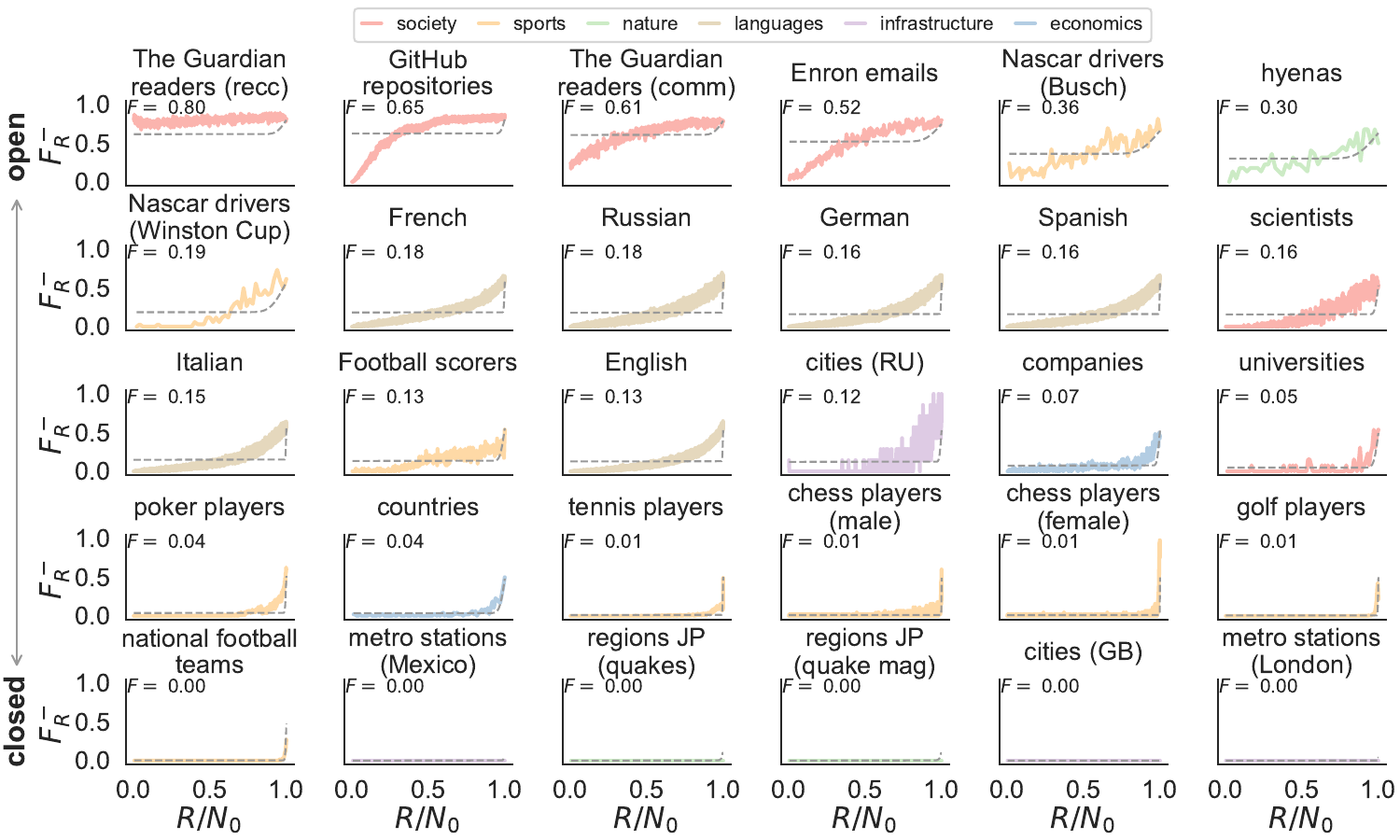}
\caption{\small {\bf Rank out-flux in open systems.} Out-flux $F^-_R$ as a
function of normalized rank $R / N_0$, defined as the probability that an
element in rank $R$ at time $t - 1$ leaves the ranking at time $t$, averaged
over time. Out-flux is shown for data (continuous lines) and the model of
\sref{sec:model} with \eref{eq:outFluxModelApprox} for $s = N$ (dashed lines)
(for parameter fitting see \tref{tab:parameters} and \sref{sec:fitting}).
Datasets are ordered from most open (upper row) to closed (lower row) according
to mean rank flux $F$ [\eref{eq:fluxMean}]. Out-flux has very similar
qualitative behavior to in-flux in all considered datasets.}
\label{fig:Oflux}
\end{figure} 

We define {\it out-flux} $F^-_R$ as the probability that an
element in rank $R$ at time $t - 1$ leaves the ranking list at time $t$,
averaged over all times $t = 1, \ldots, T-1$ (\fref{fig:Oflux}). Similarly, we
define {\it in-flux} $F^+_R$ as the probability that an element out of the
ranking list at time $t - 1$ gets rank $R$ at time $t$, averaged over time.
Out-/in-fluxes are time averages of flux that
determine what part of the ranking contributes more to the flow of elements out
of (and into) it. Even though out-/in-fluxes do not need to be equal in general, they are very
similar in all considered datasets. In most open systems the top of the ranking
(low $R$) is more stable than the bottom (high $R$), since high out-flux tends
to appear only at the bottom. This functional form is recovered by the model of
\sref{sec:model} [\eref{eq:outFluxModelApprox}], except for very open
systems [upper rows in \fref{fig:Oflux}]. On the other hand, the change in
concavity of out-flux (as a function of $R$) for some of the most open systems
is not captured by the model.

\subsection{Rank turnover} 
\label{ssec:openness}

\begin{figure}[t]
\centering
\includegraphics[width=\textwidth]{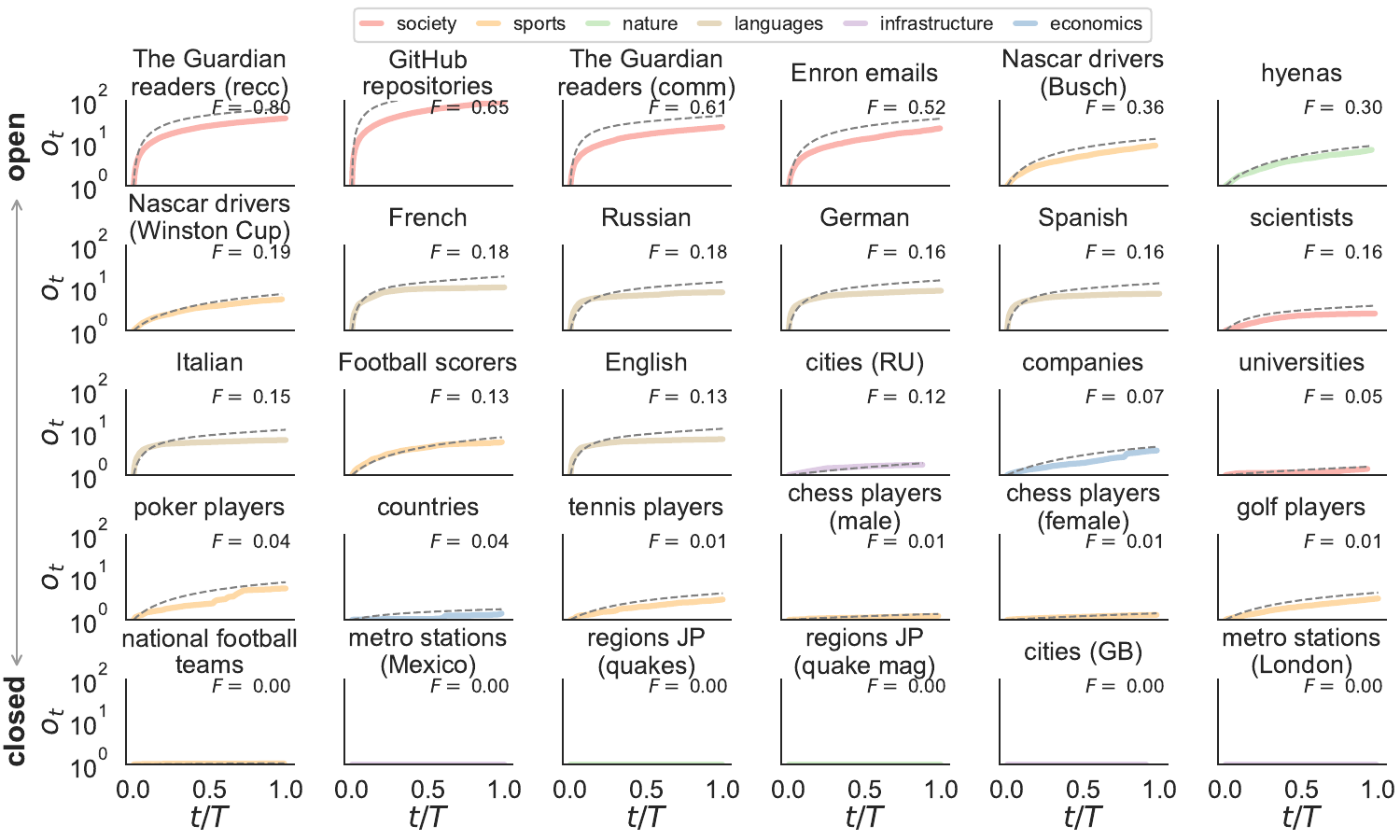}
\caption{\small {\bf Rank turnover in open to closed ranking lists.} Rank
turnover $o_t  = \Nt / N_0$ as a function of normalized time $t / T$, defined
as the number $\Nt$ of distinct elements that have been in the ranking at any
time up to $t$, in units of ranking list size $N_0$ [\eref{eq:openness}].
Turnover is shown for data (continuous lines) and the model of \sref{sec:model}
with \eref{eq:openModelSol} for $s = N$ (dashed lines) (for parameter fitting
see \tref{tab:parameters} and \sref{sec:fitting}). Datasets are ordered from
most open (upper row) to closed (lower row) according to mean rank flux $F$
[\eref{eq:fluxMean}]. Most datasets have a concave (or roughly linear) turnover
in time, while closed ranking lists have constant turnover $o_t = 1$ for all
$t$.}
\label{fig:openness}
\end{figure}

As mentioned in \sref{sec:intro}, $\Nt $ 
is the number of distinct
elements that have ever been in the ranking list up to time $t$ (i.e. at any
$t' = 0, \ldots, t$). Closed systems have $\Nt = N_0$ for any $t$, and open systems have $\Nt
> N_0$ for some $t > 0$. The ranking list
size $N_0$ is the initial
condition of the time series $\Nt$, since in our initial observation we can
only measure $N_0$ scores. Since $\Nt$ counts the elements that have visited
the ranking up until time $t$, it is a monotonically increasing function
($N_{t'} \leq \Nt \le N_{T - 1}$ for $t' \leq t$). The value $N_{T -
1}$ is an observable proxy for the (unknown) size of the system that may
increase with larger $T$, since more observations potentially mean access to
scores of new elements (see related discussion in the model definition of
\sref{ssec:modelDef}). To compare ranking lists of different size $N_0$, we
define the {\it rank turnover} of a ranking list at time $t$ as the number of
elements seen up until that time in units of ranking list size,
\begin{equation}
\label{eq:openness}
o_t = \frac{\Nt}{N_0}.
\end{equation}
In open ranking lists, the time series $o_t$ increases from $o_0 = 1$ to $o_{T-1} = N_{T-1} / N_0$ (see \tref{tab:datasets}), with average slope $\mod_t = ( o_t - o_0 ) / t$ (\fref{fig:openness}). The mean turnover rate after $T$ observations, $\mod \equiv \mod_{T-1} \in [0, 1]$, is a single number that characterises the turnover of a ranking list [just like the mean flux $F$ in \eref{eq:fluxMean}], from $\mod = 0$ for closed systems, to $\mod = 1$ for the most open system possible (where any element in the ranking list is only seen once across observations). In most open datasets considered, turnover $o_t$ has a concave (or roughly linear) functional form in $t$. Our minimal model of rank dynamics introduced in \sref{sec:model} captures the qualitative behavior of turnover and its derivative [see \sref{ssec:modelApproxOpen}, particularly \esref{eq:openModelSol}{eq:openDeriv}].

\subsection{Rank change} 
\label{ssec:change}

\begin{figure}[t]
\centering
\includegraphics[width=\textwidth]{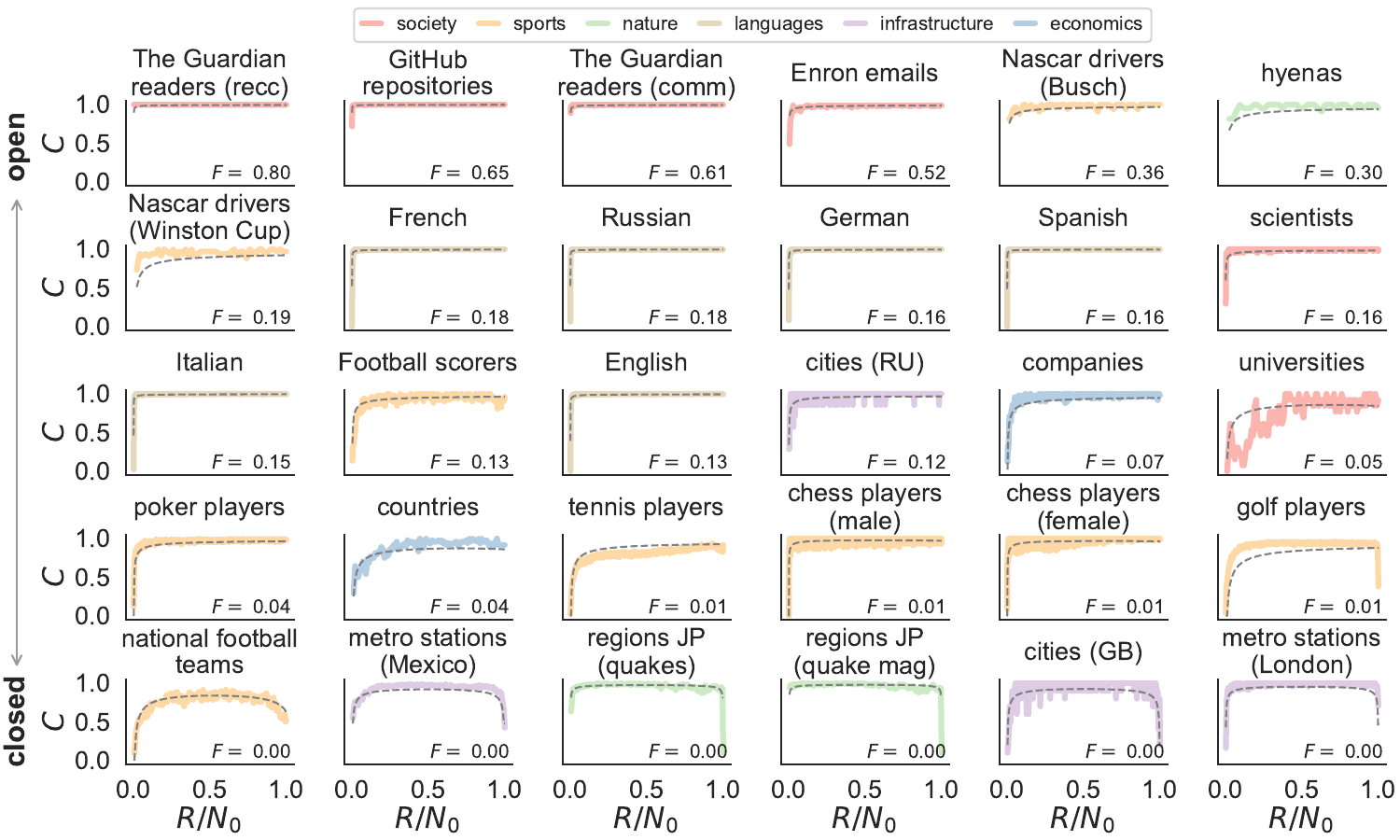}
\caption{\small {\bf Rank change in open to closed ranking lists.} Rank change
$\Cr$ as a function of normalized rank $R / N_0$, calculated as the probability
that elements in rank $R$ at times $t - 1$ and $t$ are not the same, averaged
over time. Rank change is shown for data (continuous lines) and the model of
\sref{sec:model} with \eref{eq:changeModelApprox} for $s = N$ (dashed lines)
(for parameter fitting see \tref{tab:parameters} and \sref{sec:fitting}).
Datasets are ordered from most open (upper row) to closed (lower row) according
to mean rank flux $F$ [\eref{eq:fluxMean}]. In relatively open ranking lists,
rank change is asymmetric: $\Cr$ is lower at the top of the ranking (low $R$)
than at the bottom (high $R$). In progressively more closed ranking lists, rank
change turns roughly symmetric in $R$, with low $\Cr$ also at the bottom of the
ranking.}
\label{fig:change}
\end{figure}

In order to measure the stability of rank dynamics, we introduce {\it rank
change} $\Cr$ as the probability that elements in rank $R$ at times $t - 1$ and
$t$ are not the same, averaged over all times $t = 1, \ldots, T-1$
(\fref{fig:change}). In most open ranking lists, rank change is an asymmetric
function of $R$, with low values at the top of the ranking (low $R$) and high
values elsewhere. Closed ranking lists, in turn, have symmetric $\Cr$, with low
rank change also at the bottom of the ranking (high $R$). In other words, most
systems (except the most open datasets studied) have a stable ranking at the
top where elements keep their rank most of the time. Yet closed ranking lists
are also stable at the bottom, while open ranking lists have an unstable
ranking at the bottom due to a constant flow of elements in and out of the
system (\fsref{fig:flux}{fig:Oflux}).

The qualitative behavior of $\Cr$ is similar to that of {\it rank diversity}
$d_R$, defined as the number of distinct elements occupying rank $R$ over all
times $t = 1, \ldots, T-1$, normalized by $T$. Rank diversity has been
previously introduced and studied by us in detail, particularly in the case of
ranking lists in language and
sports~\cite{cocho2015,morales2016,sanchez2018,morales2018}. While both rank
change and diversity capture the qualitative difference in ranking behavior
between open and closed systems, rank change is more analytically tractable in
our model (see \sref{sec:model}). Following \eref{eq:changeModelApprox}, the
model captures the functional shape and symmetry of $\Cr$ for
both open and closed ranking lists.

\subsection{Rank inertia} 
\label{ssec:success}

\begin{figure}[t]
\centering
\includegraphics[width=\textwidth]{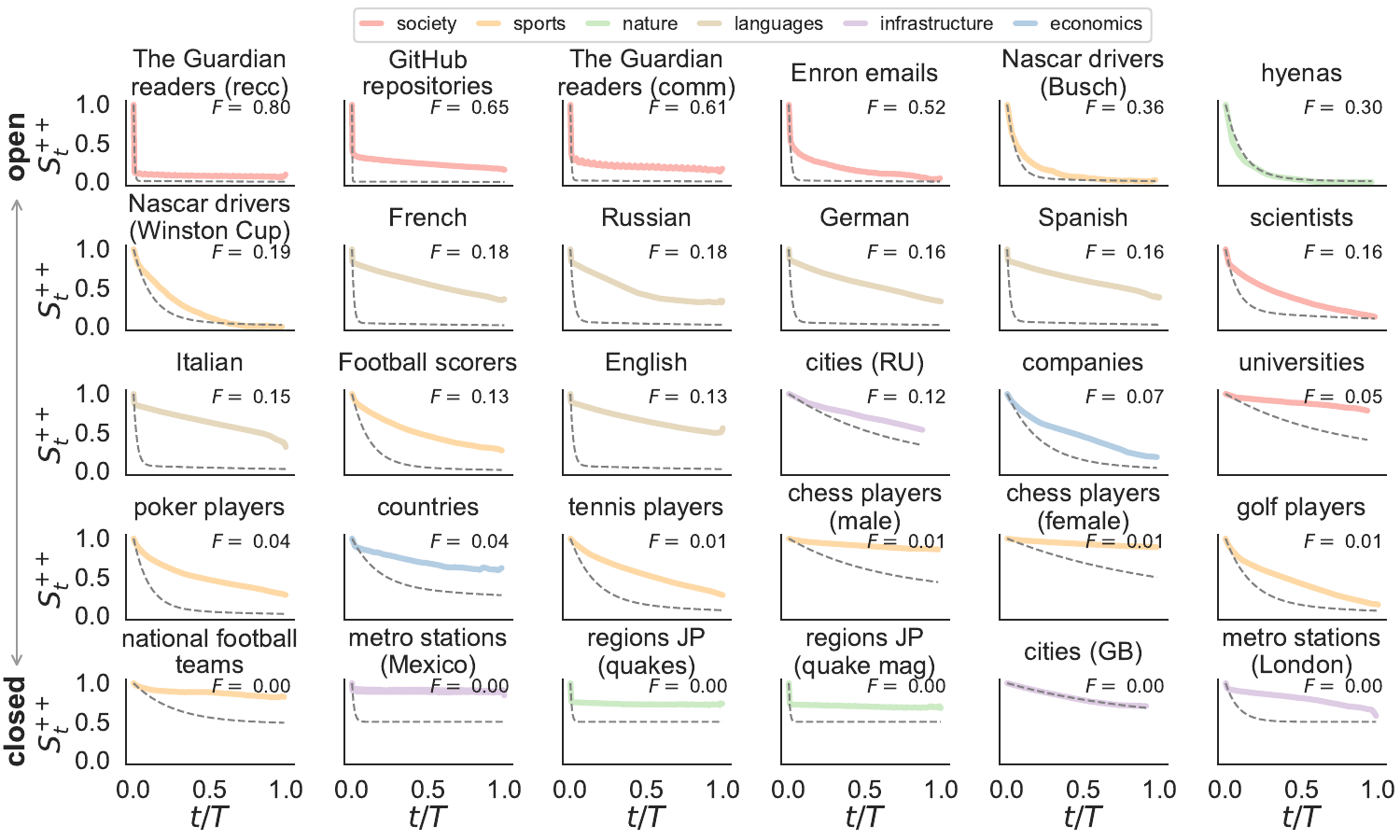}
\caption{\small {\bf Inertia in open to closed ranking lists.} Inertia
$\St$ as a function of normalized time $t / T$, calculated as the
probability that an element in the top of the ranking at time $t'$ will stay
there after $t$ observations, averaged over $t'$. The top of the ranking is
defined as all ranks lower than an arbitrary threshold ($R = 1, \ldots, c
N_0$), set here to $c = 1/2$. Inertia is shown for data (continuous lines) and
the model of \sref{sec:model} with \eref{eq:successModel} for $s = N$ (dashed
lines) (for parameter fitting see \tref{tab:parameters} and
\sref{sec:fitting}). Datasets are ordered from most open (upper row) to closed
(lower row) according to mean rank flux $F$ [\eref{eq:fluxMean}]. In all
ranking lists, the probability of staying in the top longer times decreases
with $t$.}
\label{fig:success}
\end{figure}

We further analyze the flux of elements within the ranking list by dividing it
into two regions via an arbitrary threshold $c \in (0, 1)$: we define the {\it
top} ($+$) of the ranking as all ranks lower than a threshold ($R \le c N_0$),
and the {\it bottom} ($-$) as the rest of the elements in the ranking. 
We define the
matrix element
$S^{ij}_t$ as the probability that an element in region $i$ will move to region
$j$ after $t$ observations (averaged over all compatible times $t' = 0, \ldots,
T - t - 1$), with $i, j \in \{ +, -\}$. The matrix element $S^{ij}_t$
characterizes the flux of elements within the ranking list over time periods of
size $t$. We focus on {\it rank inertia} $\St$, the probability that an element
stays in the top of the ranking list over a time $t$.
Inertia $\St$ is a decreasing function of $t$ for all datasets: while
in some ranking lists the probability of staying in the top is roughly
constant, for most systems it decays linearly, exponentially, or with a long
tail (\fref{fig:success}). In all cases, rank inertia show the same qualitative
behavior for intermediate values of the arbitrary threshold $c$ (defining the
relative sizes of the top and bottom of the ranking), although for some ranking
lists, $c$ has an effect on the functional form of $S^{ij}_t$. The model of
\sref{sec:model} captures the decay of $\St$ with lag
[\eref{eq:successModel}], although there are deviations from the empirical
values for most datasets.

\section{Model of rank dynamics} 
\label{sec:model}

Here we introduce and explore in detail a minimal model for the dynamics of
ranking lists. We derive approximate master equations and their solutions for
both the probability of an element changing rank and some of the rank measures
introduced in \sref{sec:measures}. Our aim is to show how closely numerical
simulations of the model (and their analytical approximations) emulate
microscopic and macroscopic properties of the ranking dynamics of the datasets
described in \sref{sec:data}.
\subsection{Model definition} 
\label{ssec:modelDef}

\begin{figure}[t]
\centering
\includegraphics[scale=1]{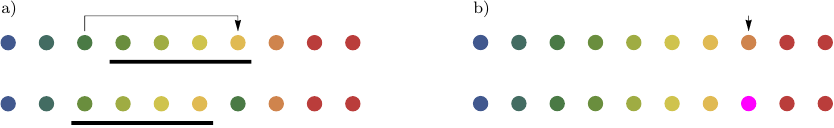}
\caption{
\small {\bf Model of rank dynamics.} At each time step of the model dynamics we perform two independent rank updates. {\bf (a)} In the first update ({\it random displacements of elements}), an element (selected uniformly at random) is reallocated to a random position, also chosen uniformly at random. This element has a (potentially long) displacement in rank space, while the elements between the previous and new position of the selected element (here underlined) move only one rank. Other elements (i.e. in the boundaries of rank space) do not change rank. {\bf (b)} In the second update ({\it random replacement of elements}), we replace an element selected uniformly at random by a new element not previously in the system.}
\label{fig:null}
\end{figure}

Our model is simple and corresponds to the null hypothesis that rank dynamics
is driven by two processes: i) {\it random displacements of elements} across
the ranking list, leading to Lévy flight- and diffusion-like movement in
rank (tuned by a model parameter $\ptau \in [0, 1]$);
and ii) {\it random replacement of elements} by new ones, leading to an
increase in rank flux and turnover (tuned by a model parameter $\pnu \in [0,
1]$). We assume that the model system is a list of $N \geq N_0$ elements (with
$N$ a parameter of the model), and define the normalized rank $r = R \Delta r =
\Delta r, 2 \Delta r, \ldots, \p0, \ldots, 1$, with $\Delta r = 1/ N$ and $\p0
= N_0 / N$.
If $N > N_0$ the system is larger than its ranking list, in which case we choose to
`ignore' the elements in ranks $R = N_0 + 1, \ldots, N$ (despite their
continuing dynamics), emulating lack of access to their scores in empirical
data.
Picturing rank space as a discrete line,
negative/positive displacement means movement to the left/right of a rank in the
line (i.e. to more/less important ranks in the ranking list; see \fref{fig:null})~\footnote{Note that the discrete rank
space $r = \Delta r, \ldots, 1$ becomes the continuous interval $[0, 1]$ in
the limit $N \to \infty$.}.
To compare time scales between model and data, we assume that the real
time interval between observations at times $t$ and $t + 1$ corresponds to $N$
time steps in the model, defining the time step variable $s = 0, 1, \ldots, N,
\ldots, N (T - 1)$, and thus $t = s \Delta r = 0, \Delta r, \ldots, 1, \ldots,
T-1$. In other words, $t$ is a macroscopic time variable that describes the (finite number of) observations of the ranking list in the data, while $s$ is a microscopic time variable that describes the dynamics in the model in the time between empirical observations.

At each time step $s $ we perform two independent rank updates. The first
update, happening with probability $\ptau$, will cause Lévy flight- and diffusion-like dynamics. The second update, happening with probability $\pnu$,
will lead to the replacement of elements in the ranking. 

For the first update, we select an element uniformly at random, remove it from the list
and place it in one of the $N$ spaces to the right or left of the remaining
$N-1$ elements. More precisely, we choose rank $R_{\mathrm{old}} = 1, \ldots,
N$ uniformly at random and take its element temporarily out of the system (thus
making the rank change $R \mapsto R - 1$ for elements previously in ranks $R =
R_{\mathrm{old}} + 1, \ldots, N$). Then
we re-introduce this element to the rank $R_{\mathrm{new}} = 1, \ldots, N$
chosen uniformly at random (making the rank change $R \mapsto R + 1$ for
elements previously in ranks $R = R_{\mathrm{new}}, \ldots, N-1$). In this way,
elements initially between (and including) ranks $R_{\mathrm{old}}$ and
$R_{\mathrm{new}}$ will change rank (if $R_{\mathrm{old}} \neq
R_{\mathrm{new}}$), while elements outside this range will keep their
rank~\footnote{For a mathematical treatment of similar shuffling models,
see~\cite{aldous1986}.} (\fref{fig:null}). 

In the second update, we choose a rank $R_{\mathrm{rep}} = 1,
\ldots, N$ uniformly at random and replace its element with a new element from
outside the system, leaving the rest of the ranks untouched. The old element is
removed from the system and ceases to have any dynamics, while the new element
participates in the rank updates of time step $s+1$ and beyond (\fref{fig:null}).

Each empirical dataset allow us to fix $N_0$ and $T$ in the model (see
\tref{tab:datasets}). In what follows we will derive analytical approximations
of the model dynamics for all values of the free parameters $\ptau$,
$\pnu$ and $N$ (or, alternatively, $\ptau$,
$\pnu$ and $\p0$). Still, it's good to highlight the two parameter ranges that
are sufficient to emulate the dynamics of empirical open and closed ranking
lists as described in \sref{sec:measures}. For {\it closed} ranking lists we
take $N = N_0$ (i.e. $\p0 = 1$) and $\pnu = 0$, so the only free parameter is $\ptau$. In
this parameter range the ranking list covers the whole model system, the ranks
of all elements are known at all times, no new elements enter the system, and
the dynamics of rank is purely driven by Lévy flights and diffusion (as we
see in \sref{ssec:modelDispProb} and \sref{ssec:modelApproxDispProb}). For {\it
open} ranking lists we take $N = N_{T-1}$ (i.e. the model system size is the
observable size of the empirical system, $\p0 = 1 / o_{T-1}$), so the free parameters are $\ptau$
and $\pnu$. In this parameter range the ranking list does not cover the
whole model system ($N_{T-1} > N_0$), so elements may enter/leave the ranking
list from/to the unobserved ranks $R = N_0 + 1, \ldots, N$ due to Lévy
flights and diffusion, and new elements may replace old ones. Our model assumes
that the flow of elements into and out of an open ranking list has two sources:
one coming from a lack of score data outside the ranking list,
and another due to birth/death of elements.

\subsection{Displacement probability} 
\label{ssec:modelDispProb}

At each time step of the dynamics, with probability $\pnu \Delta r$
the element in rank $R$ is replaced by a new element, thus leaving the system.  Alternatively (with
probability $1 - \pnu \Delta r$), the element in rank $R$ is displaced to a
new rank $X$ in the system with probability $\ptau$, either by being
randomly selected with the displacement update rule of the model ({\it Lévy
flight}), or because of the rank change of another element ({\it diffusion}).
Like with $R$, we can define the normalized rank $x = X \Delta r = \Delta r, 2
\Delta r, \ldots, \p0, \ldots, 1$. We explore the
temporal evolution of the model by formulating a master equation for the
displacement probability $\Prxt$, the probability that an element in the
normalized rank $r = R \Delta r$ moves to the normalized rank $x = X \Delta r$
after a time $t = s \Delta r$.

We first consider the displacement probability after a single time step of the
dynamics takes place ($s = 1$), $ P^r_x \equiv P_{x, \Delta r}$. An element in rank $r$ (that is not replaced) moves to
rank $x$ with probability $\ptau$ either by performing a Lévy flight (with
probability $L$) or due to the rank change of another element (with probability
$D^r_x$), or stays in place with probability $1 - \ptau$. Then we have
\begin{equation}
\label{eq:disp1step}
P^r_x = ( 1 - \pnu \Delta r ) 
          \big[ \ptau ( L + D^r_x ) + ( 1 - \ptau ) \delta^r_x \big],
\end{equation}
with $\delta^r_x$ a Kronecker delta. Note that for $\ptau = 1$ and $\pnu
= 0$ (in a closed ranking list with maximum number of displacements in a given
time interval),
 the single-step
displacement probability reduces to $P^r_x = L + D^r_x$. The Lévy probability
$L$ is straightforward to calculate: the element in rank $r$ gets picked up by
the dynamics with probability $\Delta r$, and is placed back in rank $x$ with
the same probability $\Delta r$, since both processes are uniformly random in
rank space. Thus,
\begin{equation}
\label{eq:Levi1step}
L =
\begin{cases}
\Delta r^2 &\quad x \in \{ \Delta r, 2\Delta r,\dots, 1 \}  \\
0 &\quad \text{otherwise}
\end{cases}.
\end{equation}
\eref{eq:Levi1step} explains our motivation to use the term Lévy flight: a Lévy
flight is a random walk where displacement length follows a heavy-tailed
probability distribution; since \eref{eq:Levi1step} does not depend on the
normalized displacement $d = x - r$, the probability $L$ can be thought of as a
discrete power law with exponent 0 (and cut-off due to finite system
size)~\cite{kleinberg2000,li2010}. In other words, our model implements
maximally heavy-tailed Lévy flights where all displacements are equally likely,
beyond diffusion.  If we would extend the model to allow for a
non-uniform sampling of elements and ranks in the mechanisms of displacement
and replacement, we expect that this exponent would become tunable.

The diffusion probability $D^r_x$ is slightly more involved. The element in
rank $r$ moves one step to the right ($x = r + \Delta r$) if we choose an
element to its right  (with probability $1 -
x + \Delta r$) and place it back to
its left (with probability $x - \Delta r$). Conversely, the element in rank $r$
moves one step to the left ($x = r - \Delta r$) if we choose an element to its
left (with probability $x$) and place it back to its right (with probability $1
- x$). The element at rank $r$ can stay in place ($x = r$) only if another
element is picked and placed back to its left [with probability $(x - \Delta
r)^2$] or to its right [with probability $(1 - x)^2$]. Joining these
expressions we obtain
\begin{equation}
\label{eq:diff1step}
D^r_x =
\begin{cases}
x (1 - x) &\quad x = r -\Delta r \\
(x - \Delta r)^2 + (1 - x)^2 &\quad x = r \\
(x - \Delta r) ( 1 - x + \Delta r ) &\quad x = r + \Delta r \\
0 &\quad \text{otherwise}
\end{cases}.
\end{equation}
Note that $\sum_x D^r_x = 1 - \Delta r$. In other words, the probability loss from the diffusive
process in \eref{eq:diff1step} goes to the uniform contribution made by the Lévy term
in \eref{eq:Levi1step}, meaning that the continuous limit of 
\eref{eq:diff1step} does not exactly correspond to a diffusion equation. 
Moreover, $\sum_x P^r_x = 1 - \pnu \Delta r< 1$ (for $\pnu>0$),
which implies a global loss of probability reflecting the fact 
that we replace elements in the dynamics.

Moving forward in the dynamics ($t = s \Delta r \geq 0$), we write a recurrence relation for
the displacement probability $\Prxt$. We observe that, for an element in
rank $r$ to move to rank $x$ after $s$ time steps, it needs to have moved to an
arbitrary rank $x' = \Delta r, \ldots, 1$ over the first $s - 1$ time steps,
and then from $x'$ to $x$ in the last time step, i.e.,
\begin{equation}
\label{eq:dispRecRel}
\Prxt = \sum_{x'} 
P_{x', t - \Delta r}
P^{x'}_{x} 
\end{equation}
with a Kronecker delta as initial condition ($P_{x, 0} = \delta^r_x$), since
at the start of the dynamics, an element in rank $r$ can only be in that rank
($x = r$). Note that \eref{eq:dispRecRel} for $s = 1$ recovers $P_{x, \Delta r} =
P^r_x$ in \eref{eq:disp1step}, as expected.

Let's focus for a moment on the cumulative
$\Prt = \sum_x \Prxt$, the probability that, after a time $t$, the element in rank
$r$ moves to any other rank in the system. Summing up over $x$ in
\eref{eq:dispRecRel} gives the recurrence relation $\Prt = ( 1 - \pnu
\Delta r ) P_{t-\Delta r}$, with initial condition $P_0 = 1$ (due to $P_{x, 0} =
\delta^r_x$) and $P_{\Delta r} = 1 - \pnu \Delta r$. Solving the master
equation gives
\begin{equation}
\label{eq:dispCumRecRel}
\Prt = ( 1 - \pnu \Delta r )^s \simeq e^{- \pnu s \Delta r } = e^{- \pnu t }.
\end{equation}
The approximation in \eref{eq:dispCumRecRel} comes from using the time scale
relation $t = s \Delta r$, the binomial theorem, and the power series of the
exponential, and becomes exact in the limit $N \to \infty$. In other words, the
probability of an element staying in the system under a dynamics of replacement
($\pnu > 0$) decays exponentially with time. When there is only element
displacement ($\pnu = 0$),  $\Prt = 1$ and the ranking list is indeed
closed. Consistently, \eref{eq:dispCumRecRel} is equal to the
probability that an element has not been replaced at any previous time, which decays geometrically or exponentially with time (in the discrete and continuous cases, respectively).

Inserting \esref{eq:disp1step}{eq:diff1step} into \eref{eq:dispRecRel} we obtain a master equation for the displacement probability after $s$ time steps,
\begin{multline}
\label{eq:dispMastEq}
\Prxt = ( 1 - \pnu \Delta r ) \Big\{ P_{x, t-\Delta r} + \ptau \Big[  \Delta r^2 ( 1 - \pnu \Delta r )^{s-1} + x (1 - x) P_{x + \Delta r, t - \Delta r} \\
 - 2 \Big[ x (1 - x + \Delta r) - \frac{1}{2} \Delta r^2 \Big] P_{x, t - \Delta r} + (x - \Delta r) (1 - x + \Delta r) P_{x - \Delta r, t - \Delta r} \Big] \Big\}.
\end{multline}
\eref{eq:dispMastEq} is reminiscent of a discretized diffusion equation in one
dimension (the rank space $x$). The multiplicative factor $1 - \pnu \Delta
r$ represents the dynamics of elements being replaced, thus decreasing their
probability of staying in the system as $s$ increases. The second factor on the
right hand side accounts for elements being selected by the displacement
dynamics of the model and moving (possibly long) distances in rank space as
Lévy flights. The last three terms on the right hand side describe a
(rank-dependent) local, diffusive movement of elements in the system due to
other elements performing Lévy flights. The time scale of both Lévy flights and
diffusion-like movement is regulated by $\ptau$.

\begin{figure}[t]
\centering
\includegraphics[width=\textwidth]{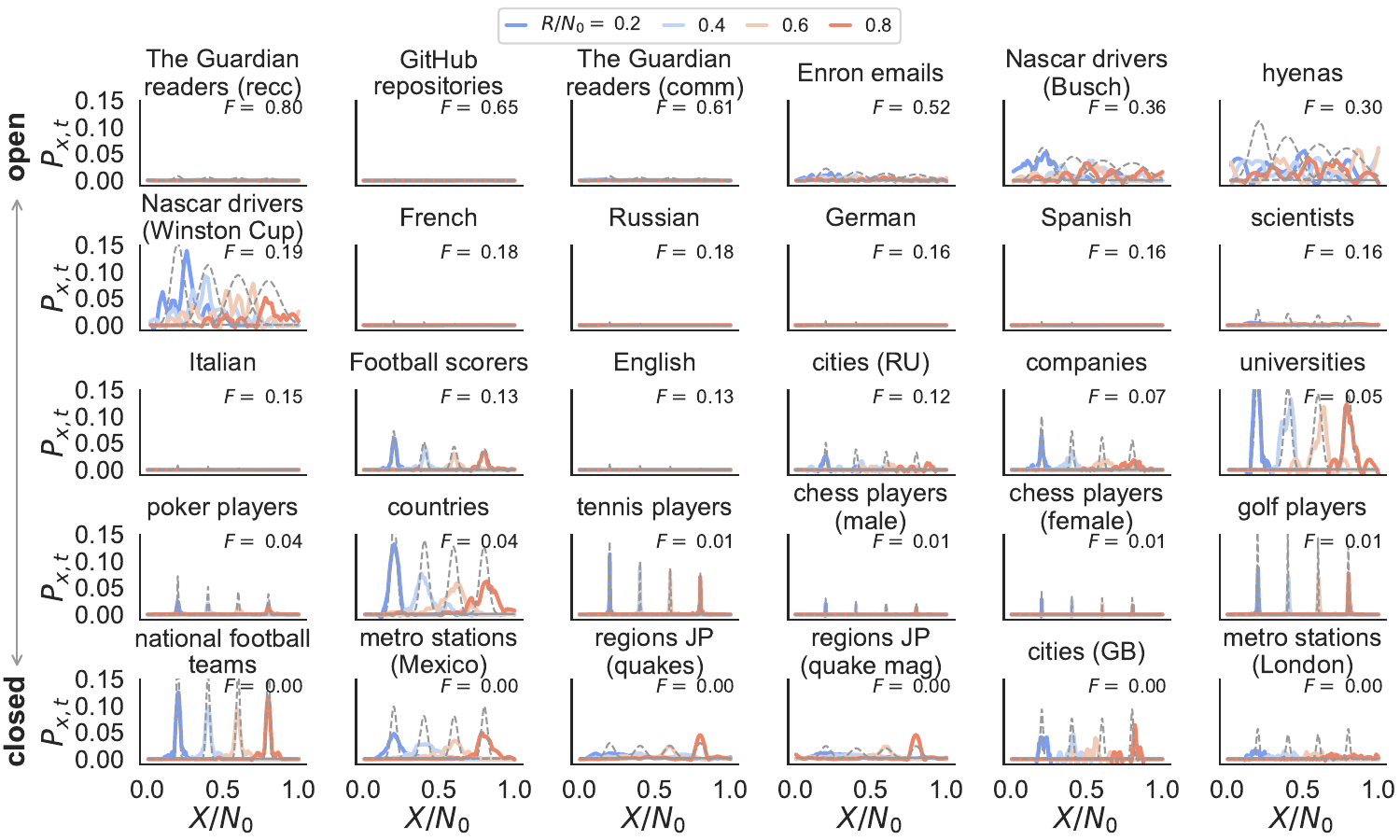}
\caption{\small {\bf Displacement probability in open to closed ranking lists.}
Displacement probability $\Prxt$ for $t=1$ as a function of normalized rank $X / N_0$ for
various $R$ values (colored lines), calculated as the probability that the
element at rank $R$ at time $t'-1$ moves to rank $X$ at time $t'$, averaged over
all times $t'$. $\Prxt$ is shown for data (continuous lines) and the model of
\sref{sec:model} with \eref{eq:aproxDisp} for $s=N$ (dashed lines) (for
parameter fitting see \tref{tab:parameters} and \sref{sec:fitting}). Datasets
are ordered from most open (upper row) to closed (lower row) according to mean
rank flux $F$ [\eref{eq:fluxMean}]. In some datasets $\Prxt$ is roughly flat,
reminiscent of a uniform Lévy sea, while in others the displacement
probability resembles a diffusion peak (see \sref{ssec:modelApproxDispProb}).
To smooth the curves, empirical data is passed through a Savitzky–Golay smoothing filter
(with polynomial order 2 and filter window length ranging from $N_0/200$ to
$N_0/10$).}
\label{fig:displacement}
\end{figure}

We can actually measure the displacement probability $\Prxt$ for $t=1$ directly from empirical data as
the probability that the element at rank $R$ at time $t'-1$ moves to
rank $X$ at time $t'$, averaged over all times $t' = 1, \ldots, T-1$ 
(\fref{fig:displacement}). The use of an average over time is supported by
the stationarity of
the empirical rank dynamics as seen in the roughly constant flux time series
of most datasets (see \fref{fig:flux}).
As we will see in
\sref{ssec:modelApproxDispProb} below, the empirical displacement probability
resembles either a flat Lévy sea or a diffusion peak, respectively
corresponding to late and early times of the rank dynamics in the model. By
calculating $\Prxt$ in
the model for $s = N$ (i.e. between times $t = 0$ and $t = 1$), we see that the model qualitatively reproduces the empirical displacement probability for most datasets and any rank [see \eref{eq:aproxDisp}].

As seen in \eref{eq:dispCumRecRel}, the probability $\Prt = \sum_x \Prxt$ of an element staying in the system for $\pnu > 0$ decays exponentially with time, i.e. $\Prt$ is not conserved. In order to approximately solve \eref{eq:dispMastEq} with a diffusion ansatz (see \sref{ssec:modelApproxDispProb}), we can renormalize $\Prxt$ by introducing the probability distribution
\begin{equation}
\label{eq:renormDispPob}
\Qrxt = \frac{\Prxt}{ ( 1 - \pnu \Delta r )^s } \simeq \Prxt e^{\pnu t },
\end{equation}
which is indeed conserved, since $\sum_x \Qrxt = 1$ for all $t$. As before, the approximation holds for $t = s \Delta r$ and is exact for $N \to \infty$. Substituting \eref{eq:renormDispPob} into \eref{eq:dispMastEq} gives
\begin{align}
\label{eq:renormDispMastEq}
\Qrxt = Q_{x, t-\Delta r} + \ptau \Big\{ & \Delta r^2 + x (1 - x) Q_{x + \Delta r, t - \Delta r} \nonumber \\
& - 2 \Big[ x (1 - x + \Delta r) - \frac{1}{2} \Delta r^2 \Big] Q_{x, t - \Delta r} + (x - \Delta r) (1 - x + \Delta r) Q_{x - \Delta r, t - \Delta r} \Big\}.
\end{align}
\eref{eq:renormDispMastEq} is a simplified master equation for the displacement
probability that effectively decouples the dynamics of displacement
($\ptau$) and replacement ($\pnu$) in the model. We can first look for
an approximate solution $\Qrxt$ of the displacement dynamics governed by
\eref{eq:renormDispMastEq}, and then consider the replacement dynamics
explicitly with $\Prxt = \Qrxt e^{- \pnu t }$. 

\subsection{Approximation for displacement probability} 
\label{ssec:modelApproxDispProb}
Rather than solving \eref{eq:renormDispMastEq} explicitly, we derive an approximate expression for the renormalized displacement probability $\Qrxt$ that becomes more accurate as $N \to \infty$. Considering the shape of \eref{eq:disp1step} (i.e. for $s = 1$), we propose an ansatz of $\Qrxt$ for any $t$ separated into Lévy flight and diffusive components,
\begin{equation}
\label{eq:dispAnyStep}
\Qrxt = \Lt + \Drxt.
\end{equation}
As we will see below, the probability $\Lt$ of moving across rank space in Lévy
flights is rank- and displacement-independent and grows slowly in time (a
uniform {\it Lévy sea}), while the probability $\Drxt$ of diffusing in rank
space due to the movement of other elements is approximately
Gaussian-distributed in $x$ (a {\it diffusion peak} that widens and decreases
in height as time goes by) (\fref{fig:dispProb}).

First, we write a recurrence relation for $\Lt$ by noting that after a time $t = s \Delta r$,
an element initially in rank $r$ can only move due to diffusion within the $2s
+ 1$ ranks around $r$, so $\Qrxt > \Lt$ for $|x-r| \leq t$. Conversely,
the only way for an element to move more than $s$ ranks to each side of $r$ is
by Lévy flight, meaning $\Qrxt = \Lt$ for $|x-r| > t$. Rewriting
\eref{eq:renormDispMastEq} for $|x-r| > t$ we obtain
\begin{equation}
\label{eq:LeviRecRel}
\Lt = \ptau \Delta r^2 + (1 - \ptau \Delta r) L_{t - \Delta r},
\end{equation}
a recurrence relation for $\Lt$ with initial condition $L_0 = 0$ (since at the
start of the dynamics no Lévy flights have yet occurred), which consistently
gives $L_{\Delta r} = \ptau L$ as in \esref{eq:disp1step}{eq:Levi1step}.
\eref{eq:LeviRecRel} can be directly solved and gives an exact expression for
the probability of moving across rank space in Lévy flights,
\begin{equation}
\label{eq:LeviSea}
\Lt = 
  \Delta r [ 1 - (1 - \ptau \Delta r)^s ] \simeq \Delta r ( 1 - e^{- \ptau t } ),
\end{equation}
with $t = s \Delta r$ and the approximation improving as $N \to \infty$. In
other words, the Lévy sea in \eref{eq:LeviSea} is a uniform probability (both
in ranks $r$, $x$ and displacement $d = x - r$) that increases asymptotically
from $\ptau \Delta r^2$ to $\Delta r$ as $t \to \infty$.

By inserting \esref{eq:dispAnyStep}{eq:LeviRecRel} into \eref{eq:renormDispMastEq}, we find a master equation for the probability $\Drxt$ of diffusing in rank space due to the movement of other elements,
\begin{equation}
\label{eq:diffMastEq}
\Drxt = D_{x, t - \Delta r} + \ptau \Big\{ x (1 - x) D_{x + \Delta r, t - \Delta r}
-2 \Big[ x (1 - x + \Delta r) - \frac{1}{2} \Delta r^2 \Big] D_{x, t - \Delta r}
+ (x - \Delta r) (1 - x + \Delta r) D_{x - \Delta r, t - \Delta r} \Big\},
\end{equation}
with initial condition $D_{x, 0} = \delta^r_x$. Just like \eref{eq:renormDispMastEq}, \eref{eq:diffMastEq} is difficult to solve exactly for any $t$. We can, however, write an expression for $\At = \sum_x \Drxt$, the area under the diffusion peak [disregarding the Lévy sea; see \eref{eq:dispAnyStep}]. First, we know that $\sum_x \Qrxt = 1$ due to its definition in \eref{eq:renormDispPob}. Then, by plugging \eref{eq:LeviSea} into this normalization condition we obtain
\begin{equation}
\label{eq:diffPeak}
\At = \sum_{x} \Drxt = (1 - \ptau \Delta r)^s \simeq e^{- \ptau t},
\end{equation}
with $t = s \Delta r$ and the approximation getting better as $N \to \infty$.
In other words, $\Lt = \Delta r (1 - \At)$. Intuitively, the area $\At$ of the
diffusion peak decreases exponentially as time goes by, `leaking probability'
into a Lévy sea that increases in height with $t$.

\subsubsection{Continuum limit for diffusion peak} 
\label{sssec:modelContLimDispProb}

We may find an explicit, approximate solution for the diffusion peak $\Drxt$ by analyzing the continuum limit of \eref{eq:diffMastEq} as $N \to \infty$.
As mentioned in \sref{ssec:modelDef}, in order to match the time scales between
model ($s$) and data ($t$) we take $t = s \Delta r$ with $\Delta r = 1 / N$. We
also use the ansatz $\Qrxt = \Lt + \Drxt$ with the Lévy sea $\Lt$
given by \eref{eq:LeviSea}. Since the diffusion peak leaks probability into the
Lévy sea as time goes by [see \eref{eq:diffPeak}], we further propose the
ansatz
\begin{equation}
\label{eq:diffApprox}
\Drxt = D(x, s \Delta r) \Delta r \simeq D(x, t) dx,
\end{equation}
where $D(x, t)$ is an unknown probability density, $x$ is taken as a continuous
variable, and the approximation improves for larger $N$. Inserting
\eref{eq:diffApprox} into \eref{eq:diffMastEq} we obtain a master equation for
$D(x, t)$,
\begin{equation}
\label{eq:diffEq}
\frac{\partial D}{\partial t} = \alpha x(1 - x) \frac{\partial^2 D}{\partial x^2}.
\end{equation}
\eref{eq:diffEq} is a diffusion-like equation with a quadratic, rank-dependent
diffusion coefficient $\alpha x(1 - x)$ and $\alpha \simeq \ptau \Delta r$
(for large but finite $N$). Note that the probability density $D(x, t)$
decreases in time as the L{\'e}vy sea increases, so \eref{eq:diffEq} is not a
standard diffusion equation. If we would modify \eref{eq:diff1step} to force
$\At = 1$ by, say, adding a term $\Delta r$ in $D^r_r$, we would obtain a
diffusion equation in which the rank-dependent coefficient is inside the first
rank derivative. The initial condition $D_{x, 0} = \delta^r_x$ in
\eref{eq:diffMastEq} leads to the initial condition $D(x, 0) = \delta (x -
x_0)$ (with $x_0 \equiv r \in [0, 1]$ the initial rank of the element) and to
the Dirichlet boundary conditions $D(0, t) = D(1, t) = 0$.
Notice that \eref{eq:diffEq} is the Wright–Fisher equation \cite{muirhead2016}, a continuous model
of genetic drift that has been extensively
studied from a mathematical point of view (see \cite{tran2013,chen2010,epstein2010}
and references therein).

We can solve \eref{eq:diffEq} exactly by separation of variables. Proposing the
ansatz $D(x, t) = u(x) v(t)$ leads to $v(t) = v(0) e^{-\lambda t}$ with
$\lambda$ a separation
constant, while $u(x)$ needs to satisfy the eigenvalue equation
\begin{equation}
\label{eq:eigenEq}
\alpha x(1 - x)u'' + \lambda u = 0.
\end{equation}
By using the Frobenius method and the boundary conditions of \eref{eq:diffEq},
we determine the allowed values of $\lambda$ implicitly and the associated
eigenfunctions $u(x)$ as infinite series, over which the initial condition
$D(x, 0)$ can be expanded to obtain the particular solution of \eref{eq:diffEq}
we are looking for.

\begin{figure}[t]
\centering
\includegraphics[width=0.9\textwidth]{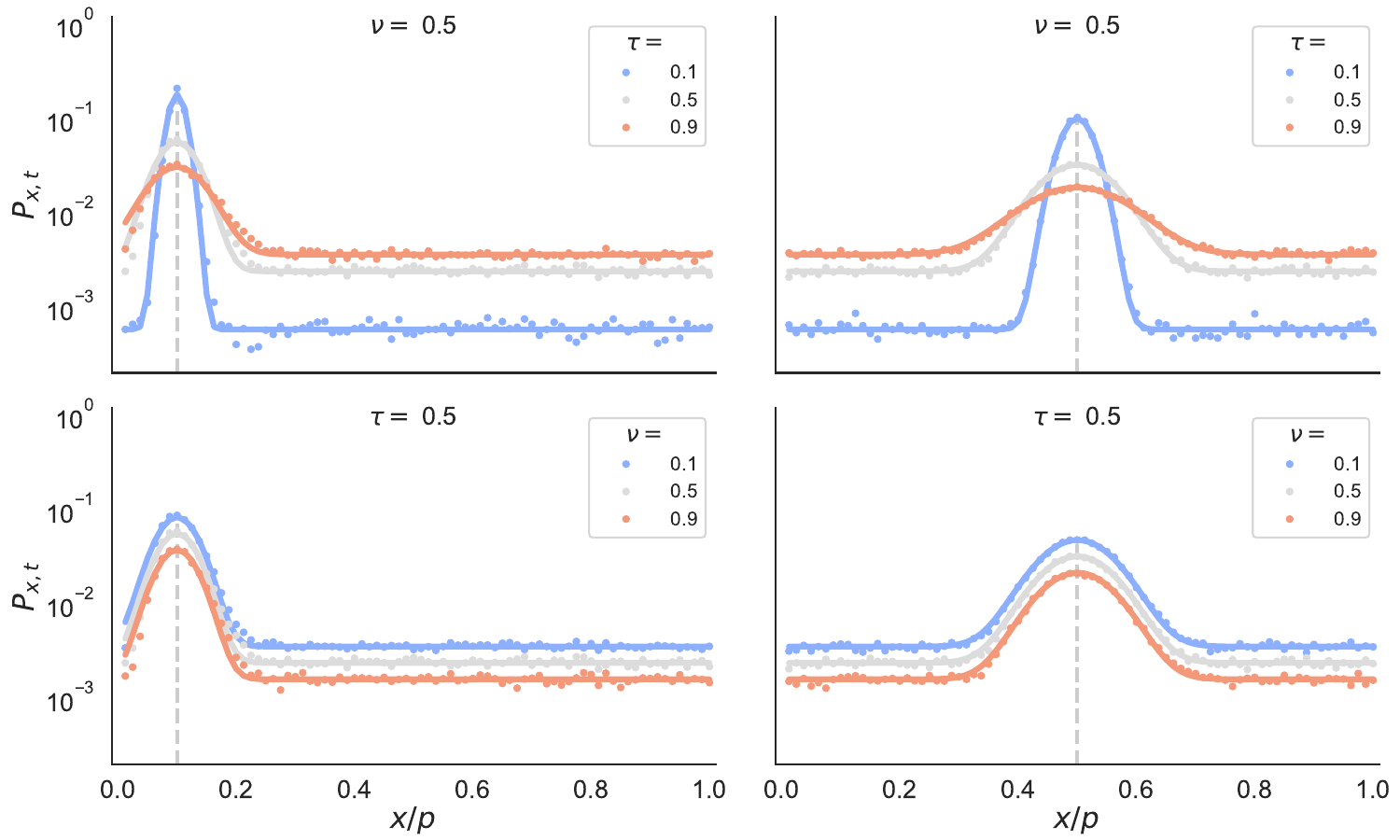}
\caption{\small {\bf Displacement probability in model.}  Displacement
probability $\Prxt$ as a function of normalized rank $x / \p0$ for
$t = 1$ ($s = N$), with $N = 10^2$ and $\p0 = 0.8$, in numerical
simulations of the model (dots) and the analytical approximation of
\eref{eq:aproxDisp} (lines). {\bf (top row)} Dependence of $\Prxt$ on
$\ptau$ for fixed $\pnu = 0.5$, with initial rank $r = 0.08$ (left) and $0.4$ (right). As $\ptau$ increases, the diffusion
peak around $x = r$ (dashed line) widens and decreases in height, while the uniform Lévy sea
grows according to \eref{eq:LeviSea}. {\bf (bottom row)} Dependence of $\Prxt$
on $\pnu$ for fixed $\ptau = 0.5$, with initial rank $r = 0.08$ (left) and $0.4$ (right). As $\pnu$ increases, the
displacement probability is exponentially lower due to element replacement [see
\eref{eq:dispCumRecRel}]. \eref{eq:aproxDisp} is accurate even for relatively
low $N$. Note that the probability peak is less diffusive at the extremes of rank space (left) than at the center (right). Simulations are averages over $10^5$ realizations (using $T = 10$).
}
\label{fig:dispProb}
\end{figure}

However, it may be more instructive to find a closed-form approximation for
$D(x, t)$ that lets us qualitatively understand the behaviour of the
displacement probability in rank and time. In a small enough interval around
the initial rank $x_0 = r$, we can approximate $x(1 - x) \simeq x_0 (1 - x_0)$
in \eref{eq:diffEq}, which leads to the standard diffusion equation with
diffusion coefficient $\alpha x_0 (1 - x_0)$. Then, the fundamental solution
(in the infinite rank domain $-\infty < x < \infty$) for $D(x, 0) = \delta (x -
x_0)$ is $G( x_0, \sigma_t^2 )$, the Gaussian distribution with mean $x_0$ and time-dependent standard deviation $\sigma_t = \sqrt{2 \alpha x_0
(1 - x_0) t}$, which spreads symmetrically around $x_0$ as time goes
by~\footnote{Note that the Gaussian approximation fails progressively as we
approach the `tails' of the solution at $x \simeq 0, 1$, both due to
the failing assumption $x(1 - x) \simeq x_0 (1 - x_0)$ and to the mismatch
between the boundary conditions of \eref{eq:diffEq} and the infinite rank
domain of \eref{eq:diffGaussSol}. The overall effect is more accuracy for intermediate values of $x_0 = r$ rather than at the extremes of the interval $[0, 1]$ (\fref{fig:dispProb}).}. This allows us to write
\begin{equation}
\label{eq:diffGaussSol}
D(x, t) \simeq e^{- \ptau t} G( x_0, \sigma_t^2 ) = e^{- \ptau t} \frac{1}{ \sigma_t \sqrt{2 \pi}} \exp \left[ -\frac{1}{2} \left( \frac{x - x_0}{ \sigma_t } \right)^2 \right],
\end{equation}
where the term $e^{- \ptau t}$ is necessary for the Gaussian approximation to comply with \eref{eq:diffPeak}.

Finally, we derive an approximate expression for the displacement probability
$\Prxt = \Qrxt e^{- \pnu t }$ that becomes more accurate as $N \to
\infty$. Inserting \eref{eq:diffGaussSol} into \eref{eq:diffApprox}, and using
the ansatz of \eref{eq:dispAnyStep} alongside \eref{eq:LeviSea} we obtain
\begin{equation}
\label{eq:aproxDispSimple}
\Prxt \simeq e^{- \pnu t} \left[ \Lt + e^{- \ptau t} G( r, \sigma_t^2 ) \Delta r \right],
\end{equation}
with $t = s \Delta r$, or writing everything explicitly,
\begin{equation}
\label{eq:aproxDisp}
\Prxt \simeq e^{- \pnu t} \left[ \Delta r ( 1 - e^{- \ptau t} ) +
\sqrt{ \frac{ \Delta r }{4 \pi \ptau r (1 - r) t }} \exp \left( -\frac{ (x-r)^2 }{4 \ptau r (1 - r) t \Delta r} - \ptau t \right) \right].
\end{equation}
\eref{eq:aproxDispSimple} captures the approximate behavior of the displacement
probability $\Prxt$ intuitively: a Gaussian diffusion peak $G( r, \sigma_t^2 )$
widening in time and leaking probability to the uniform Lévy sea
$\Lt$, all regulated by an exponential loss in probability due to new elements.
\eref{eq:aproxDisp} is a good approximation for $\Prxt$ even for relatively low
$N = 10^2$, as we can see by comparing with numerical simulations of the model
described in \sref{ssec:modelDef} for $s = N$ (i.e. $t = 1$)
(\fref{fig:dispProb}). \eref{eq:aproxDisp} also shows that the model
captures the displacement probability of the empirical
datasets (see \fref{fig:displacement}).

\subsection{Approximation for rank flux} 
\label{ssec:modelApproxFlux}

Beyond the microscopic description of the dynamics given by the displacement
probability $\Prxt$, we also explore the temporal evolution of the
model
by approximating the rank measures introduced in
\sref{sec:measures} with closed expressions. We start with the mean rank flux
$F$, measured for data as the probability that any element in the ranking list
at time $t - 1$ leaves the ranking at time $t$, averaged over all recorded (observable) elements in
the ranking list and over time [see \sref{ssec:flux} and \eref{eq:fluxMean}].
To find $F$ in the model, we define the time-dependent flux $\Ft$ as the probability that a given element in rank $r
\le \p0$ leaves any of these ranks after time $t = s \Delta r$ (either
by displacement or by replacement).
Following \eref{eq:renormDispPob}, flux is given by
\begin{equation}
\label{eq:OutFluxDef}
\Ft = 1 - \sum_{x = \Delta r}^{\p0} \Prxt = 1 - ( 1 - \pnu \Delta r )^s \sum_{x = \Delta r}^{\p0} \Qrxt.
\end{equation}
where the step size in the sum is $\Delta r$, i.e. $x = \Delta r, 2 \Delta r, \ldots, \p0$.

We now find a master equation for the partial cumulative $\Qrabt
\equiv \sum_{x=a}^b \Qrxt$ for arbitrary $a, b = \Delta r, \ldots, 1$ and $b>a$ \footnote{This general notation will be useful to determine both the flux $\Ft$ via \eref{eq:OutFluxDef} by setting $a = \Delta r$ and $b = \p0$, as well as the inertia $\St$ via \eref{eq:successDef} by setting $a = \Delta r$ and $b = c \p0$ (see \sref{ssec:modelApproxSuccess} below).}.
Using \eref{eq:renormDispMastEq}, summing over $x$, and changing
dummy indices, we obtain
\begin{align}
\label{eq:CumulQMastEq}
& \Qrabt = (1 - \ptau \Delta r) Q_{[a, b], t-\Delta r} \nonumber \\
& + \ptau \Big[ (b - a + \Delta r) \Delta r + (a - \Delta r) (1 - a +\Delta r) ( Q_{a - \Delta r, t-\Delta r} - Q_{a, t-\Delta r} ) + b (1 - b) ( Q_{b + \Delta r, t-\Delta r} - Q_{b, t-\Delta r} ) \Big].
\end{align}
\eref{eq:CumulQMastEq} is not closed on the cumulative $\Qrabt$, since
it depends directly on the renormalized displacement probability (last two
terms on the right side of the equation). To close the master equation we make
the approximations $Q_{a - \Delta r, t} \simeq Q_{a, t}$ and $Q_{b +
\Delta r, t} \simeq Q_{b, t}$ for all $t$ (accurate as long as the diffusion
peak of $\Prxt$ is far enough from $a$ or $b$), which leads to
\begin{equation}
\label{eq:CumulQMastEqApprox}
\Qrabt = (1 - \ptau \Delta r) Q_{[a, b], t-\Delta r} + \ptau (b - a + \Delta r) \Delta r.
\end{equation}

\begin{figure}[t]
\centering
\includegraphics[width=0.9\textwidth]{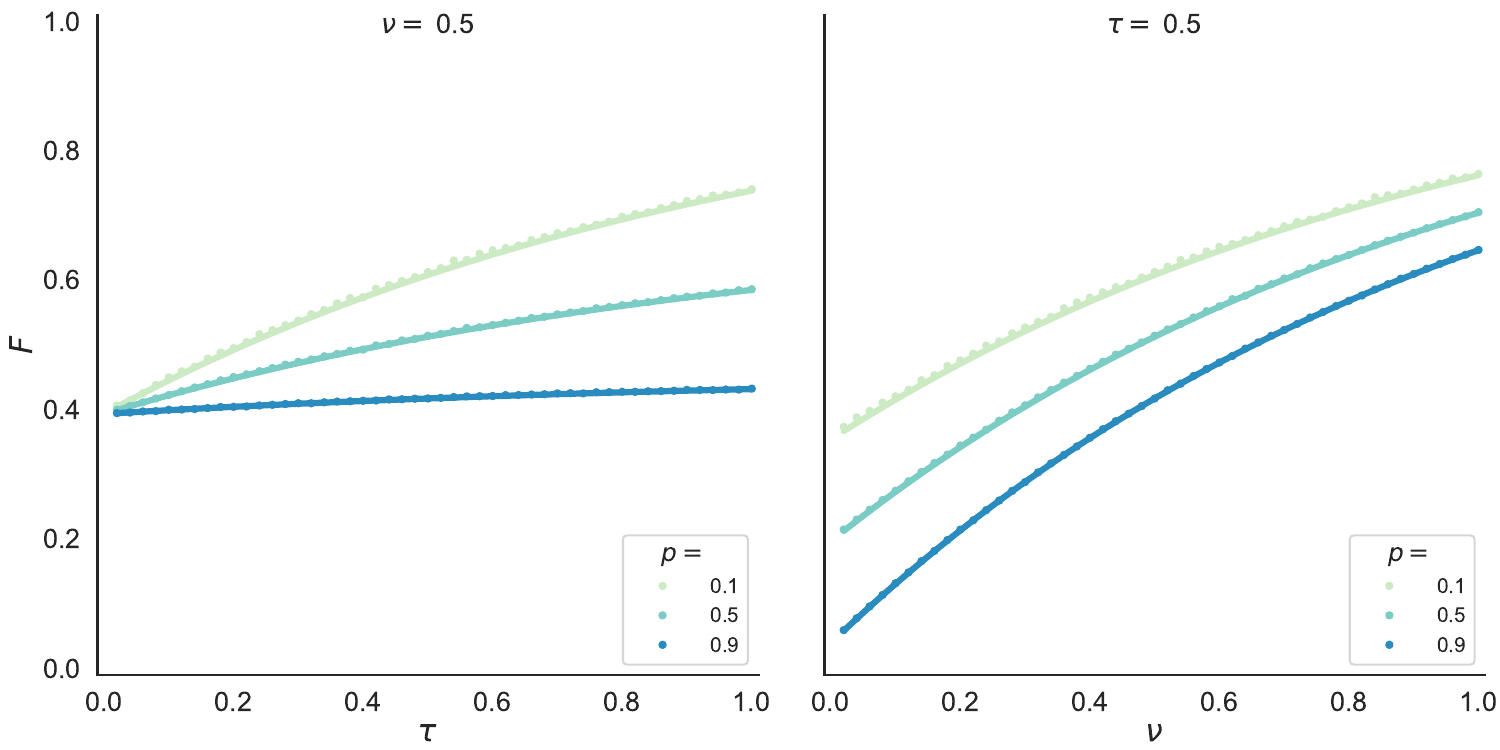}
\caption{\small {\bf Rank flux in model.} Rank flux $F$ as a function of
$\ptau$ for fixed $\pnu$ {\bf (left)}, and as a function of $\pnu$ for
fixed $\ptau$ {\bf (right)}, in both numerical simulations of the model
(dots) and the analytical approximation of \eref{eq:fluxModel} (lines) for
varying ranking list size $\p0$. Rank flux increases with both $\ptau$ and
$\pnu$, since the displacement/replacement dynamics takes elements out of
the ranking list. A relatively longer ranking list (large $\p0$) makes it less
likely that elements leave the ranking list. Simulations are averages over
$10^2$ realizations (with $T = 10$) in a system of size $N = 10^4$.}
\label{fig:fluxModel}
\end{figure}

We go back to the particular case of the flux $\Ft$. With $a = \Delta r$, $b = \p0$, and the initial condition $Q_{[\Delta r,
\p0], 0} = 1$ (since the element starts off within the ranking list), we can
solve \eref{eq:CumulQMastEqApprox} using common formulas for geometric series to obtain
\begin{equation}
\label{eq:CumulQsol}
\sum_{x = \Delta r}^{\p0} \Qrxt = \p0 + ( 1 - \p0 ) ( 1 - \ptau \Delta r )^s \simeq \p0 + ( 1 - \p0 ) e^{- \ptau t}
\end{equation}
for $t = s \Delta r$ and sufficiently large $N$. 
Due to the approximation in \eref{eq:CumulQMastEqApprox}, $\Qrabt$ and
$\Ft$
no longer depend explicitly on $r$.
For $t=1$ (i.e. $s = N$) we finally obtain
\begin{equation}
\label{eq:fluxModel}
F \simeq 1 - e^{- \pnu} [ \p0 + ( 1 - \p0 ) e^{- \ptau} ].
\end{equation}
Additionally, we define $F = 0$ for $\p0 = 0$ for consistency.
\eref{eq:fluxModel} is an analytical approximation
for the probability that any element in the ranking list at time $t - 1$ leaves
the ranking at time $t$, according to our model. Intuitively, flux $F$
increases with $\ptau$ and $\pnu$, since more displacements or
replacements make it more likely that elements will leave the ranking list.
Conversely, larger $\p0$ makes the ranking list longer (relative to system
size) and thus less likely to lose elements due to out-flux.
\eref{eq:fluxModel} is a good approximation of $F$ for all parameters
considered, as we can see by comparing with numerical simulations of the model
described in \sref{ssec:modelDef} (\fref{fig:fluxModel}). \eref{eq:fluxModel}
also shows that the model
reproduces the empirical flux of most
datasets (see \fref{fig:flux}).

\begin{figure}[t]
\centering
\includegraphics[width=0.9\textwidth]{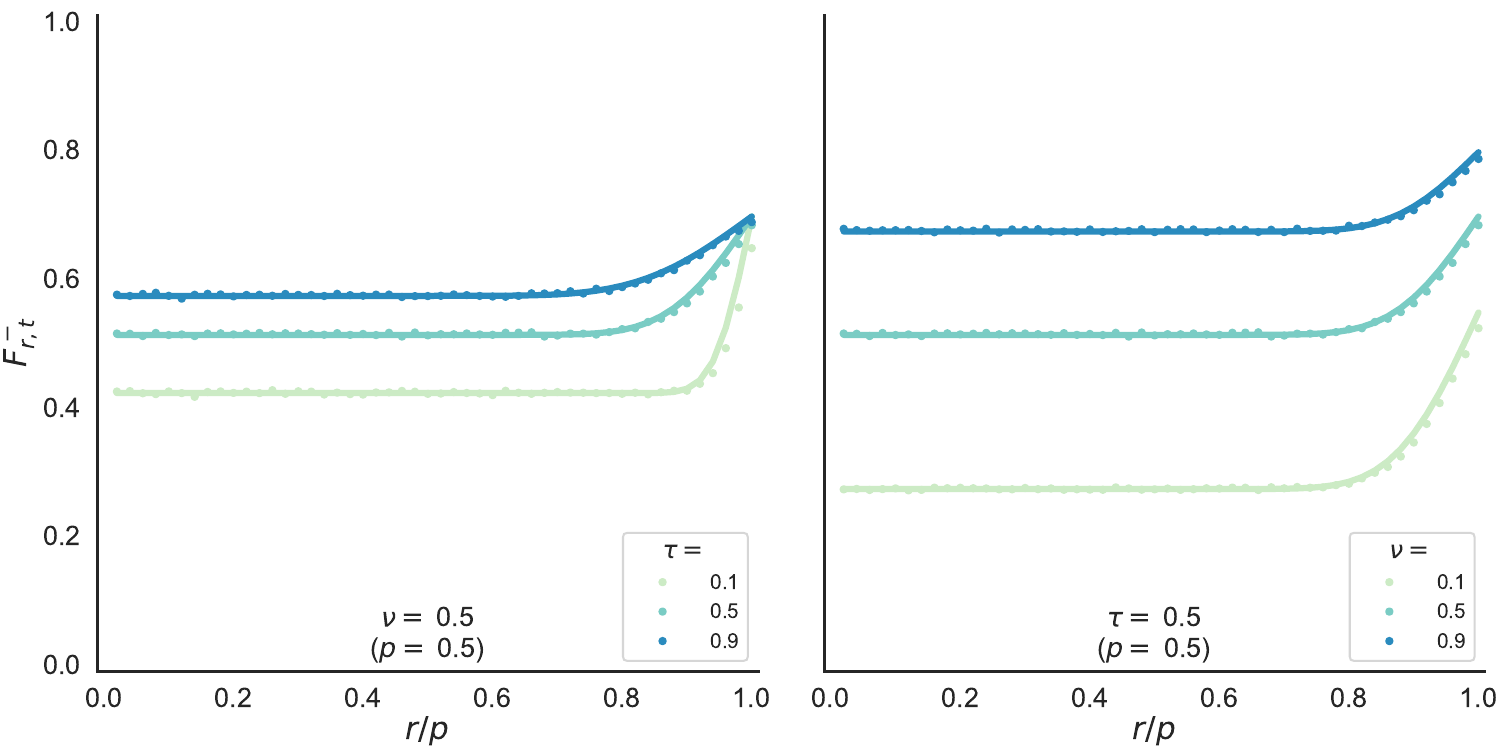}
\caption{\small {\bf Rank out-flux in model.}
Rank out-flux $\Fmrt$ as a function of normalized rank $r / \p0$ for $t = 1$ ($s =
N$) in both numerical simulations of the model (dots) and the
analytical approximation of \eref{eq:outFluxModelApprox} (lines) for fixed
ranking list size $\p0$. Results correspond to varying $\ptau$ for fixed
$\pnu$ {\bf (left)}, and to varying $\pnu$ for fixed $\ptau$ {\bf
(right)}. Rank out-flux increases with both $\ptau$ and $\pnu$, since the
displacement/replacement dynamics takes elements out of the ranking list.
$\Fmrt$ is also relatively constant across the ranking list, except for
bottom ranks ($r \sim \p0$) where out-flux increases. Simulations are averages
over $10^5$ realizations (with $T = 10$) in a system of size $N = 10^2$.}
\label{fig:OfluxModel}
\end{figure}

As described in \sref{ssec:flux}, in the data we can also measure what part of a
ranking list contributes most to the flow of elements out of it by calculating
the out-flux $F^-_R$: the probability that the element in rank $R$ at time $t -
1$ leaves the ranking list at time $t$, averaged over all observed times.
Similarly, in the model we define out-flux $\Fmrt$ as the probability that
the element in the normalized rank $r$ leaves the ranking list after time $t = s \Delta r$. We can compare $F^-_R$ in the data with $\Fmrt$ for $t = 1$ (i.e. $s = N$) in the model.
An element
contributes to out-flux either because it is replaced with a new element, or
because it moves out of the ranking list via the displacement dynamics. Thus,
\begin{equation}
\label{eq:outFluxModelDef}
\Fmrt = 1 - ( 1 - \pnu \Delta r )^s + \sum_{x = \p0 + \Delta r}^1 \Prxt,
\end{equation}
where the displacement probability $\Prxt$ is given explicitly by
\eref{eq:aproxDisp} in the limit of large $N$. Approximating the
sum in \eref{eq:outFluxModelDef} by an integral in the interval $x \in [\p0,
\infty]$, changing variables and integrating by parts, we obtain
\begin{equation}
\label{eq:outFluxModelApprox}
\Fmrt \simeq 1 - e^{- \pnu t} \left\{ 1 - (1 - \p0) (1 - e^{- \ptau t}) - \frac{1}{2} e^{- \ptau t} \left[ 1 - \text{erf} \left( \frac{ \p0 - r }{ 2 \sqrt{ \ptau r(1 - r) t \Delta r } } \right) \right] \right\},
\end{equation}
with $t = s \Delta r$ and erf the error function. In the case $\ptau = 0$,
$P^r_x = ( 1 - \pnu \Delta r ) \delta^r_x$ in \eref{eq:disp1step} leads to
$\Fmrt \simeq 1 - e^{- \pnu t}$, i.e. a constant out-flux with respect
to rank $r$. \eref{eq:outFluxModelApprox} is a very good approximation for
out-flux in the model, showing how $\Fmrt$ increases with both $\ptau$
and $\pnu$ (due to enhanced displacement and replacement dynamics), see
\fref{fig:OfluxModel}. Out-flux is also relatively constant across the
ranking list, except for bottom ranks ($r \sim \p0$) where it increases as we
approach the end of the ranking list. \eref{eq:outFluxModelApprox}
reproduces the empirical out-flux of several datasets (\fref{fig:Oflux}).

\subsection{Approximation for rank turnover} 
\label{ssec:modelApproxOpen}

Here we derive an approximate, closed expression for the rank turnover $o_t =
\Nt / N_0$,
where $\Nt$ is the number
of distinct element that have been in the ranking list up to (and including)
time $t$ [see \eref{eq:openness}]. Similarly, in the model $\Nt$ is
the number of distinct elements that have been in the ranking  list ($r =
\Delta r, \ldots, \p0$) at any time step $s' \leq s$ (with $t = s \Delta r$). 
Thus, our task is to find an explicit expression for $\Nt$. We start
by introducing the probability $\pt$ that a randomly chosen element
has been in the ranking list at any time $t' \leq t$,
\begin{equation}
\label{eq:OpenPsDef}
\pt = \frac{\Nt}{\Mt},
\end{equation}
where $\Mt$ is the number of distinct elements that have been in the whole
system at any time $t' \leq t$. Since the replacement dynamics adds one new
element every time step with probability $\pnu$, on average we have $\Mt \simeq N + \pnu
s$,
which leads to
\begin{equation}
\label{eq:OpenNs}
\Nt \simeq (N +  \pnu s) \pt.
\end{equation}
In order to find $\pt$, we write a master equation for the probability $\pxt$
that the element in any rank $x$ of the model system \footnote{$x =
\Delta r, \ldots, 1$.} has been in
the ranking list at any time $t' \leq t$, which we average to obtain $\pt =
\Delta r \sum_x \pxt$.

We first note that the initial condition for $\pxt$ is
\begin{equation}
\label{eq:openIniCond}
p_{x, 0} =
\begin{cases}
1 &\quad x = \Delta r, \ldots, \p0 \\
0 &\quad \text{otherwise}
\end{cases},
\end{equation}
a step function over rank space. Within the ranking list, $\pxt$ is always 1 and does not change in time, i.e., $\pxt = p_{x, t - \Delta r} = 1$ for $x = \Delta r, \ldots,
\p0$ and $t \geq \Delta r$. In the rest of the system, $\pxt$ increases from its initial condition 0.
We now write the corresponding equation for elements
outside of the ranking list, in a similar way to \eref{eq:dispRecRel}. 
For an element in rank $x$ to have ever been in the ranking list, it must have belonged to the ranking list at some point in its dynamics, and finally moved from some rank $r$ to rank $x$ in one time step.
We can thus write
\begin{equation}
\label{eq:openRecRel}
\pxt = \sum_{r = \Delta r}^1 P^r_x p_{r, t - \Delta r}, \qquad \forall x > \p0.
\end{equation}
With \esref{eq:disp1step}{eq:diff1step} and some simplification
we obtain a master equation for $\pxt$,
\begin{multline}
\pxt = ( 1 - \pnu \Delta r ) \Big\{ p_{x, t-\Delta r} + \ptau \Big[  p_{t-\Delta r} \Delta r + x (1 - x) p_{x + \Delta r, t-\Delta r} \\
 - 2 \Big[ x (1 - x + \Delta r) - \frac{1}{2} \Delta r^2 \Big] p_{x, t-\Delta r} + (x - \Delta r) (1 - x + \Delta r) p_{x - \Delta r, t-\Delta r} \Big] \Big\}, \qquad \forall x > \p0.
\label{eq:openMastEq}
\end{multline}
\eref{eq:openMastEq} is a discrete, diffusion-like equation similar to
\eref{eq:dispMastEq}, except that a term with $p_{t-\Delta r}$ couples the equations
for all $x$. The temporal evolution of $\pxt$ starts from a step function
in rank space at $t = 0$ [\eref{eq:openIniCond}], after which the front
$p_{\p0, t} = 1$
propagates to larger ranks $x > \p0$, making $\pxt$ increase.
At the same time, the replacement dynamics (regulated by $\pnu$)
decreases the probability that elements visit the ranking list. The asymptotic
final state of the dynamics ($\pxt$ for $t \to \infty$) depends on the parameters $\ptau$, $\pnu$, and $\p0$ [see
\eref{eq:openAsympt} below], as well as on the rank $x$.

\begin{figure}[t]
\centering
\includegraphics[width=0.9\textwidth]{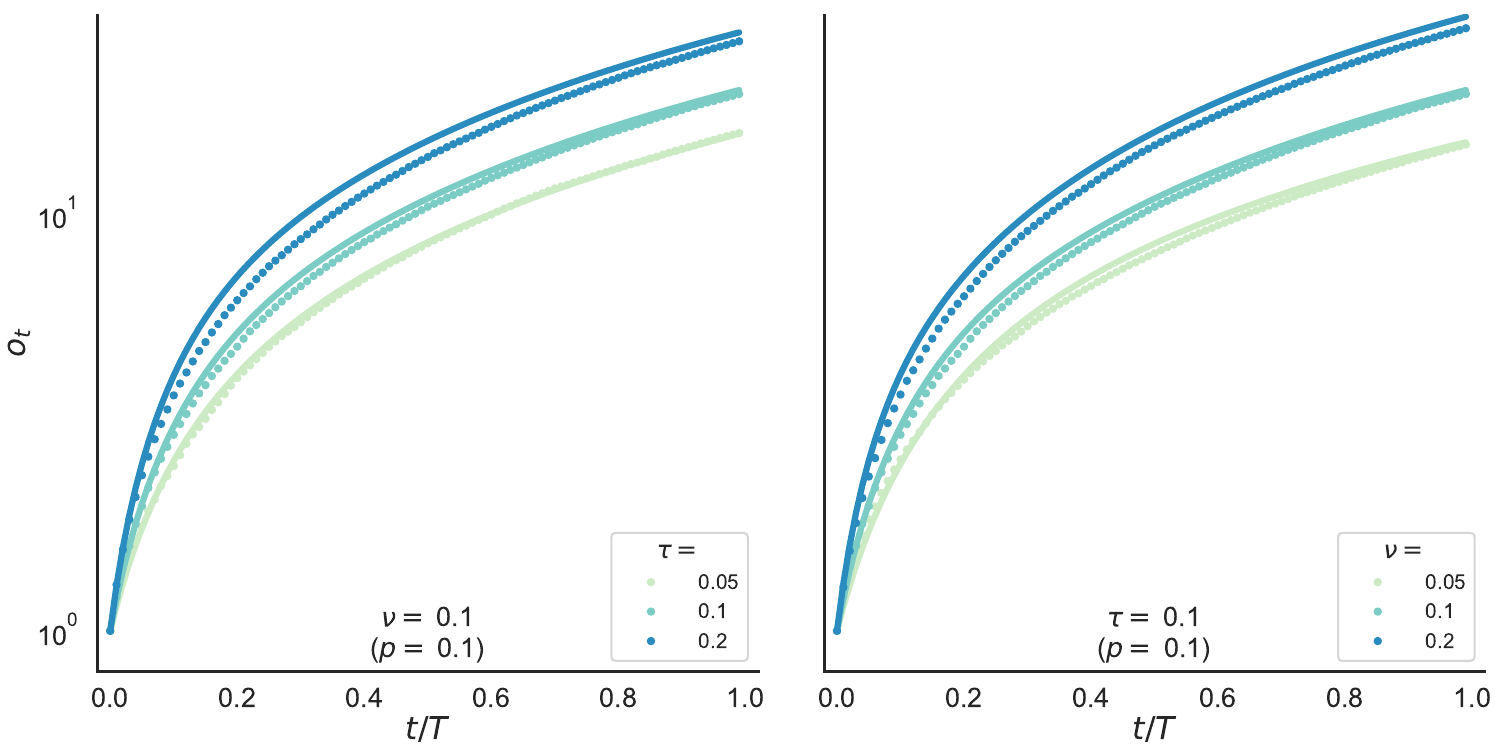}
\caption{\small {\bf Rank turnover in model.}
Rank turnover $o_t = \Nt / N_0$ as a function of normalized time $t / T$ in both numerical simulations of the model (dots) and the analytical approximation of \eref{eq:openModelSol} (lines) for fixed ranking list size $\p0$. Results correspond to varying $\ptau$ for fixed $\pnu$ {\bf (left)}, and to varying $\pnu$ for fixed $\ptau$ {\bf (right)}. Rank turnover has a concave shape for early times and becomes asymptotically linear. The approximation of \eref{eq:openModelSol} is more accurate for small $\ptau$ or $\pnu$ (shown here), or for $\p0$ close to 1. Simulations are averages over $10^4$ realizations (with $T = 100$) in a system of size $N = 10^2$.
}
\label{fig:opennessModel}
\end{figure}

Instead of solving \eref{eq:openMastEq} exactly, we find an approximate analytical solution for the corresponding master equation of $\pt$. Summing up over $x$ both \eref{eq:openMastEq} and the trivial solution for $x = \Delta r, \ldots, \p0$ [\eref{eq:openIniCond}], we obtain
\begin{equation}
\label{eq:openMasterEqAprox}
\pt = \p0 \Big[ 1 - (1 - \pnu \Delta r) (1 - \ptau \Delta r) \Big] + (1 - \pnu \Delta r) (1 - \p0 \ptau \Delta r) p_{t-\Delta r},
\end{equation}
where we have made the approximation $1 = p_{\p0, t} \simeq p_{\p0 + \Delta r, t}$
in order to close the equation. We explicitly solve
\eref{eq:openMasterEqAprox} by recursion from the initial condition $\p0$ via a
geometric series,
\begin{equation}
\label{eq:openFinite}
\frac{\pt}{\p0} = e^{-(\pnu + \p0 \ptau)t} + \frac{\pnu + \ptau}{\pnu + \p0 \ptau} \Big( 1 - e^{-(\pnu + \p0 \ptau)t} \Big),
\end{equation}
which leads to the asymptotic value
\begin{equation}
\label{eq:openAsympt}
\frac{\pt}{\p0} = \frac{\pnu + \ptau}{\pnu + \p0 \ptau}, \qquad t \to \infty.
\end{equation}
\eref{eq:openAsympt} clarifies the range of accuracy of the approximation
$p_{\p0 + \Delta r, t} \simeq 1$. Depending only on the parameters of the model
and as long as $t$ is sufficiently large, if the fraction in the right hand
side of \eref{eq:openAsympt} is close to $1 / \p0$, then
\eref{eq:openMasterEqAprox} is an accurate master equation for $\pt$. This
typically happens for small $\ptau$ or $\pnu$ (see
\fref{fig:opennessModel}), or for $\p0$ close to 1.

With \eref{eq:OpenNs}, \eref{eq:openness}, and the explicit expression for $\pt$ in \eref{eq:openFinite},
we finally obtain
\begin{equation}
\label{eq:openModelSol}
o_t \simeq (1 + \pnu t) \left[ e^{-(\pnu + \p0 \ptau)t} + \frac{\pnu + \ptau}{\pnu + \p0 \ptau} \Big( 1 - e^{-(\pnu + \p0 \ptau)t} \Big) \right],
\end{equation}
with $t = s \Delta r$ and sufficiently large $N$, an analytical approximation for the fraction of elements in the model that visit its ranking list as time goes by. \eref{eq:openModelSol} recovers turnover in numerical simulations of the model (\fref{fig:opennessModel}). In agreement with empirical measurements of $o_t$ (see \fref{fig:openness}), \eref{eq:openModelSol} starts off with a concave shape from its initial condition $o_0 = 1$, and has a linear behavior for long times,
\begin{equation}
\label{eq:openLongTimes}
o_t \simeq \frac{\pnu + \ptau}{\pnu + \p0 \ptau} (1 + \pnu t), \qquad t \gg 0.
\end{equation}
where the slope of the line is regulated by the same fraction we see in \eref{eq:openAsympt}. Moreover, the mean turnover rate after $t$ observations, $\mod_t = (o_t - o_0) / t$, has an asymptotic value $\mod \equiv \mod_{T-1}$ given by
\begin{equation}
\label{eq:openDeriv}
\mod \simeq \pnu \frac{\pnu + \ptau}{\pnu + \p0 \ptau}, \qquad T \gg 0.
\end{equation}

\subsection{Approximation for rank change} 
\label{ssec:modelApproxChange}

\begin{figure}[t]
\centering
\includegraphics[width=0.9\textwidth]{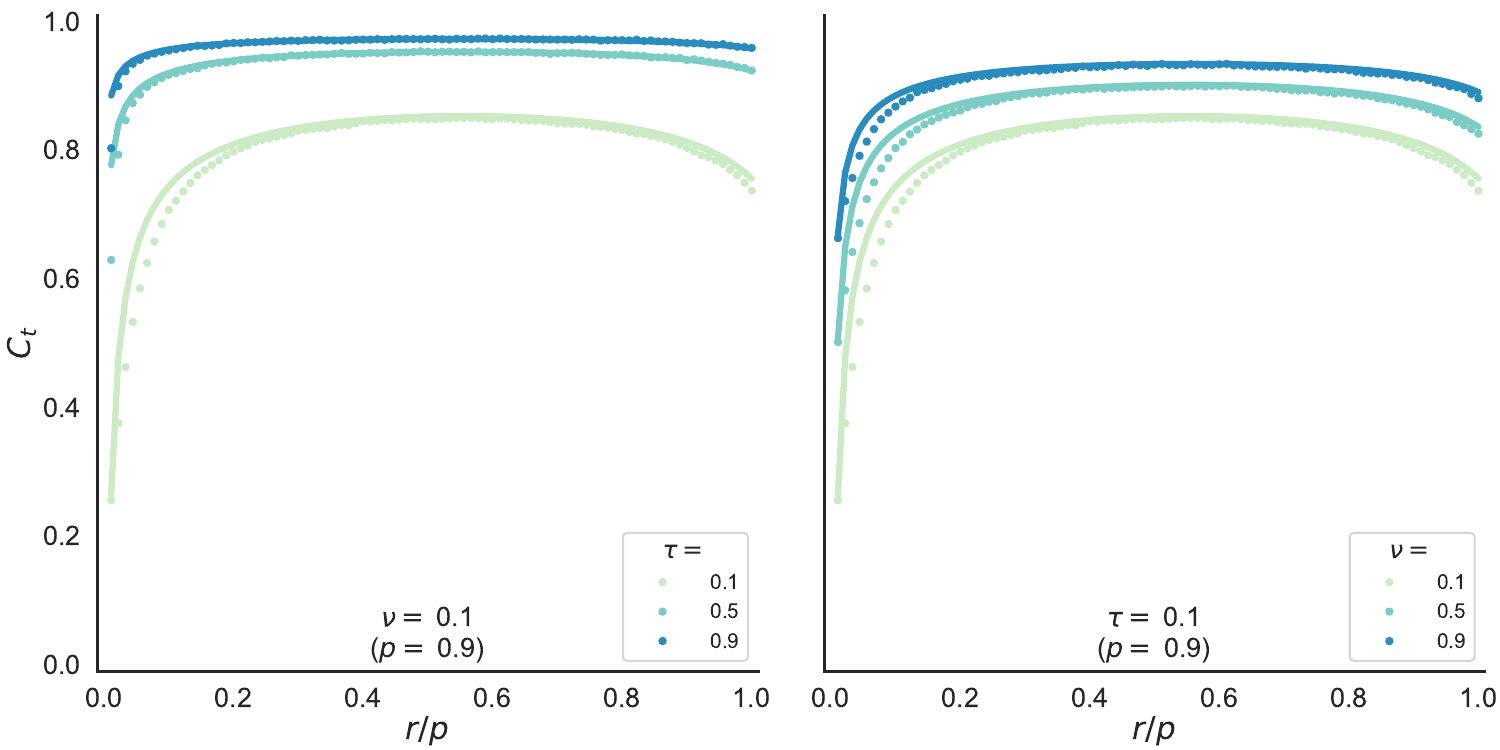}
\caption{\small {\bf Rank change in model.}
Rank change $\Ct$ as a function of normalized rank $r / \p0$ for $t = 1$ ($s = N$) in both numerical simulations of the model (dots) and the analytical approximation of \eref{eq:changeModelApprox} (lines) for fixed ranking list size $\p0$. Results correspond to varying $\ptau$ for fixed $\pnu$ {\bf (left)}, and to varying $\pnu$ for fixed $\ptau$ {\bf (right)}. Rank change increases with both $\ptau$ and $\pnu$, since the displacement/replacement dynamics moves elements away from their previous ranks. $\Ct$ has a symmetric concave shape across the ranking list, which looks asymmetric if the system is open ($\p0 < 1$). Simulations are averages over $10^5$ realizations (with $T = 10$) in a system of size $N = 10^2$.
}
\label{fig:changeModel}
\end{figure}

The approximation for the displacement probability $\Prxt$ in
\eref{eq:aproxDisp} leads us directly to an explicit expression for the rank
change $\Cr$, defined for data as the probability that elements in rank $R$ at
times $t - 1$ and $t$ are not the same, averaged over all times (see
\sref{ssec:change} and \fref{fig:change}). In the model we define $\Ct$ as
the probability that an element located in rank $r$ changes place after time $t$ (we will finally compare $\Cr$ with $\Ct$ for $t = 1$, i.e. $s = N$).
Since $\Ct = 1 - P_{r, t}$, we have
\begin{equation}
\label{eq:changeModelApprox}
\Ct \simeq 1 - e^{- \pnu t} \left[ \Delta r \left( 1 - e^{- \ptau t} \right) + \sqrt{ \frac{\Delta r}{ 4 \pi \ptau r(1 - r) t } } e^{- \ptau t} \right]
\end{equation}
for $t = s \Delta r$ and sufficiently large $N$. \eref{eq:changeModelApprox} is
a good approximation for rank change in the model that fails slightly at the
extremes of the ranking list (\fref{fig:changeModel}). In agreement with
empirical data (see \fref{fig:change}), $\Ct$ has a symmetric concave shape
as a function of $r$, which looks asymmetric if the system is open ($\p0 < 1$).
Thus, in closed systems, the probability of changing rank is larger for
elements in the middle of the ranking list, and lower at its extremes. In open
systems, the bottom of the ranking list ($r \sim \p0$) is also unstable.

\subsection{Approximation for rank inertia} 
\label{ssec:modelApproxSuccess}

\begin{figure}[t]
\centering
\includegraphics[width=0.9\textwidth]{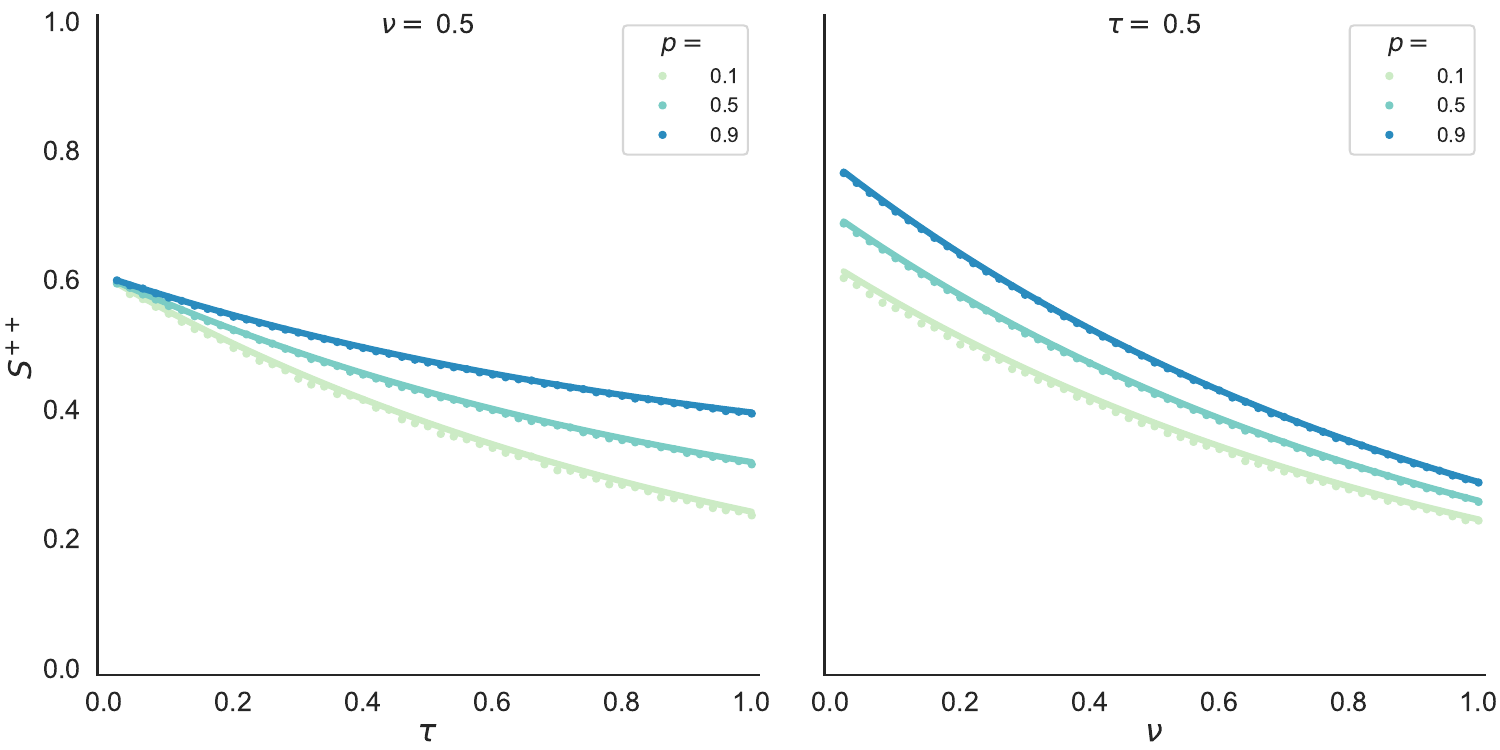}
\caption{\small {\bf Rank inertia in model.}
Inertia $S^{++}$ as a function of $\ptau$ for fixed $\pnu$ {\bf (left)}, and as a function of $\pnu$ for fixed $\ptau$ {\bf (right)}, in both numerical simulations of the model (dots) and the analytical approximation of \eref{eq:successModel} (lines) for varying ranking list size $\p0$, calculated for time $t = 1$ ($s = N$) and threshold $c = 0.5$. Inertia decreases as both $\ptau$ and $\pnu$ increase, since the displacement/replacement dynamics takes elements out of the top of the ranking list. A relatively longer ranking list (large $\p0$) makes it less likely that elements leave the top of the ranking list. Simulations are averages over $10^2$ realizations (with $T = 10$) in a system of size $N = 10^4$.
}
\label{fig:successModel}
\end{figure}

Here we derive an approximate, closed expression for the inertia measure
$\St$ introduced in \sref{ssec:success}.
Similarly to the case of empirical data, in the model we define the matrix element
$S^{ij}_t$ as the probability that an element in region $i$ will move to region
$j$ after a time $t = s \Delta r$.
Regions $i$, $j$ are divided by
an arbitrary threshold $c = \Delta r / \p0, \ldots, 1$ into the top ($+$) of
the ranking ($r = \Delta r, \ldots, c \p0$) and the bottom ($-$) of the ranking
($r = c \p0 + \Delta r, \ldots, \p0$).
Inertia $\St$ is then given by
\begin{equation}
\label{eq:successDef}
\St = \sum_{x = \Delta r}^{c \p0} \Prxt = (1 - \pnu \Delta r)^s \sum_{x = \Delta r}^{c \p0} \Qrxt.
\end{equation}
with $r = \Delta r, \ldots, c \p0$. The sum in the right hand side of
\eref{eq:successDef} is a particular case of the partial cumulative $\Qrabt = \sum_{x=a}^b \Qrxt$ we introduced in \sref{ssec:modelApproxFlux},
with $a = \Delta r$ and $b = c \p0$. We can  solve directly the associated
recurrence relation [\eref{eq:CumulQMastEqApprox}] with initial condition
$Q_{[\Delta r, c \p0], 0} = 1$ using simple formulas for  geometric series.
Substituting the solution into \eref{eq:successDef} we finally obtain
\begin{equation}
\label{eq:successModel}
\St \simeq e^{-\pnu t } \left[ c \p0 + (1 - c \p0) e^{-\ptau t}  \right],
\end{equation}
for $t = s \Delta r$
and sufficiently large $N$. \eref{eq:successModel} is an
analytical approximation for the probability that elements will stay in the top
of the ranking across arbitrarily long times, according to our model. Inertia
$\St$ has an exponential decay with time $t$, regulated by both $\ptau$
and $\pnu$, which matches with numerical simulations of the model.
Denoting $S^{++} \equiv \St$ for $t = 1$ (i.e. $s = N$), we see that
inertia decreases as $\ptau$ and $\pnu$ increase (since there are more
displacements and replacements of elements), and conversely, $S^{++}$ is larger
for a ranking list of longer size $\p0$ (\fref{fig:successModel}).
\eref{eq:successModel} captures the decay of inertia with lag in empirical data, although with some deviations (see \fref{fig:success}).

Rank flux $F$ [\eref{eq:fluxModel}] and inertia $S^{++}$
[\eref{eq:successModel} for $t = 1$] have similar shapes, since they both
measure the chance that elements leave, or stay, in (part of) the
ranking list. Combining these equations gives a linear constitutive relation
between flux and inertia,
\begin{equation}
\label{eq:successFlux}
S^{++} = \beta (1 - F), \qquad \beta = \frac{ c \p0 + (1 - c \p0) e^{-\ptau} }{ \p0 + (1 - \p0) e^{-\ptau} }.
\end{equation}
Intuitively, the probability of staying in the top of the ranking list {\it
decreases linearly} with the flux of elements out of (or into) the ranking
list. For $\p0 = 0$ (a closed system) we get a slope $\beta = 1$, while for
$\p0 = 1$ and large $\ptau$ (an open system with displacement dynamics) we
get a slope regulated by the inertia threshold parameter, $\beta \simeq c$.
This defines a region in $(F, S^{++})$-space where the model can reproduce
pairs of $(F, S^{++})$ values coming from the empirical datasets, an area that
becomes broader as $c$ decreases.

\section{Fitting data with model} 
\label{sec:fitting}
Here we describe the process of fitting the rank measures of all considered
datasets with the minimal model of rank dynamics introduced in
\sref{sec:model}. First, we set $N_0$, $T$ and the time series $\Nt$ with the
empirical values of each dataset listed in \tref{tab:datasets}. As described in
\sref{sec:intro} and \sref{ssec:modelDef}, we classify ranking lists as closed
if $\Nt = N_0$ (for all $t$) and as open otherwise. Then, closed ranking lists
have flux $\Ft = 0$, out-flux $F^-_R = 0$, and turnover $o_t = 1$ for all $t$
and $R$, while open ranking lists have non-zero flux and out-flux, and turnover
that increases in time. We find five strictly closed systems [Metro stations
(London and Mexico), Cities (GB), and Regions JP (quake mag and quakes)], while
the rest have varying degrees of turnover, quantified as a pair of scalars by
the mean flux $F \in [0, 1]$ and the mean turnover rate $\mod \in [0, 1]$. In terms of model parameters, for closed ranking
lists we choose $N = N_0$ (i.e. $\p0 = 1$) and $\pnu = 0$, so $\ptau$ is
the only free parameter. For open ranking lists we take $N = N_{T-1}$ (i.e.
$\p0 = N_0 / N_{T-1}$), so both $\ptau$ and $\pnu$ are free parameters.

\begin{figure}[t]
\centering
\includegraphics[width=0.9\textwidth]{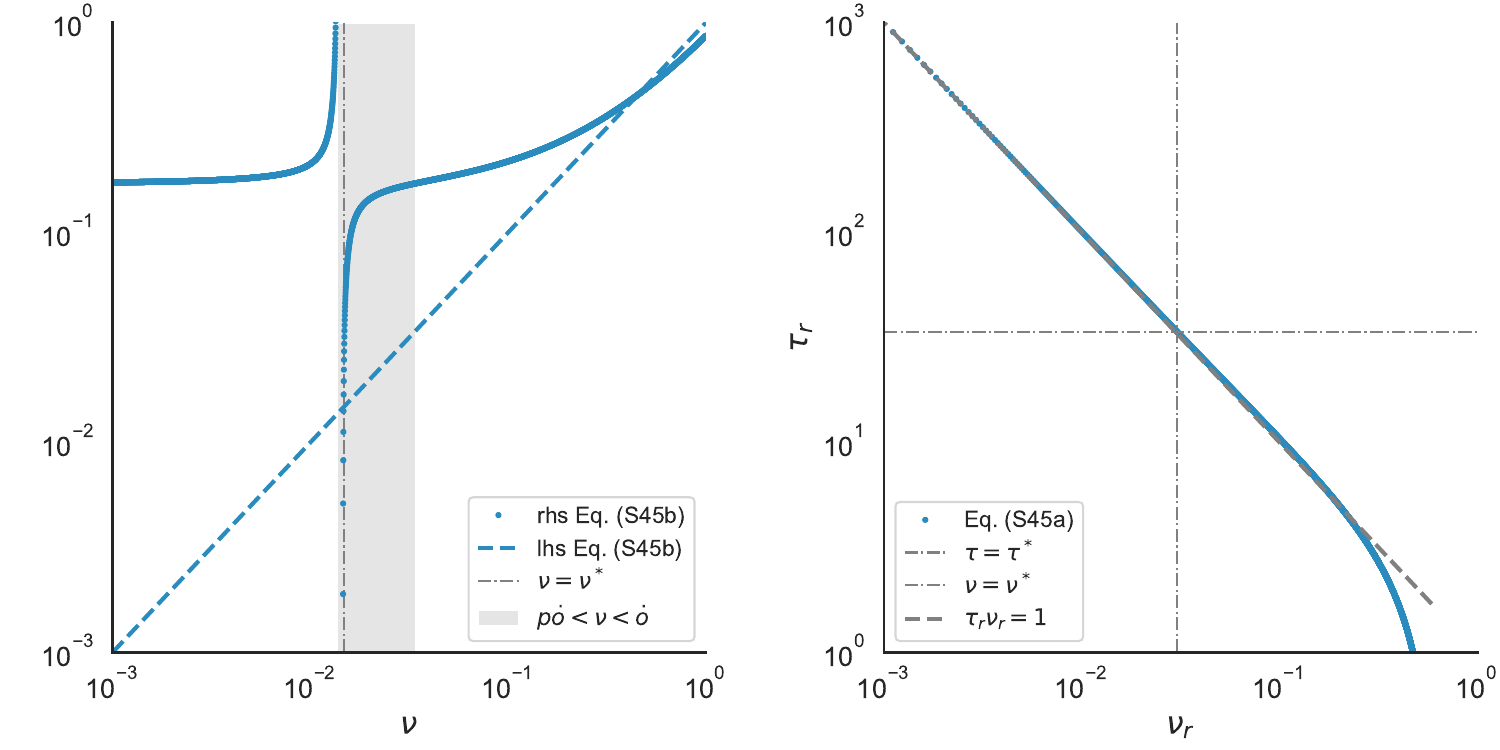}
\caption{\small {\bf Fitting data with model in open ranking lists.} {\bf
(left)} Graphical solution of the transcendental equation for $\pnu$
defined by \eref{eq:paramsSystEqs}. The right hand side (rhs) of
\eref{eq:PnuSsystEqs} has a singularity at $\pnu = \p0 \mod$ and crosses the identity in the left hand side (lhs) of
\eref{eq:PnuSsystEqs} in two fixed points. We only consider the solution
$\pnu^*$ in the interval that ensures $\ptau > 0$ in
\eref{eq:PtauSsystEqs} (shaded area). {\bf (right)}
\eref{eq:PtauSsystEqs} for
rescaled $\prtau$ and $\prnu$ [see \eref{eq:rescParams}], which allows us to find $\ptau^*$. The
solution $(\ptau^*, \pnu^*)$ collapses to the universal curve $\prtau
\prnu = 1$ when rescaling by the empirical values of ranking list size
$\p0$, flux $F$, and mean turnover rate $\mod$.
Data corresponds to the Scientists dataset~\cite{sinatra2015} (see
\tref{tab:datasets} and \sref{sec:data}), but the graphical solution is
qualitatively the same for all open ranking lists.}
\label{fig:fitting}
\end{figure}

We consider open ranking lists first. From \sref{ssec:modelApproxFlux} and
\sref{ssec:modelApproxOpen} we have explicit expressions for $F$
[\eref{eq:fluxModel}] and $\mod$ [\eref{eq:openDeriv}] in
the model, which become more accurate as $N$ and $T$ increase. These relations
lead to a non-linear system of equations for $\ptau$ and $\pnu$,
\begin{subequations}
\label{eq:paramsSystEqs}
\begin{align}
\ptau &= \pnu \frac{ \mod - \pnu }{ \pnu - \p0 \mod }, \label{eq:PtauSsystEqs} \\
\pnu &= \ln \left( \frac{ \p0 + (1 - \p0) e^{- \ptau } }{ 1 - F } \right), \label{eq:PnuSsystEqs}
\end{align}
\end{subequations}
determined by the specific values of ranking list size, flux, and turnover of
each dataset. Solving for $\pnu$ (or $\ptau$) leads to a transcendental
equation in $\pnu$ (or $\ptau$) with no explicit solution. Still, we can
gain insight by applying a graphical method and find the solution $(\ptau^*,
\pnu^*)$ numerically (\fref{fig:fitting}). From \eref{eq:PtauSsystEqs} we
first note that, in order to ensure $\ptau > 0$, $\pnu$ has to be in the
interval $\p0 \mod < \pnu < \mod$.
Since $\mod$ is relatively low for all studied datasets
(less than $\sim 0.3$; see \tref{tab:parameters}), we only need to consider
small $\pnu$ values in the fitting process. Within this range we find a
single solution $\pnu^*$ by graphically equating the two sides of
\eref{eq:PnuSsystEqs} (left panel in \fref{fig:fitting}). This value can be
inserted into \eref{eq:PtauSsystEqs} to find the remaining solution
$\ptau^*$ (right panel in \fref{fig:fitting}).

\begin{table}[!ht]
\small
\noindent\makebox[\textwidth]{
\begin{tabular}{l | r r r | r r r | r }
\toprule
& \multicolumn{3}{c}{Data measure} & \multicolumn{4}{c}{Model parameter} \\
\cmidrule(l){2-8}
Dataset & $F$ & $\dot{o}$ & $S^{++}$ & $p$ & $\tau$ & $\nu$ & $\nu / \ell$ (days$^{-1}$) \\
\midrule
{\bf Society} & & & & & & & \\
\cmidrule(l){1-1}
GitHub repositories~\cite{github2018} & 0.6461 & 0.1287 & 0.4339 & 0.0106 & 1.0000 & 0.0015 & 0.001550 \\
The Guardian readers (recc)~\cite{guardian2018} & 0.7996 & 0.2085 & 0.1288 & 0.0258 & 1.0000 & 0.0061 & 0.006102 \\
The Guardian readers (comm)~\cite{guardian2018} & 0.6118 & 0.1283 & 0.3764 & 0.0413 & 1.0000 & 0.0060 & 0.006027 \\
Enron emails~\cite{enron2015} & 0.5215 & 0.2158 & 0.5394 & 0.0443 & 0.7747 & 0.0129 & 0.001850 \\
Scientists~\cite{sinatra2015,sinatra2016} & 0.1578 & 0.0343 & 0.8083 & 0.3982 & 0.2767 & 0.0147 & 0.000040 \\
Universities~\cite{shanghai2016} & 0.0477 & 0.0308 & 0.9532 & 0.7143 & 0.0907 & 0.0238 & 0.000065 \\
\midrule
{\bf Languages} & & & & & & & \\
\cmidrule(l){1-1}
Russian~\cite{google2018,michel2011,cocho2015,morales2018} & 0.1796 & 0.0331 & 0.8436 & 0.1262 & 0.2244 & 0.0048 & 0.000013 \\
Spanish~\cite{google2018,michel2011,cocho2015,morales2018} & 0.1603 & 0.0245 & 0.8585 & 0.1361 & 0.2006 & 0.0037 & 0.000010 \\
German~\cite{google2018,michel2011,cocho2015,morales2018} & 0.1605 & 0.0292 & 0.8677 & 0.1159 & 0.1957 & 0.0039 & 0.000011 \\
French~\cite{google2018,michel2011,cocho2015,morales2018} & 0.1849 & 0.0255 & 0.8338 & 0.0967 & 0.2238 & 0.0027 & 0.000008 \\
Italian~\cite{google2018,michel2011,cocho2015,morales2018} & 0.1489 & 0.0234 & 0.8684 & 0.1495 & 0.1875 & 0.0039 & 0.000011 \\
English~\cite{google2018,michel2011,cocho2015,morales2018} & 0.1272 & 0.0181 & 0.8928 & 0.1426 & 0.1566 & 0.0029 & 0.000008 \\
\midrule
{\bf Economics} & & & & & & & \\
\cmidrule(l){1-1}
Companies~\cite{fortune2005} & 0.0734 & 0.0558 & 0.9323 & 0.2639 & 0.0689 & 0.0260 & 0.000071 \\
Countries~\cite{atlas2018,hidalgo2009,hausmann2014} & 0.0372 & 0.0084 & 0.8967 & 0.7122 & 0.1137 & 0.0061 & 0.000017 \\
\midrule
{\bf Infrastructure} & & & & & & & \\
\cmidrule(l){1-1}
Cities (RU)~\cite{cottineau2016} & 0.1221 & 0.1073 & 0.8824 & 0.5711 & 0.1230 & 0.0793 & 0.000018 \\
Metro stations (London)~\cite{murcio2015} & 0.0000 & 0.0000 & 0.9207 & 1.0000 & 0.1728 & 0.0000 & 0.000000 \\
Cities (GB)~\cite{edwards2016} & 0.0000 & 0.0000 & 0.9517 & 1.0000 & 0.1015 & 0.0000 & 0.000000 \\
Metro stations (Mexico) & 0.0000 & 0.0000 & 0.9100 & 1.0000 & 0.1984 & 0.0000 & 0.000000 \\
\midrule
{\bf Nature} & & & & & & & \\
\cmidrule(l){1-1}
Hyenas~\cite{ilany2015} & 0.3023 & 0.2748 & 0.5289 & 0.1419 & 0.3213 & 0.0911 & 0.000249 \\
Regions JP (quake mag)~\cite{junec2018,karsai2012} & 0.0000 & 0.0000 & 0.7293 & 1.0000 & 0.7797 & 0.0000 & 0.000000 \\
Regions JP (quakes)~\cite{junec2018,karsai2012} & 0.0000 & 0.0000 & 0.7625 & 1.0000 & 0.6444 & 0.0000 & 0.000000 \\
\midrule
{\bf Sports} & & & & & & & \\
\cmidrule(l){1-1}
Chess players (male)~\cite{fide2018} & 0.0106 & 0.0051 & 0.9877 & 0.8148 & 0.0354 & 0.0042 & 0.000138 \\
Chess players (female)~\cite{fide2018} & 0.0099 & 0.0068 & 0.9929 & 0.7667 & 0.0193 & 0.0055 & 0.000182 \\
Poker players~\cite{poker2018} & 0.0377 & 0.0203 & 0.9714 & 0.1832 & 0.0402 & 0.0058 & 0.000824 \\
Tennis players~\cite{tennis2018} & 0.0112 & 0.0050 & 0.9916 & 0.3338 & 0.0137 & 0.0021 & 0.000302 \\
Golf players~\cite{golf2018} & 0.0058 & 0.0028 & 0.9904 & 0.3166 & 0.0068 & 0.0012 & 0.000168 \\
Football scorers~\cite{football2018} & 0.1340 & 0.0960 & 0.9091 & 0.1669 & 0.1377 & 0.0305 & 0.004339 \\
NASCAR drivers (Busch)~\cite{nascar2018} & 0.3648 & 0.2392 & 0.6348 & 0.1124 & 0.4744 & 0.0455 & 0.000124 \\
NASCAR drivers (Winston Cup)~\cite{nascar2018} & 0.1865 & 0.1306 & 0.8145 & 0.1838 & 0.2052 & 0.0422 & 0.000115 \\
National football teams~\cite{teams2018} & 0.0031 & 0.0007 & 0.9728 & 0.9524 & 0.0532 & 0.0000 & 0.000000 \\
\bottomrule
\end{tabular}}
\caption{\small {\bf Data measures and fitted model parameters}. Values of empirical measures used in the fitting process (mean flux $F$, mean turnover rate $\mod$, and inertia $S^{++}$) and of fitted model parameters (relative ranking list size $\p0$, displacement probability $\ptau$, and replacement probability $\pnu$) for all considered datasets (see \tref{tab:datasets} and \sref{sec:data}). For open ranking lists, we fit the model to empirical data by setting $N = N_{T-1}$ (i.e. $\p0 = N_0 / N_{T-1}$, see \sref{ssec:modelDef}) and by computing $\ptau$ and $\pnu$ numerically from \eref{eq:paramsSystEqs} in terms of $F$ and $\mod$ (\fref{fig:fitting}). For closed ranking lists, we set $N = N_0$ and $\pnu = 0$, and obtain $\ptau$ explicitly from \eref{eq:fittingClosedSyst} in terms of $S^{++}$ (for time $t=1$ and threshold $c = 0.5$). We also show the rate of element replacement $\nu / \ell$, with $\ell$ the real time between observations (see \tref{tab:datasets}). Datasets are classified by an (arbitrary) system type based on the nature of the elements in the ranking list.}
\label{tab:parameters}
\end{table}

The fitting process for closed ranking lists is, in turn, quite straightforward. Closed ranking lists have trivial values of flux and turnover, so we cannot use \eref{eq:paramsSystEqs} for fitting. Instead we consider the explicit expression of inertia $S^{++}$ (for time $t=1$) derived in \sref{ssec:modelApproxSuccess}. $S^{++}$ is reminiscent of $F$ in the sense that it measures the flux of elements out of (into) the top of the ranking list into (from) the bottom, where top and bottom are arbitrarily separated by the threshold $c$. Solving for $\ptau$ in \eref{eq:successFlux} gives
\begin{equation}
\label{eq:fittingClosedSyst}
\ptau = \ln \left( \frac{ 1 - c \p0 }{ S^{++} - c \p0 } \right), \qquad F = 0,
\end{equation}
which allows us to find $\ptau$ for all closed ranking lists considered.

\tref{tab:parameters} summarizes the results of the fitting process described
above. Our minimal model of rank dynamics fits most rank measures of many
datasets (see \fsref{fig:flux}{fig:success}). In the model, rank flux $\Ft$ is
a constant over time that recovers the empirical mean with high precision,
apart from fluctuations (\fref{fig:flux}). Out-flux $F^-_R$ is relatively
constant for most $R$ values, and in open systems has a sharp increase for
large $R$. This is only a good fit for relatively closed ranking lists, since
in very open systems out-flux is a more gradually increasing function of $R$
(\fref{fig:Oflux}). Turnover $o_t$ is concave and monotonically
increasing in time, which follows empirical data (\fref{fig:openness}). Rank
change $\Cr$ recovers the asymmetry of most open systems (low values only in
the top of the ranking list) and the rough symmetry in closed systems (low
values in both the top and bottom of the ranking list), although empirical
closed systems are not as symmetric as the model predicts (\fref{fig:change}).
Inertia $\St$ is a decaying function of time $t$,
in agreement with
several datasets. Even if the difference in
qualitative behavior between open and closed ranking lists is less pronounced
for $\St$ than for other rank measures, we still see that inertia
tends to decay faster in open systems (\fref{fig:success}).

\subsection{Dynamical regimes of open ranking lists} 
\label{ssec:dynamRegimes}

Having explicit expressions in \eref{eq:paramsSystEqs} lets us further analyze
the solution $(\ptau^*, \pnu^*)$. We introduce the rescaled parameters
\begin{equation}
\label{eq:rescParams}
\prtau = \frac{\ptau}{ \p0 (1 - \p0) \mod }, \quad \text{and} \quad \prnu = \frac{ \pnu - \p0 \mod }{ \mod },
\end{equation}
which allows us to rewrite \eref{eq:PtauSsystEqs} for sufficiently small $\pnu$ (relative to $\p0$ and $1 - \p0$),
\begin{equation}
\label{eq:paramsConstEq}
\prtau \prnu \simeq 1.
\end{equation}
\eref{eq:paramsConstEq} shows that, when fitting the model to datasets with
given values of $\p0$, $F$, and $\mod$, the probabilities
$\ptau$ and $\pnu$ can be rescaled to collapse into a universal curve, a
power law with exponent 1 (\fref{fig:fitting}). In order to reproduce the
dynamics of empirical ranking lists, is it enough to consider an inverse
relationship between the rates of displacement and replacement of elements in
the model, meaning that the considered datasets have either an active dynamics
of Lévy flights and diffusion plus a few replacements, or less
displacements with a larger exchange of elements. This makes the fitting
process effectively one-dimensional across open ranking lists, even if the
model can be studied for arbitrary $\ptau$ and $\pnu$.

\begin{figure}[t]
\centering
\includegraphics[width=\textwidth]{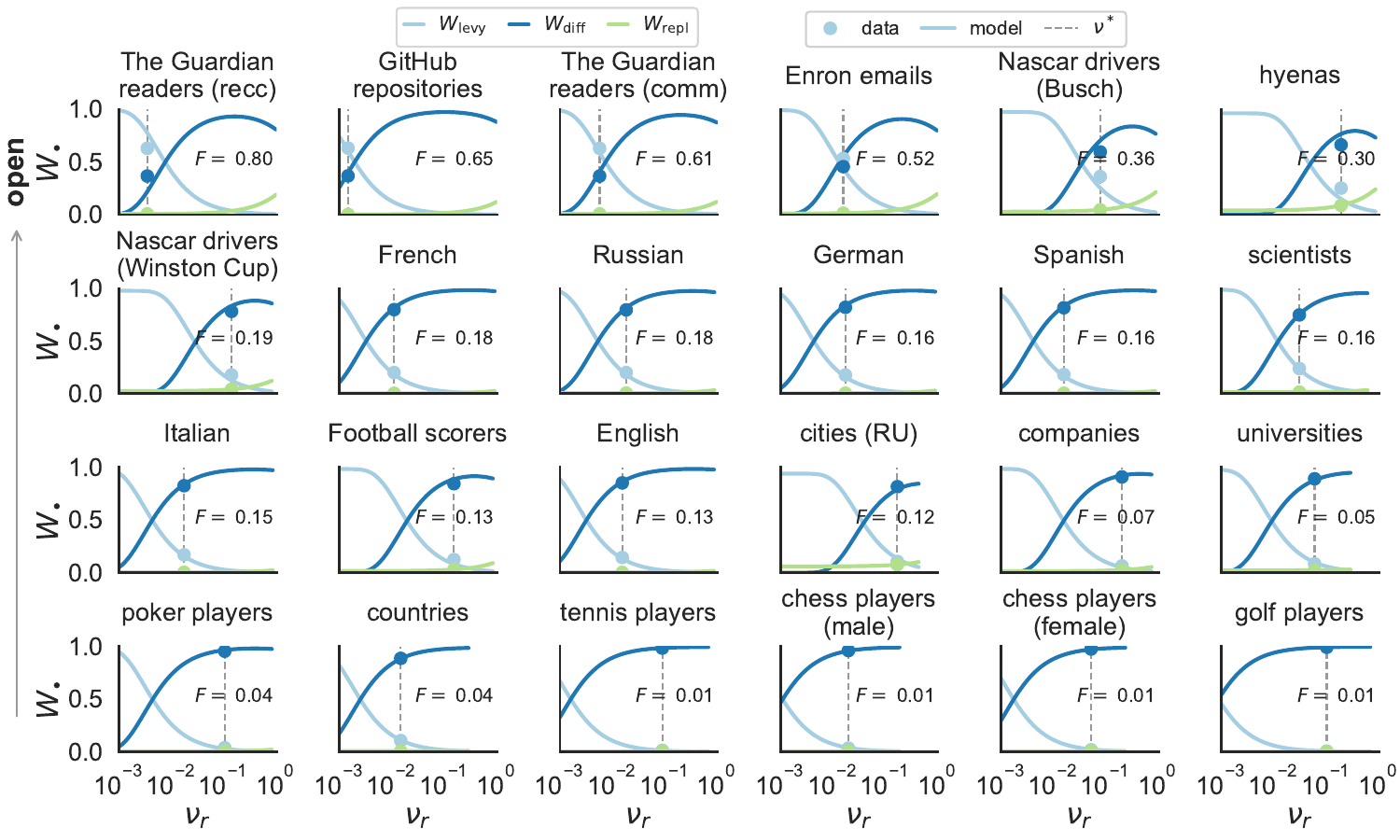}
\caption{\small {\bf Regimes of dynamical behavior in open ranking lists.}
Average probability that an element changes rank by Lévy
walk ($\Wlevy$), diffusion ($\Wdiff$), or is replaced ($\Wrepl$) between
consecutive observations in the data. Probabilities shown both for all datasets
with $\mod > 10^{-3}$ (dots), and for the model moving along
the curve $\prtau \prnu = 1$ [\eref{eq:paramsConstEq}] with the same
$\p0$ and $\mod$ as the data (lines). Datasets are ordered
from most open (upper row) to least open (lower row) according to mean rank
flux $F$ [\eref{eq:fluxMean}]. The model reveals a crossover in empirical
ranking lists between Lévy walk ($\Wlevy > \Wdiff, \Wrepl$) and diffusion
($\Wdiff > \Wlevy, \Wrepl$) regimes. Although not seen in data, there is
also a third regime driven by replacement ($\Wrepl > \Wlevy, \Wdiff$).}
\label{fig:probabilities}
\end{figure}

The universal curve in \eref{eq:paramsConstEq} displays three regimes in the dynamics of open ranking lists, as measured by the average probabilities that, between consecutive observations in the data (i.e. between times $t$ and $t+1$), an element performs either a Lévy walk,
\begin{equation}
\label{eq:AvgProbLevi}
\Wlevy = e^{-\pnu} (1 - e^{-\ptau} ),
\end{equation}
changes rank by diffusion
\begin{equation}
\label{eq:AvgProbDiff}
\Wdiff = e^{-\pnu} e^{-\ptau},
\end{equation}
or is replaced,
\begin{equation}
\label{eq:AvgProbRepl}
\Wrepl = 1 - e^{-\pnu},
\end{equation}
where $\Wlevy + \Wdiff + \Wrepl = 1$. In the most open systems we study, elements mostly change rank via long jumps, forming a Lévy walk regime where $\Wlevy > \Wdiff, \Wrepl$ (\fref{fig:probabilities}). Here, long-range rank changes take elements in and out of a small ranking list within a big system (low $\p0$), thus generating large mean flux $F$. The rest of the datasets belong to a diffusion regime with $\Wdiff > \Wlevy, \Wrepl$. A local, diffusive rank dynamics is the result of elements smoothly changing their scores and overcoming their neighbors in rank space. Although not observed in empirical data, the model predicts a third regime dominated by replacement ($\Wrepl > \Wlevy, \Wdiff$). The curves in \fref{fig:probabilities} show how close a ranking list might be to a change of regime. If a dataset has values of $\p0$ and $\mod$ that make it fall close to the boundaries of these regimes, small changes in the parameters regulating the displacement ($\ptau$) and replacement ($\pnu$) can change the dynamical regime of the ranking list.

\subsection{Fluctuations, estimation bias, and goodness of fit} 
\label{ssec:fluctuations}

The relationship between rescaled parameters $\prtau$ and $\prnu$ in
\eref{eq:paramsConstEq} is a consequence of fitting average values of flux $F$
and turnover rate $\mod$ in empirical data with asymptotic approximations
coming from the model. Here we investigate the fluctuations around such average
values arising in numerical simulations of the model,  which reveal a small but nonzero bias in the estimation of $\ptau$. Despite this bias, simulations recover the empirical values of system-wide quantities ($F$, $\mod$, and inertia $S^{++}$), showcasing the ability of the model to reproduce rank dynamics in data. We further quantify the goodness of fit by showing that, for most systems considered, data is closer to the universal curve [\eref{eq:paramsConstEq}] than model simulations.

We start by taking a dataset with given values of $T$, $\p0 = N_0 / N_{T-1}$, and fitted
parameters $\ptau$ and $\pnu$ (see \tsref{tab:datasets}{tab:parameters}). We
first run a large number of realizations of the model with the same values of
$N_0$, $T$, $\ptau$,  and $\pnu$,  plus an arbitrary but smaller system size $\Nsim < N_{T-1}$,
leading to a certain number of elements ever seen in the ranking list of each simulation,  which we denote by $\NT1sim$. Since the model is stochastic,
$\NT1sim$ varies across realizations, with average $\langle \NT1sim \rangle$. We choose the $\Nsim$ that minimizes the quantity $| N_{T-1} - \langle \NT1sim \rangle |$, ensuring that $\psim = N_0 / \NT1sim$ is close to the empirical $\p0$, beyond fluctuations.
With $\Nsim$ fixed, we again run a large number of realizations of the model with the same values of
$N_0$, $T$, $\ptau$, and $\pnu$.
Each model realization will have estimated values of flux $\Fsim$, turnover rate $\modsim$, and inertia $\Ssim$ similar
to those in the original dataset (apart from fluctuations), which we use to fit the model
to itself [via \eref{eq:paramsSystEqs} or \eref{eq:fittingClosedSyst}] and
obtain estimated parameters $\ptausim$ and $\pnusim$.
This constitutes a parametric bootstrap of the model, allowing us to explore expected variations in the fitted parameters $\ptau$ and $\pnu$.

\begin{figure}[t] 
\centering
\includegraphics[width=\textwidth]{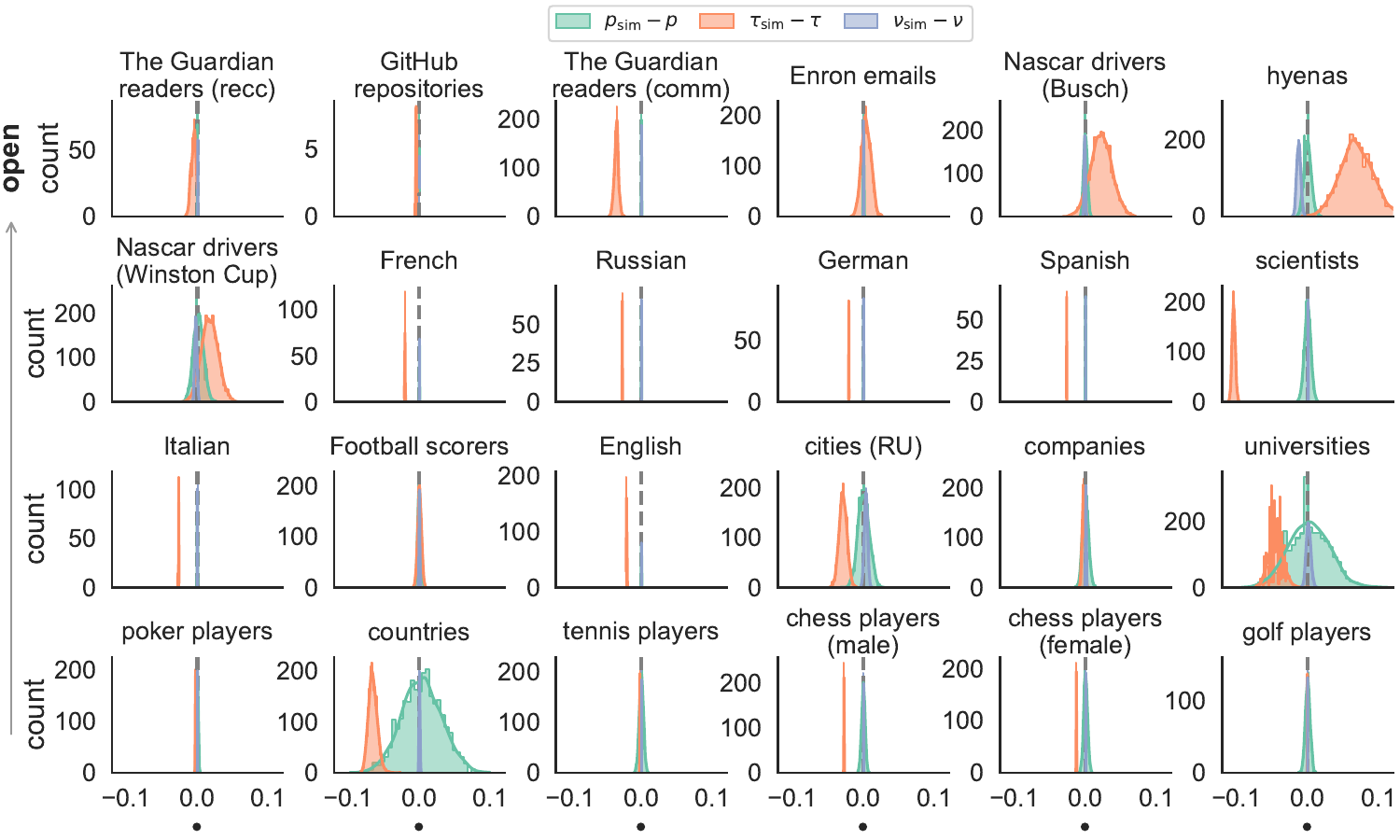}
\caption{\small {\bf Model fluctuations and estimation bias in open ranking lists.}
Distributions of bootstrapped parameter differences $\Delta \p0 = \psim - \p0$, $\Delta \ptau = \ptausim - \ptau$, and $\Delta \pnu = \pnusim - \pnu$ between fitted parameters
$\p0$, $\ptau$, and $\pnu$ of a dataset, and the corresponding $\psim$, $\ptausim$, and $\pnusim$
when fitting the model to itself. Distributions are plotted as
binned counts (histograms) and kernel density estimates (continuous lines). The dashed line at 0 indicates perfect agreement between fitted parameters and bootstrapped estimates. Number of simulations varies
from 30 to 2500 per
system. Datasets (with $\mod > 10^{-3}$) are ordered from most open (upper
row) to least open (lower row) according to mean rank
flux $F$ [\eref{eq:fluxMean}].}
\label{fig:devs_params}
\end{figure} 

We compare the distributions of $\psim$, $\ptausim$, and $\pnusim$ over model simulations
with the fitted parameters $\p0$, $\ptau$, and $\pnu$ of open ranking lists by
calculating the differences $\Delta \p0 = \psim - \p0$, $\Delta \ptau = \ptausim - \ptau$, and $\Delta \pnu = \pnusim - \pnu$ (\fref{fig:devs_params}).
We obtain $\langle \Delta \p0 \rangle \simeq \langle \Delta \pnu \rangle \simeq 0$ for all datasets (apart from hyenas, where $\pnu$ is slightly underestimated), meaning that the fitting process consistently recovers the values of $\p0$ and $\pnu$ in simulations, without bias.
Fluctuations around $\langle \Delta \p0 \rangle$, $\langle \Delta \ptau \rangle$, and $\langle \Delta \pnu \rangle$ (measured by
the spread of their associated distributions) are only noticeable for some of
the smallest datasets (Enron emails, hyenas, Nascar drivers, cities [RU],
universities, and countries).
Even when accounting for fluctuations, we find a small bias in $\ptau$ ($|\langle \Delta \ptau \rangle | < 0.1$) for several datasets [The Guardian readers (comm), hyenas, languages, scientists, cities (RU), universities, countries, and chess], indicating that the fitting process systematically under-
($\Delta \ptau < 0$) or over- ($\Delta \ptau > 0$)
estimates the displacement probability in simulations, particularly in systems with low $N_{T-1}$ or
$T$.

\begin{figure}[t]
\centering
\includegraphics[width=\textwidth]{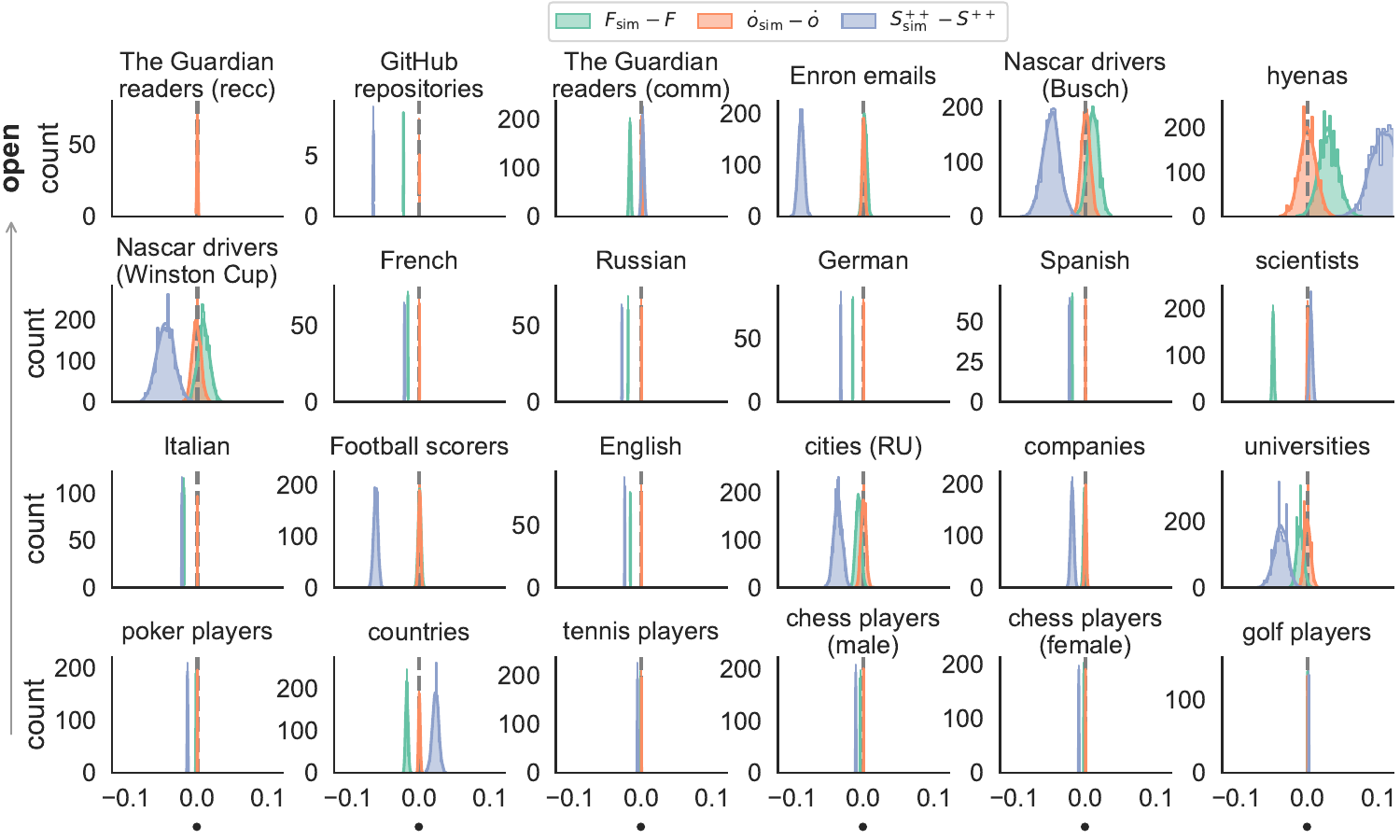}
\caption{\small {\bf Simulations recover properties of open ranking lists.}
Distributions of bootstrapped differences $\Delta F = \Fsim - F$, $\Delta \mod = \modsim - \mod$, and $\Delta S^{++} = \Ssim - S^{++}$ between flux $F$, turnover rate $\mod$, and inertia $S^{++}$ in a dataset, and the corresponding $\Fsim$, $\modsim$, and $\Ssim$ in simulations of the model with fitted model parameters (see \tsref{tab:datasets}{tab:parameters}). Distributions are plotted as binned counts (histograms) and kernel density estimates (continuous lines). The dashed line at 0 indicates perfect agreement between empirical properties and bootstrapped estimates. Number of simulations varies from 30 to 2500 per system. Datasets (with $\mod > 10^{-3}$) are ordered from most open (upper
row) to least open (lower row) according to mean rank
flux $F$ [\eref{eq:fluxMean}].
}
\label{fig:devs_measures}
\end{figure}

Despite the small bias in $\ptau$ (\fref{fig:devs_params}), numerical simulations with fitted model parameters recover aggregate properties of empirical data, which we quantify via the differences $\Delta F = \Fsim - F$, $\Delta \mod = \modsim - \mod$, and $\Delta S^{++} = \Ssim - S^{++}$ (\fref{fig:devs_measures}). Turnover rate is perfectly reproduced in all datasets ($\langle \Delta \mod \rangle \simeq 0$), apart from fluctuations, while flux has a small disagreement ($| \langle \Delta F \rangle | < 0.1$) for some systems [GitHub repositories, The Guardian readers (comm), hyenas, languages, scientists, and countries]. Simulations even recover, to some extent, quantities not involved in the fitting of open ranking lists: inertia is perfectly reproduced ($\langle \Delta S^{++} \rangle \simeq 0$) in The Guardian, scientists, tennis, chess, and golf datasets, and has a small disagreement ($| \langle \Delta S^{++} \rangle | < 0.1$) for the rest. 

\begin{figure}[t]
\centering
\includegraphics[width=\textwidth]{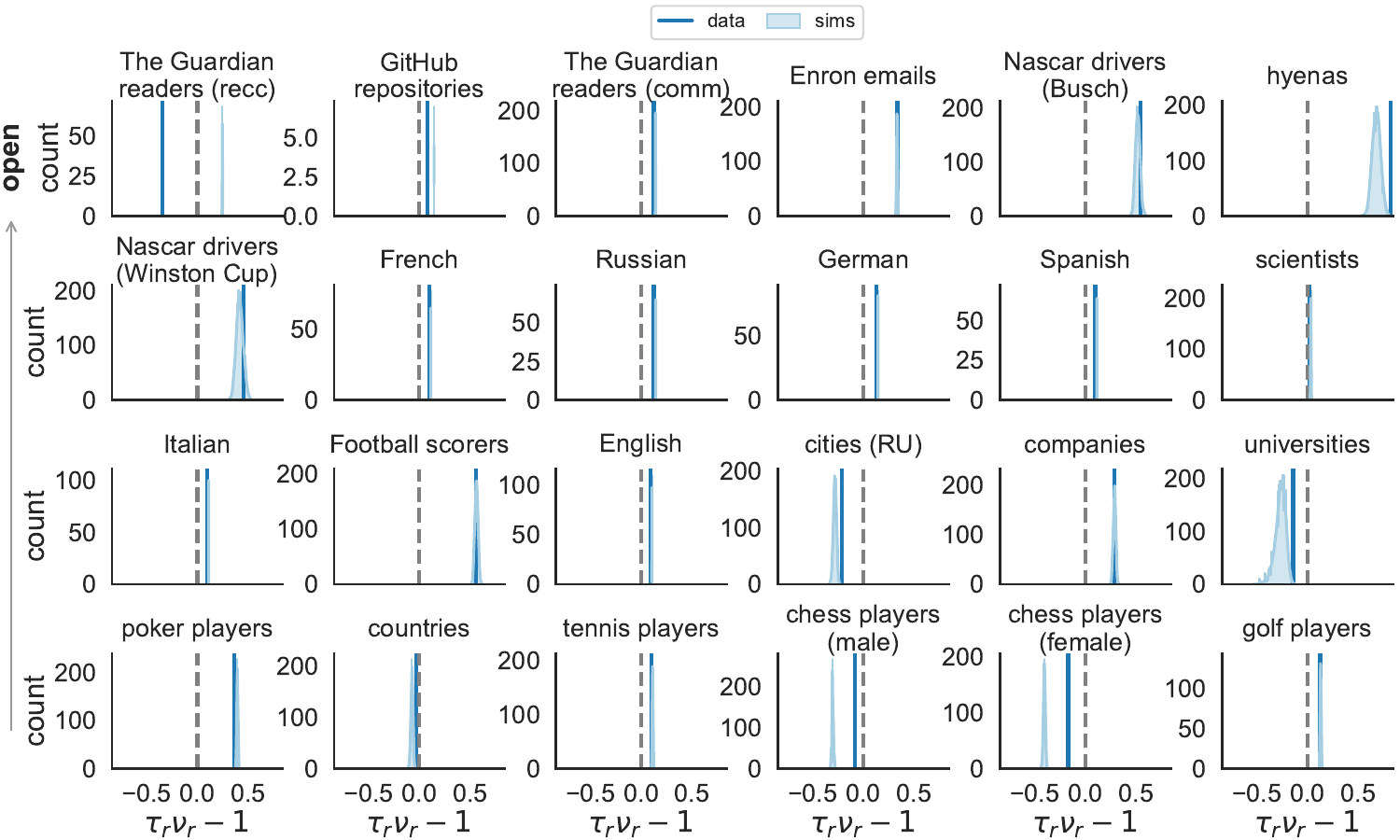}
\caption{\small {\bf Distance to universal curve in open ranking lists.}
Distribution of relative difference $\prtau \prnu - 1$ between rescaled parameters ($\prtau$ and $\prnu$) and universal curve [$\prtau \prnu = 1$; see \eref{eq:paramsConstEq}], both for the fitted parameters $\ptau$ and $\pnu$ of a dataset (data; shown as vertical line), and for the corresponding $\ptausim$ and $\pnusim$ when fitting the model to itself (sims). Distributions are plotted as binned counts (histograms) and kernel density estimates (continuous lines). The dashed line at 0 indicates perfect agreement between fitted parameters / bootstrapped estimates and the universal curve in \eref{eq:paramsConstEq}. Number of simulations varies from 30 to 2500 per system. Datasets (with $\mod > 10^{-3}$) are ordered from most open (upper
row) to least open (lower row) according to mean rank
flux $F$ [\eref{eq:fluxMean}].}
\label{fig:devs_curve}
\end{figure}

In order to further quantify the goodness of fit, we explore how the rescaled fitted parameters and bootstrapped estimates fall around the universal curve of \eref{eq:paramsConstEq}. For a dataset with given values of $\p0$, $\mod$, and fit $\ptau$ and $\pnu$, the rescaled parameters $\prtau$ and $\prnu$ in \eref{eq:rescParams} have the same relative difference from the expected universal behavior ($\prtau \prnu = 1$) alongside either axis,
\begin{equation}
\label{eq:testStatistic}
\frac{ \prtau - 1/\prnu }{ 1/\prnu } = \frac{ \prnu - 1/\prtau }{ 1/\prtau } = \prtau \prnu - 1.
\end{equation}
We measure $\prtau \prnu - 1$ for all datasets, as well as the distribution of the corresponding quantity $\prtausim \prnusim - 1$ over model simulations, calculated with $\psim$, $\modsim$, and the bootstrapped fit $\ptausim$ and $\pnusim$~\footnote{Explicitly, $\prtausim = \frac{\ptausim}{\psim (1 - \psim) \modsim}$ and $\prnusim = \frac{\pnusim - \psim \modsim}{\modsim}$, following \eref{eq:rescParams}.} (\fref{fig:devs_curve}). We do not expect rescaled parameters to follow the universal curve perfectly, due to the approximations used to derive $F$ and $\mod$ in the model [\eref{eq:fluxModel} and \eref{eq:openDeriv}; see \ssref{ssec:modelApproxFlux}{ssec:modelApproxOpen}], as well as the assumption of small $\pnu$ in the derivation of \eref{eq:paramsConstEq} itself (see \sref{ssec:dynamRegimes} and \fref{fig:fitting}). Still, in all datasets we find good agreement for fitted model parameters ($|\prtau \prnu - 1| < 0.5$) and bootstrapped estimates ($|\langle \prtausim \prnusim \rangle - 1| < 0.5$). For most systems,
the dataset has $\prtau$ and $\prnu$ closer to the universal curve than most simulations of the model itself,  suggesting than empirical ranking lists follow \eref{eq:paramsConstEq}.  Exceptions are The Guardian readers (recc), Enron emails, Nascar drivers (Busch), and hyenas.  We consider that these systems do not follow the proposed universal behavior either because of small sample size (leading to data noise), or due to a systematic failure of the model to recover their dynamics.

Finally, it's worth noting that it would be ideal to have a more formal statistical approach to model fitting than the process used here. It is indeed possible to derive maximum likelihood equations for two probability distributions: (i) The distribution of the time $t$ an element stays in the ranking list [the product $\mod S_t^{++}$ of mean turnover rate and time-dependent rank inertia according to \eref{eq:openDeriv} and \eref{eq:successModel}]; and (ii) the probability of element displacement in \eref{eq:aproxDispSimple}. The maximum likelihood equations (for $\ptau$ and $\pnu$) coming from maximizing the log-likelihoods of these two distributions are trascendental and cannot be solved analytically. We can, however, solve the equations numerically and compare their solutions to the fitted $\ptau$ and $\pnu$ from our `mean-field' fitting process. For most datasets, mean-field parameter estimates lead to lower mean-squared-error deviations between the fitted model and several measures in the data (rank flux, turnover, out-flux, change probability, rank diversity, and inertia). This means that our fitting process is able to recover aggregate features of empirical ranking data, despite it being based on asymptotic approximations of mean behavior and showing a small bias in estimating $\ptau$. As added value, the mean-field fitting process is more analytically tractable, computationally faster to obtain, and provides insight on the balance between amounts of element displacement and replacement seen in data.

\subsection{Effect of subsampling} 
\label{ssec:sampling}

Our model of ranking dynamics suggests that empirical ranking lists belong to
one of several dynamical regimes constrained by the universal curve in
\eref{eq:paramsConstEq}. Here we explore the role of sampling rate (related to
the time between observations of the ranking, see $\ell$ in
\tref{tab:datasets}) in the regime occupied by a system. We show that
subsampling observations typically makes systems go downwards along the
universal curve (right panel in \fref{fig:fitting}
and Fig. 3 in main
text), moving, for example, from a regime with many Lévy walks to one more
driven by diffusion.

For a synthetic or empirical system with $T$ observations, we implement
subsampling by only considering the ranking list at times $t = 0, k, 2k,
\ldots, T_{\mathrm{eff}}-1$ and discarding the rest of the data, with $k = 1,
2, \ldots, T-1$ a tunable parameter. The subsampling process decreases the
number of observations to $T_{\mathrm{eff}} = \lceil T / k \rceil \leq T$, such
that $k=1$ recovers the original dataset ($T_{\mathrm{eff}} = T$) and $k=T-1$
corresponds to the most subsampled system possible ($T_{\mathrm{eff}} = 2$). We
then calculate the ($k$-dependent) subsampled rank flux $F_k$, turnover rate
$\mod_k$, and inertia $S^{++}_k$, and compare them with their values $F$,
$\mod$, and $S^{++}$ in the original ranking list. Finally, we fit the model to
the subsampled data [via \eref{eq:paramsSystEqs} or
\eref{eq:fittingClosedSyst}] for each $k$ value, obtaining the subsampled
relative list size $\p0_k$, displacement $\ptau^*_k$ and replacement
$\pnu^*_k$, in contrast with the original fit $\p0$, $\ptau^*$, and $\pnu^*$.

\begin{figure}[t] 
\centering
\includegraphics[width=0.8\textwidth]{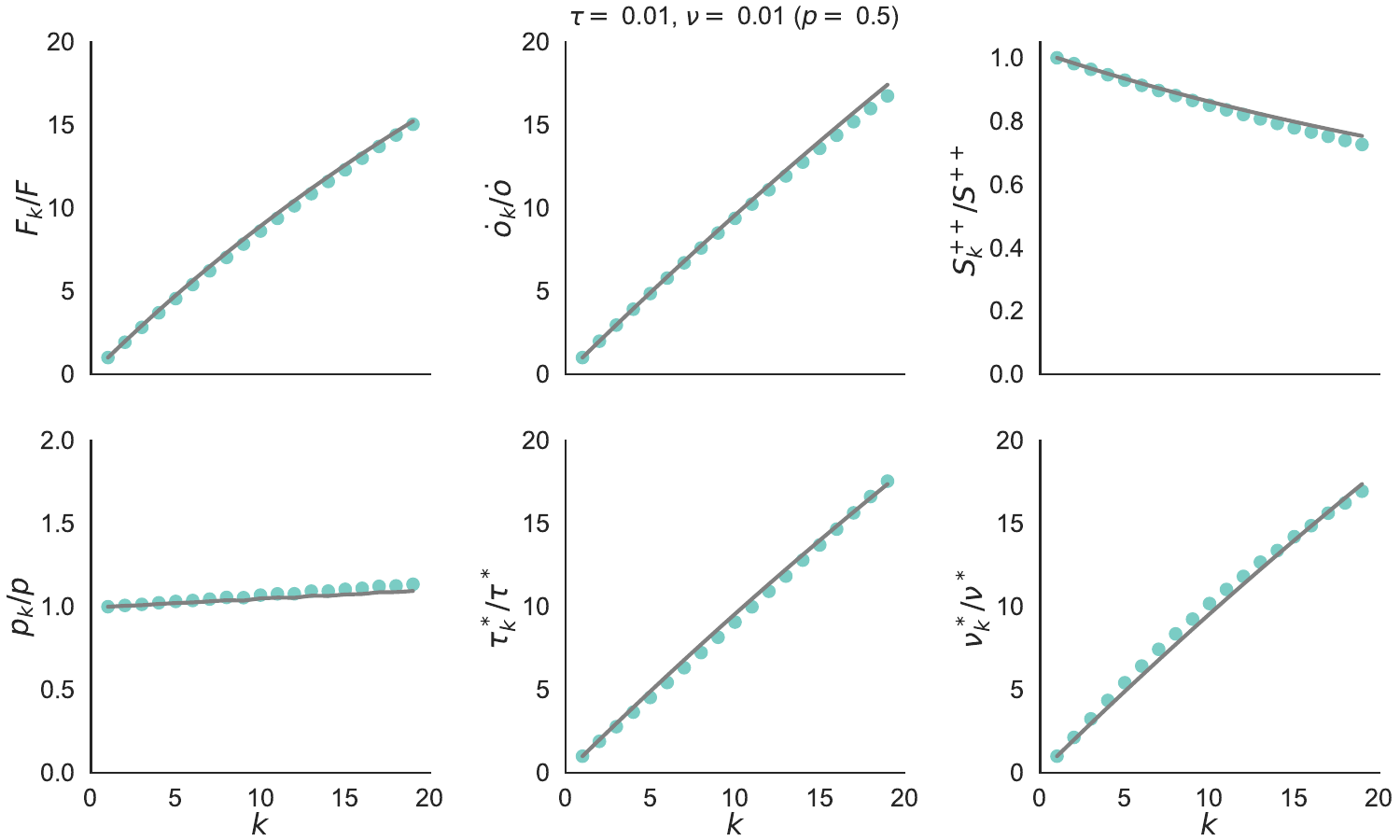}
\caption{\small {\bf Subsampling in model.}
Rank flux $F_k$, turnover rate $\mod_k$, and inertia $S^{++}_k$ (for threshold
$c=0.5$), as well as fitted parameters $\p0_k$, $\ptau^*_k$ and $\pnu^*_k$ as a
function of subsampling parameter $k$, relative to their original values ($F$,
$\mod$, $S^{++}$, $\p0$, $\ptau^*$, and $\pnu^*$) for fixed $\ptau$, $\pnu$,
and $\p0$. Results are both numerical simulations of the model (dots) and their
corresponding analytical approximations (lines), with $\ptau_k$ and $\pnu_k$
given by \eref{eq:subsampParams}. Subsampling increases the rates of
displacement and replacement, leading to larger flux and turnover, and lower
inertia. We only show $k$ values such that $T_{\mathrm{eff}} / T > 0.05$.
Simulations are averages over $10$ realizations (with $T = 10^3$) in a system
of size $N = 10^3$.}
\label{fig:model_sampling}
\end{figure} 

We can gain some insight on the role of subsampling by first analyzing
synthetic ranking lists coming from the model itself. Consider a model dynamics
(with parameters $N$, $\ptau$, and $\pnu$) and its subsampled version (with
effective parameters $\ptau_k$ and $\pnu_k$ for given $k$)
(\fref{fig:model_sampling}). The probability that an element is not replaced
between two subsampled observations of the original dynamics, $(1 - \pnu \Delta
r)^{k N}$, must be equal to the probability of not being replaced between
consecutive observations of the subsampled dynamics, $(1 - \pnu_k \Delta
r)^{N}$. A similar argument applies to $\ptau$ and $\ptau_k$. Taking the limit
$N \to \infty$ and the first-order power series of the exponential, we obtain
\begin{subequations}
\label{eq:subsampParams}
\begin{align}
\ptau_k = 1 - (1 - \ptau)^k, \\
\pnu_k = 1 - (1 - \pnu)^k,
\end{align}
\end{subequations}
meaning that subsampling increases the rates of displacement and replacement in
the model, almost linearly for small enough $\ptau$ and $\pnu$.
\eref{eq:subsampParams} allows us to write approximate analytical expressions
for the subsampled rank flux $F_k$, turnover rate $\mod_k$, and inertia
$S^{++}_k$ by replacing $(\ptau, \pnu)$ with $(\ptau_k, \pnu_k)$ in
\eref{eq:fluxModel}, \eref{eq:openModelSol}, and \eref{eq:successModel},
respectively. As $k$ grows, larger rates of displacement and replacement lead
to an increase in flux and turnover, and a corresponding lower probability to
stay in the top of the ranking. Since some elements enter and leave the ranking
list between subsampled observations, the number of elements ever seen in the
list tends to decrease, leading to a slightly larger $\p0_k$
(\fref{fig:model_sampling}). We note that the approximation of \eref{eq:subsampParams} might become less accurate for some parameter values and as $k$ increases.

\begin{figure}[t]
\centering
\includegraphics[width=\textwidth]{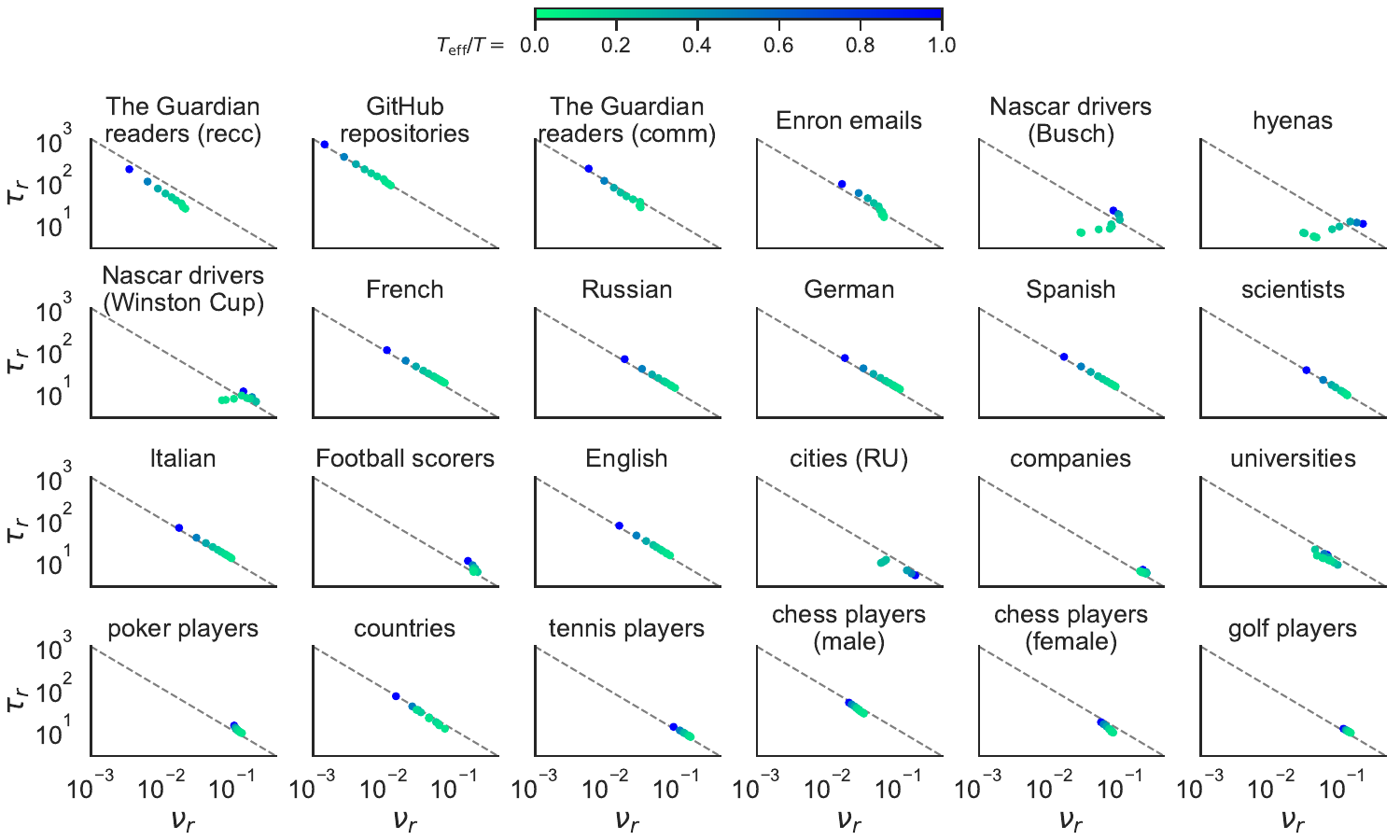}
\caption{\small {\bf Subsampling in open ranking lists.}
Rescaled parameters $\prtau$ and $\prnu$ for varying subsampling parameter $k$, color coded by the effective number of observations $T_{\mathrm{eff}} = \lceil T / k \rceil $ relative to its original value $T$. As subsampling increases, the fitted model estimates movement along the universal curve from the Lévy to the diffusion regimes. We only use $k$ values such that $T_{\mathrm{eff}} / T > 0.1$. Datasets (with $\mod > 10^{-3}$) are ordered from most open (upper row) to least open (lower row) according to mean rank flux $F$ [\eref{eq:fluxMean}].}
\label{fig:sampling}
\end{figure}

We perform the same subsampling process for all open empirical ranking lists,
and further calculate the rescaled parameters $\prtau$ and $\prnu$ for varying
$k$ [by using \eref{eq:rescParams} with the subsampled quantities $\ptau^*_k$
$\pnu^*_k$, $\p0_k$, and $\mod_k$ instead of their original values]
(\fref{fig:sampling}). As subsampling increases, most datasets go downwards
along the universal curve of \eref{eq:paramsConstEq} in $(\prnu,
\prtau)$-space, moving from a regime with a certain amount of Lévy walks to one
more driven by diffusion. Some datasets do not follow this trend [Nascar
drivers, hyenas, cities [RU], companies, universities], which follows from the
condition $\pnu^*_k \sim 0$ in \eref{eq:paramsConstEq} not being fulfilled for
large enough $k$ (see \fref{fig:model_sampling}). We also notice some overlap
between these datasets and the ones showing both large fluctuations and distance to the universal curve in \sref{ssec:fluctuations} (\fref{fig:devs_params} and \fref{fig:devs_curve}), further
suggesting that only a few empirical ranking lists have dynamics not captured
by the model and its universal curve.

\section{On the heterogeneity of rank dynamics} 
\label{sec:heteroDynam}

Our results suggest that rank dynamics are \emph{heterogeneous}: some elements change their rank slower than others, even when the dynamics of the elements themselves (i.e. the scores) is homogeneous. Soccer teams play roughly the same number of games; metro stations are open mostly at the same time of day. So why are rank dynamics heterogeneous? A possibility is that systems require additional mechanisms for maintaining a homogeneous temporality. Still, are there advantages to such a heterogeneity?
When considering open systems in close interaction with their surroundings (and for which the entirety of the associated ranking has a physical meaning, e.g. in the case of a genetic code),
there is evolutionary advantage in preserving the functionality of the most
essential elements (to maintain robustness~\cite{wagner2005robustness}), while allowing
for fast variability of less crucial components (conferring
adaptivity~\cite{heylighen2001science}). An understanding of the balance
between stability and variability has so far been mostly constrained to
critical phase transitions in homogeneous
models~\cite{lee1952statistical,Kauffman1993,aldana2003boolean}. We pose the following hypothesis: if we assume varying timescales of dynamical behavior according to rank (with
elements at the top of a ranking list changing more slowly than the rest),
systems would benefit by having robustness and adaptability at the same time,
independently of critical parameters: ``slower'' elements would provide robustness, while ``faster'' elements would provide adaptivity.
The ability of a simple model to
reproduce rank dynamics in a wide variety of phenomena, regardless of their
domain, suggests that such balanced dynamics can be achieved with random rank change.
Thus, complex systems may need much less to be evolvable than previously
thought (on the basis of homogeneous timescales only), implying that random
variation of natural selection, if heterogeneous, is enough to produce the complex adaptations
seen in evolutionary biology and computer science~\cite{wagner1996perspective}. This hypothesis remains to be explored further, but it might be useful to guide the interpretation of the results presented here.
